\documentclass[twocolumn,amssymb,secnumarabic, prl, aps, nofootinbib]{revtex4-1}

\usepackage{graphicx}
\usepackage{amsfonts}
\usepackage{braket}
\usepackage[dvipsnames]{xcolor}
\usepackage[colorlinks,linkcolor=black,citecolor=black,urlcolor=black,filecolor=black]{hyperref}

\usepackage{times}

\usepackage{amsmath, amssymb, amsthm}
\usepackage{wasysym}
\usepackage{ulem}
\usepackage{lipsum}
\usepackage{algorithm}
\usepackage{algpseudocode}
\usepackage{outlines}
\usepackage[margin=1in]{geometry}
\usepackage{enumerate}
\usepackage{setspace}

\usepackage{wrapfig}

\usepackage{xcolor}
\usepackage[export]{adjustbox}
\usepackage{tikz}
\usetikzlibrary{decorations.pathreplacing}
\usetikzlibrary{arrows,arrows.meta,calc,shapes.geometric,shapes.misc}
\tikzset{
	>=stealth',
	help lines/.style={dashed, thick},
	important line/.style={thick},
	connection/.style={thick, dotted},
}
\renewcommand\thesection{\arabic{section}}
\DeclareMathAlphabet{\mymathbb}{U}{BOONDOX-ds}{m}{n}

\newcommand{\mc}{\mathcal}

\newcommand{\MIS}{\textnormal{MIS}}
\newcommand{\PMIS}{P_\textnormal{MIS}}
\newcommand{\DMIS}{D_{|\textnormal{MIS}|}}
\newcommand{\DMISminusone}{D_{|\textnormal{MIS}|-1}}

\begin{document} 
	
\title{
	Quantum Optimization of Maximum Independent Set using Rydberg Atom Arrays
}

\author
{S. Ebadi$^{1, \ast}$, A. Keesling$^{1,2, \ast}$, M. Cain$^{1, \ast}$, T. T. Wang$^{1}$, H. Levine$^{1, \ddagger}$,
	D. Bluvstein$^{1}$, G. Semeghini$^{1}$, A. Omran$^{1,2}$,  J.-G. Liu$^{1,2}$,
	R. Samajdar$^{1}$,  X.-Z. Luo$^{2,3,4}$, B. Nash$^{5}$, X. Gao$^{1}$,
	B. Barak$^{5}$, E. Farhi$^{6,7}$, S. Sachdev$^{1,8}$,  N. Gemelke$^2$, L. Zhou$^{1,9}$,
	S. Choi$^{7}$, H. Pichler$^{10,11}$, S.-T. Wang$^{2}$, 
	M. Greiner$^{1,\dagger}$, V. Vuleti\'{c}$^{12,\dagger}$, M. D. Lukin$^{1,\dagger}$}

%\affiliation{$^1$Department of Physics, Harvard University, Cambridge, MA 02138, USA \\ $^2$Institute for Theoretical Physics, University of Innsbruck, Innsbruck A-6020, Austria \\ $^3$Institute for Quantum Optics and Quantum Information, Austrian Academy of Sciences, Innsbruck A-6020, Austria \\ $^4$ QuEra Computing Inc., Boston, MA 02135, USA \\ $^5$ Department of Physics and Research Laboratory of Electronics, Massachusetts Institute of Technology, Cambridge, MA 02139, USA}

\affiliation{\mbox{$^1$Department of Physics, Harvard University, Cambridge, MA 02138, USA}\\ \mbox{$^2$ QuEra Computing Inc., Boston, MA 02135, USA}\\ \mbox{$^3$Department of Physics and Astronomy, University of Waterloo, Waterloo N2L 3G1, Canada}\\ \mbox{$^4$Perimeter Institute for Theoretical Physics, Waterloo, Ontario N2L 2Y5, Canada}\\ \mbox{$^5$School of Engineering and Applied Science, Harvard University, Cambridge, MA 02138, USA} \\ \mbox{$^{6}$Google Quantum AI, Venice, CA 90291}\\ \mbox{$^{7}$Center for Theoretical Physics, Massachusetts Institute of Technology,Cambridge, MA 02139}\\ \mbox{$^{8}$School of Natural Sciences, Institute for Advanced Study, Princeton, NJ 08540, USA}\\ \mbox{$^9$Walter Burke Institute for Theoretical Physics, California Institute of Technology, Pasadena, CA 91125}\\ \mbox{$^{10}$Institute for Theoretical Physics, University of Innsbruck, Innsbruck A-6020, Austria}\\ \mbox{$^{11}$Institute for Quantum Optics and Quantum Information, Austrian Academy of Sciences, Innsbruck A-6020, Austria}\\ \mbox{$^{12}$Department of Physics and Research Laboratory of Electronics, Massachusetts Institute of Technology, Cambridge, MA 02139, USA}\\ \mbox{$^{\ast}$ These authors contributed equally to this work}\\ \mbox{$^{\ddagger}$ \textit{Current affiliation}: AWS Center for Quantum Computing, Pasadena, CA 91125}\\ \mbox{$^{\dagger}$ Corresponding authors}}

\begin{abstract}
	% CURRENT LENGTH 123 WORDS (125word limit) Feb15th, 12:17pm
	Realizing quantum speedup for practically relevant, computationally hard  problems is a central challenge in quantum information science. Using Rydberg atom arrays with up to 289 qubits in two spatial dimensions, we experimentally investigate quantum algorithms for solving the Maximum Independent Set problem. We use a hardware-efficient encoding associated with Rydberg blockade, realize closed-loop optimization to test several variational  algorithms, and subsequently apply them to systematically explore a class of graphs with programmable connectivity. We find the problem hardness is controlled by the solution degeneracy and number of local minima, and experimentally benchmark the quantum algorithm's performance against classical simulated annealing. On the hardest graphs, we observe a superlinear quantum speedup in finding exact solutions  in the deep circuit regime and analyze its origins. 
	
%	\textit{One sentence summary: We experimentally observe quantum speedup for solving the Maximum Independent Set problem using Rydberg atom arrays.} %limit 125 chatacter
\end{abstract}

\maketitle

\date{}

Combinatorial optimization is ubiquitous in many areas of science and technology. Many such problems have been shown to be computationally hard and form the basis for understanding complexity classes in modern computer science~\cite{sipser13}. The use of quantum machines to accelerate solving such problems has been theoretically explored for over two decades using a variety of quantum algorithms  \cite{farhi_quantum_2000,farhi_quantum_2014, lidar_albash_review_2018}. Typically, a relevant cost function is encoded in a quantum Hamiltonian \cite{Lucas_Ising_Formulation}, and its low-energy state is sought starting from a generic initial state either through an adiabatic evolution~\cite{farhi_quantum_2000} or a variational approach~\cite{farhi_quantum_2014}, via closed optimization loops \cite{Wecker_Training_Optimizer, Kokail2019}. The computational performance of such algorithms has been investigated theoretically \cite{Barahona_complexity_spin_glass, lidar_albash_review_2018,
	bapst_2012,Farhi_NP_complete, farhi_random_regular_2012, knysh_2016, young_2010} and experimentally \cite{harrigan_quantum_2021, pagano_quantum_2020, graham2022demonstration} in small quantum systems with shallow quantum circuits, or in systems lacking the many-body coherence believed to be central for quantum  advantage \cite{Lidar_Martinis_Speedup, Speedup_Spin_Glass_DWave}. However, 
these studies %make it challenging to predict
offer only limited insights into algorithms' performances in the most interesting regime involving large system sizes and high circuit depths ~\cite{farhi_2020_quantum, chou_overlap_gap_2021}.

\begin{figure*}
	\centering
	\includegraphics[width=160mm]{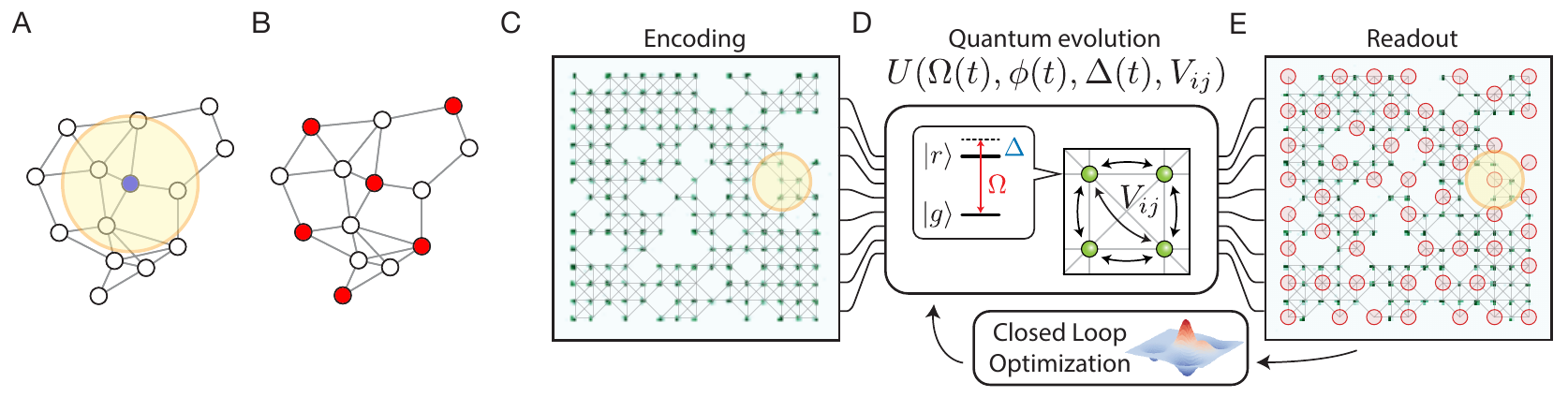}
	\caption{\textbf{Hardware-efficient encoding of the maximum independent set using Rydberg atom arrays.} \textbf{A.} An example of a unit disk graph, with any single vertex (e.g. the blue vertex) being connected to all other vertices within a disk of unit radius. \textbf{B.} A corresponding MIS solution (denoted by the red nodes). 		\textbf{C.} The MIS problem is encoded with atoms placed at the vertices of the target graph and with interatomic spacing chosen such that the unit disk radius of the graph corresponds to the Rydberg blockade radius. Shown is an example fluorescence image of atoms, with gray lines added to indicate edges between connected vertices. \textbf{D.} The system undergoes coherent quantum many-body evolution under a programmable laser drive $\left( \Omega(t),  \phi(t) , \Delta(t) \right)$ and long-range Rydberg interactions $V_{ij}$. \textbf{E.} A site-resolved projective measurement reads out the final quantum many-body state, with atoms excited to the Rydberg state (red circles) corresponding to vertices forming an independent set. A classical optimizer uses the results to update the parameters of the quantum evolution $\left( \Omega(t),  \phi(t) , \Delta(t) \right)$ to maximize a figure of merit for finding the MIS.} 
	\label{fig_1}
\end{figure*}

Here we address this challenge using a quantum device based on coherent, programmable arrays of neutral atoms trapped in optical tweezers to investigate quantum optimization algorithms for systems ranging from $39$ to $289$ qubits, and effective depths sufficient for the quantum correlations to spread across the entire graph. Specifically, we focus on Maximum Independent Set (MIS), a paradigmatic NP-hard optimization problem~\cite{garey_computers_1979}. It involves finding the largest independent set of a graph---a subset of vertices such that no edges connect any pair in the set. An important class of such MIS problems involves unit disk graphs, which are defined by vertices on a two-dimensional plane with edges connecting all pairs of vertices within a unit distance of one another (Fig.~\ref{fig_1}A,B). Such instances arise naturally in problems associated with geometric constraints that are important for many practical applications, such as modeling wireless communication networks \cite{unit_disk_network, clark_unit_1990}. While there exist polynomial-time classical algorithms to find approximate solutions to the MIS problem on such graphs \cite{FPTAS_unit_disk}, solving the problem exactly is known to be NP-hard in the worst case \cite{clark_unit_1990, SI}. 

\noindent \textbf
{Maximum Independent Set on Rydberg Atom Arrays.}
Our approach utilizes a two-dimensional atom array described previously in \cite{ebadi_quantum_2020}. Excitation from a ground state $|0\rangle$ into a Rydberg state $|1\rangle$ is utilized for hardware-efficient encoding of the unit disk MIS  problem~\cite{pichler_quantum_2018}. For a particular graph, we create a geometric configuration of atoms using optical tweezers such that each atom represents a vertex. The edges are drawn according to the unit disk criterion for a unit distance given by the Rydberg blockade radius $R_b$ (Fig.~\ref{fig_1}C), the distance 
within which excitation of more than one atom to the Rydberg state is prohibited due to strong interactions \cite{lukin_dipole_2001}. The Rydberg blockade mechanism thus restricts the evolution primarily to the subspace spanned by the states that obey the independent set constraint of the problem graph. Quantum algorithms for optimization are implemented via global atomic excitation using homogeneous laser pulses with a time-varying Rabi frequency (and a time-varying phase) $\Omega(t) e^{i\phi(t)}$ and detuning $\Delta(t)$ (Fig.~\ref{fig_1}D). The resulting quantum dynamics is governed by the Hamiltonian $H= H_q + H_{\rm cost}$, with the quantum driver $H_q$ and the cost function $H_{\rm cost}$ given by
\begin{align}\label{Eq:Hamiltonian} 
	H_q = \frac{\hbar}{2} \sum_{i}(\Omega(t) e^{i\phi(t)} \ket{0}_{i}\bra{1} + {\rm h.c.}),\nonumber \\
	 \quad  H_{\rm cost} = - \hbar \Delta(t) \sum_{i} n_i
	+ \sum_{i<j}V_{ij}n_i n_j,\end{align}
where $n_i = \ket{1}_i\bra{1}$,  and $V_{ij} = V_0/(|r_{i}-r_{j}|)^{6}$ is the interaction potential that sets the blockade radius $R_b$ and
%$R_b = (V_0/\max(\Omega))^{1/6}$ 
determines the connectivity of the graph. For a positive laser detuning $\Delta$, the many-body ground state of the cost function Hamiltonian maximizes the total number of qubits in the Rydberg state under the blockade constraint, corresponding to the MIS of the underlying unit disk graph~\cite{pichler_quantum_2018} (Fig.~\ref{fig_1}E). Remarkably, this Hamiltonian can effectively encode the MIS as the ground state even with the finite blockade energy and long-range interaction tails~\cite{pichler_quantum_2018}.

\begin{figure*}[t]
	\includegraphics[width=110mm]{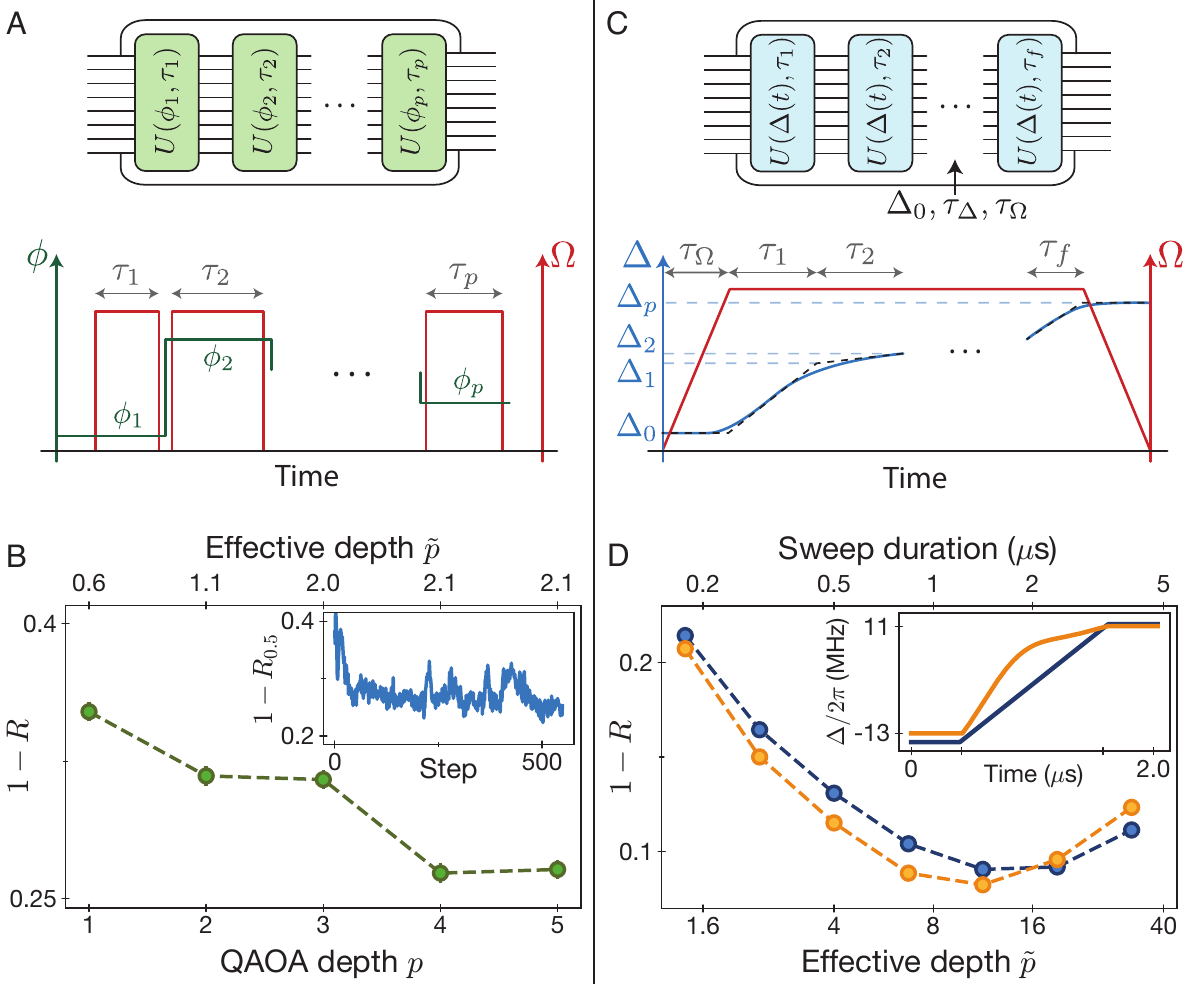}
	\caption{\textbf{Testing variational quantum algorithms.} \textbf{A.} Implementation of the quantum approximate optimization algorithm (QAOA), consisting of sequential layers of resonant pulses with variable duration $\tau_i$ and laser phase $\phi_i$. \textbf{B.} Variational optimization of QAOA parameters results in a decrease in approximation error $1-R$, up to depth $p=4$ (inset: example performance of quantum-classical closed-loop optimization at $p=5$). Approximation error calculated using the top 50 percentiles of independent set sizes ($1-R_{0.5}$) is used as the figure of merit during optimization \cite{SI}. \textbf{C.} Quantum evolution can also be parametrized as a variational quantum adiabatic algorithm (VQAA) using a quasi-adiabatic pulse with a piecewise-linear sweep of detuning $\Delta(t)$ at constant Rabi coupling $\Omega(t)$. $\Omega(t)$ is turned on and off within $\tau_\Omega$, and a low-pass filter with timescale $\tau_\Delta$ is used to smoothen the $\Delta(t)$ sweep. \textbf{D.} Performance of a rescaled piecewise-linear sweep as a function of its effective depth $\tilde{p} = (\tau_1+...+\tau_f)$/$\tau_{\pi}$. Variational optimization of a three-segment (orange) piecewise-linear pulse improves on the performance of a simple one-segment linear (blue) pulse as well as the best results from QAOA (inset: detuning sweep profiles for one-segment (blue) and three-segment (orange) optimized pulses for a total pulse duration of $2.0$~$\mu$s). Error bars for approximation ratio $R$ are the standard error of the mean here and throughout the text, and are smaller than the points.}
	\label{fig_2_new}
\end{figure*}

\newpage

\noindent \textbf
{Closed-loop Variational Optimization.}
In the experiment we deterministically prepare graphs with vertices occupying $80$\% of an underlying square lattice, with the blockade extending across nearest and next-nearest (diagonal) neighbors (Fig.~\ref{fig_1}C). This allows us to explore a class of nonplanar graphs, for which finding the exact solution of MIS is NP-hard for worst-case instances \cite{SI}. To prepare quantum states with a large overlap with the MIS solution space, we employ a family of variational quantum optimization algorithms using a quantum-classical optimization loop. We place atoms at positions defined by the vertices of the chosen graph, initialize them in state $\ket{0}$, and implement a coherent quantum evolution corresponding to the specific choice of variational parameters (Fig.~\ref{fig_1}D). Subsequently, we sample the wavefunction with a projective measurement and determine the size of the output independent set by counting the number of qubits in $\ket{1}$, utilizing classical post-processing to remove blockade violations and reduce detection errors \cite{SI} (Fig.~\ref{fig_1}E). This procedure is repeated multiple times to estimate the mean independent set size $\langle \sum_i n_i \rangle$ of the sampled wavefunction, the approximation ratio $R \equiv \langle \sum_i n_i \rangle/|\text{MIS}|$, and the probability $\PMIS$ of observing an MIS (where $|\text{MIS}|$ denotes the size of the maximum independent set of the graph). The classical optimizer tries to maximize $\langle \sum_i n_i \rangle$ by updating the variational parameters in a closed-loop hybrid quantum-classical optimization protocol \cite{SI} (Fig.~\ref{fig_1}D).

We test two algorithm classes, defined by different parametrizations of the quantum driver and the cost function in Eq.~(\ref{Eq:Hamiltonian}). The first approach consists of resonant ($\Delta = 0$) laser pulses of varying durations $\tau_i$ and phases $\phi_i$ (Fig.~\ref{fig_2_new}A). This algorithm closely resembles the canonical Quantum Approximate Optimization Algorithm (QAOA)~\cite{farhi_quantum_2014}, but instead of exact single-qubit rotations, resonant driving generates an effective many-body evolution within the subspace of independent sets associated with the blockade constraint \cite{SI}. Phase jumps between consecutive pulses implement a global phase gate~\cite{mckay_efficient_2017}, with a phase shift proportional to the cost function of the MIS problem in the subspace of independent sets. Taken together, these implement the QAOA, where each pulse duration $\tau_i$ and phase $\phi_i$ are used as a variational parameters. 

The performance of QAOA as a function of depth $p$ (the number of pulses) is shown in Fig.~\ref{fig_2_new}B for an instance of a $179$-vertex graph embedded in a $15\times 15$ lattice. We find that the approximation ratio grows as a function of the number of pulses up to $p=4$, and increasing the depth further does not appear to lead to better performance (Fig.~\ref{fig_2_new}B). As discussed in \cite{SI}, we attribute these performance limitations to the difficulty of finding the optimal QAOA parameters for large depths within a limited number of queries to the experiment, leakage out of the independent set subspace during resonant excitation due to imperfect blockade associated with the finite interaction energy between next-nearest neighbors, as well as laser pulse imperfections. 

\begin{figure*}[t]
	\centering
	\includegraphics[width = 110mm]{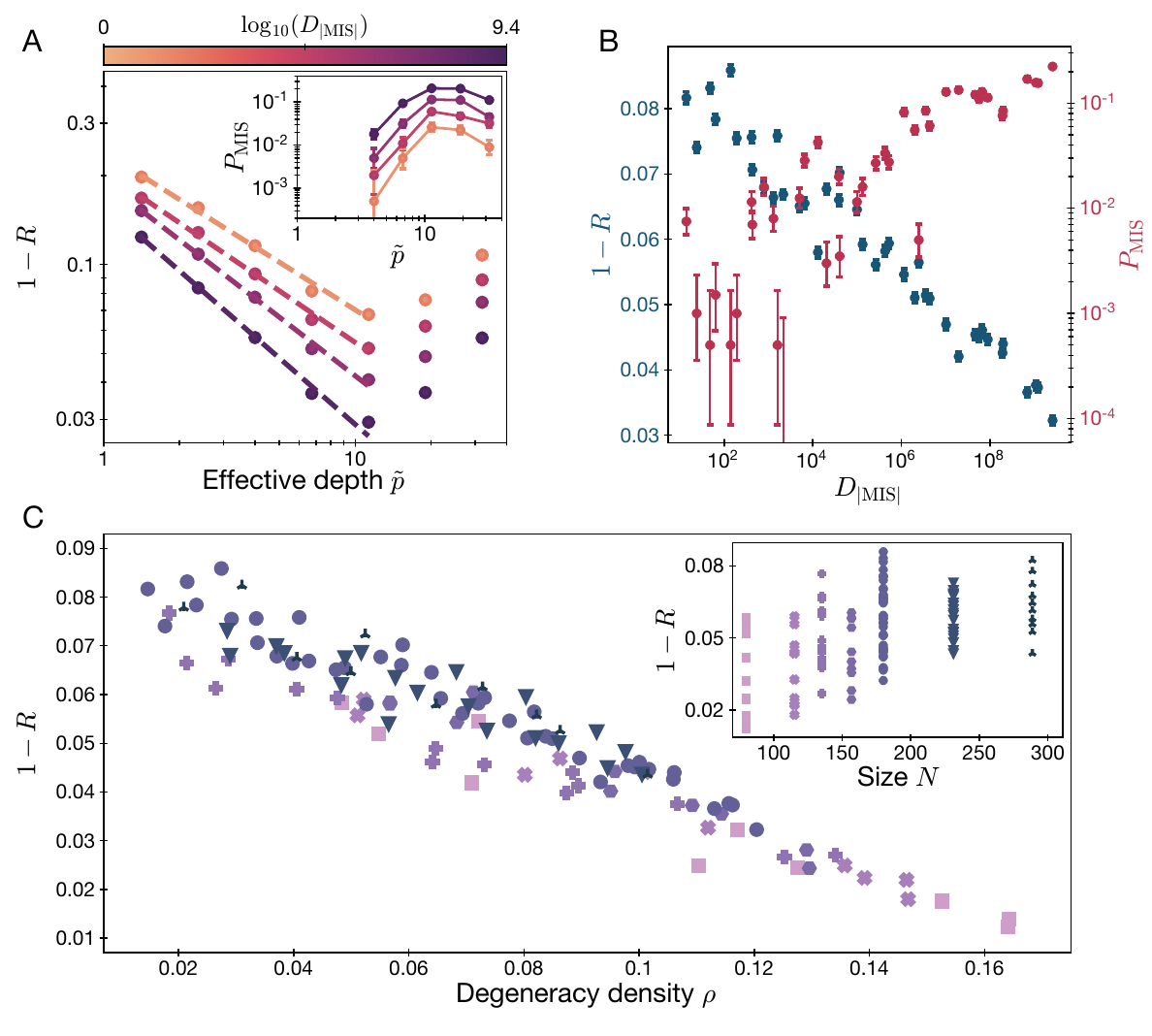}
	\caption{\textbf{Quantum algorithm performance across different graphs.} 
		\textbf{A.} The approximation error $1 - R$ for an optimized quasi-adiabatic sweep plotted as a function of effective depth $\tilde{p}$ on four graphs of the same size ($N = 180$ vertices), showing strong dependence on the number of MIS solutions (MIS degeneracy) $\DMIS$ (inset: corresponding MIS probability $\PMIS$ vs. $\tilde{p}$). \textbf{B.} At a fixed depth $\tilde{p} = 20$, $1-R$ and $\PMIS$ for various $180$-vertex graphs are strongly correlated with $\DMIS$. \textbf{C.} At the same effective depth $\tilde{p} = 20$, $1-R$ for $115$ graphs of different sizes ($N = 80$--$289$) and MIS degeneracies $\DMIS$ exhibit universal scaling with the degeneracy density $\rho \equiv \log(\DMIS)/N$ (inset: data plotted as a function of $N$). Error bars for $\PMIS$, here and throughout the text, denote the $68\%$ confidence interval.
	}
	\label{fig_2}
\end{figure*}

The second approach is a variational quantum adiabatic algorithm (VQAA) \cite{farhi_quantum_2000, schiffer2021adiabatic}, related to methods previously used to prepare quantum many-body ground states \cite{ebadi_quantum_2020, semeghini_probing_2021, Scholl2021}.  In this approach, we sweep the detuning $\Delta$ from an initial negative detuning $\Delta_0$ to a final large positive value $\Delta_f$ at constant Rabi frequency $\Omega$, along a piecewise-linear schedule characterized by a total number of segments $f$, the duration $\tau_i$ of each, and the end detuning $\Delta_i$ of each segment. Moreover, we turn on the coupling $\Omega$ in duration $\tau_\Omega$ and smoothen the detuning sweep using a low-pass filter with a characteristic filter time $\tau_\Delta$ (Fig.~\ref{fig_2_new}C), both of which minimize nonadiabatic excitations and serve as additional variational parameters. For this evolution, we define an effective circuit depth $\Tilde{p}$ as the duration of the sweep ($T= \tau_1+...+\tau_f$) in units of the $\pi$-pulse time $\tau_{\pi}$, which is the time required to perform a spin flip operation. 

We find that with only $3$ segments optimized for an effective depth of $\tilde{p} = 10$ (Fig.~\ref{fig_2_new}D inset), the optimizer converges to a pulse  that substantially outperforms the QAOA approach described above. Furthermore, the optimized pulse shows a better performance compared to a linear (one-segment) detuning sweep of the same $\tilde{p}$ (Fig.~\ref{fig_2_new}D). We find that similar pulse shapes produce high approximation ratios for a variety of graphs (see e.g., Fig.~S8C), consistent with theoretical predictions of pulse shape concentration \cite{brandao_2018, zhou_quantum_2020, chou_overlap_gap_2021, SI}. At large sweep times ($\Tilde{p}>15$), we observe a turn-around in the performance likely associated with decoherence \cite{SI}. For the remainder of this work, we focus on the quantum adiabatic algorithm for solving the MIS problem.\\

\noindent \textbf
{Quantum Optimization on Different Graphs.}
The experimentally optimized quasi-adiabatic sweep (depicted in Fig.~\ref{fig_2_new}D) was applied to $115$ randomly generated graphs of various sizes ($N$ = $80$--$289$ vertices). For graphs of the same size ($N = 180$), the approximation error $1-R$ decreases and the probability of finding an MIS solution $\PMIS$ increases with the effective circuit depth at early times, with the former showing a power-law relation (Fig.~\ref{fig_2}A).
%For graphs of the same size ($N = 180$), the approximation error $1-R$  decreases and the probability to find an optimal solution $\PMIS$ initially increase with the effective circuit depth, with the former showing a power-law relation (Fig.~\ref{fig_2}A).
We find a strong correlation between the performance of the quantum algorithm on a given graph and its total number of MIS solutions, which we refer to as the MIS degeneracy $\DMIS$. This quantity is calculated classically using a novel tensor network algorithm \cite{SI, liu_tensor_2021} and varies by nine orders of magnitude across different $180$-vertex graphs. We observe a clear logarithmic relation between $\DMIS$ and the approximation error $1-R$, accompanied by a nearly three-orders-of-magnitude variation of $\PMIS$ at a fixed depth ${\Tilde p}= 20$  (Fig.~\ref{fig_2}B). Note that $\PMIS$ does not scale linearly with the MIS degeneracy, as would be the case for a naive algorithm that samples solutions at random. Figure~\ref{fig_2}C shows the striking collapse of $1-R$ as a function of the logarithm of the MIS degeneracy normalized by the graph size, $\rho \equiv \log(\DMIS)/N$. This quantity, a measure of MIS degeneracy density, determines the hardness in approximating solutions for the quantum algorithm at shallow depths.

%Alternate versions of this are moved to the bottom of the

These observations can be modeled as resulting from a Kibble-Zurek-type mechanism where the quantum algorithm locally solves the graph in domains whose sizes are determined by the evolution time and speed at which quantum information propagates \cite{zurek_dynamics_2005, Lieb_Robinson_Bound}. In \cite{SI}, we show that the scaling of the approximation error with depth can originate from the conflicts between local solutions at the boundaries of these independent domains. In graphs with a large degeneracy density $\rho$, there may exist many MIS configurations that are compatible with the local ordering in these domains. This provides a possible mechanism to reduce domain walls at their boundaries (Fig.~S14) and decrease the approximation error. Such a scenario would predict a linear relation between $1-R$ and $\rho$ at a fixed depth, which is consistent with our observations (Figs.~\ref{fig_2}C,~S15).\\

\noindent \textbf
{Benchmarking Against Simulated Annealing.}
To benchmark the results of the quantum optimization against a classical algorithm, we use simulated annealing (SA), a general-purpose algorithm widely used in solving combinatorial optimization problems \cite{Kirkpatrick_SA}. SA seeks to minimize the energy of a cost Hamiltonian by thermally cooling a system of classical spins while maintaining thermal equilibrium. Our highly optimized variant of SA stochastically updates local clusters of spins using the Metropolis-Hastings \cite{Metropolis} update rule, rejecting energetically unfavorable updates with a probability dependent on the energy cost and the instantaneous temperature  \cite{SI}. We use collective updates under the MIS Hamiltonian cost function (Eq.~S15), which applies an optimized uniform interaction energy to each edge, penalizing states that violate the independent set criterion \cite{SI}. The annealing depth $p_{\text{SA}}$ is defined as the average number of attempted updates per spin.

\begin{figure*}
	\includegraphics[width=150mm]{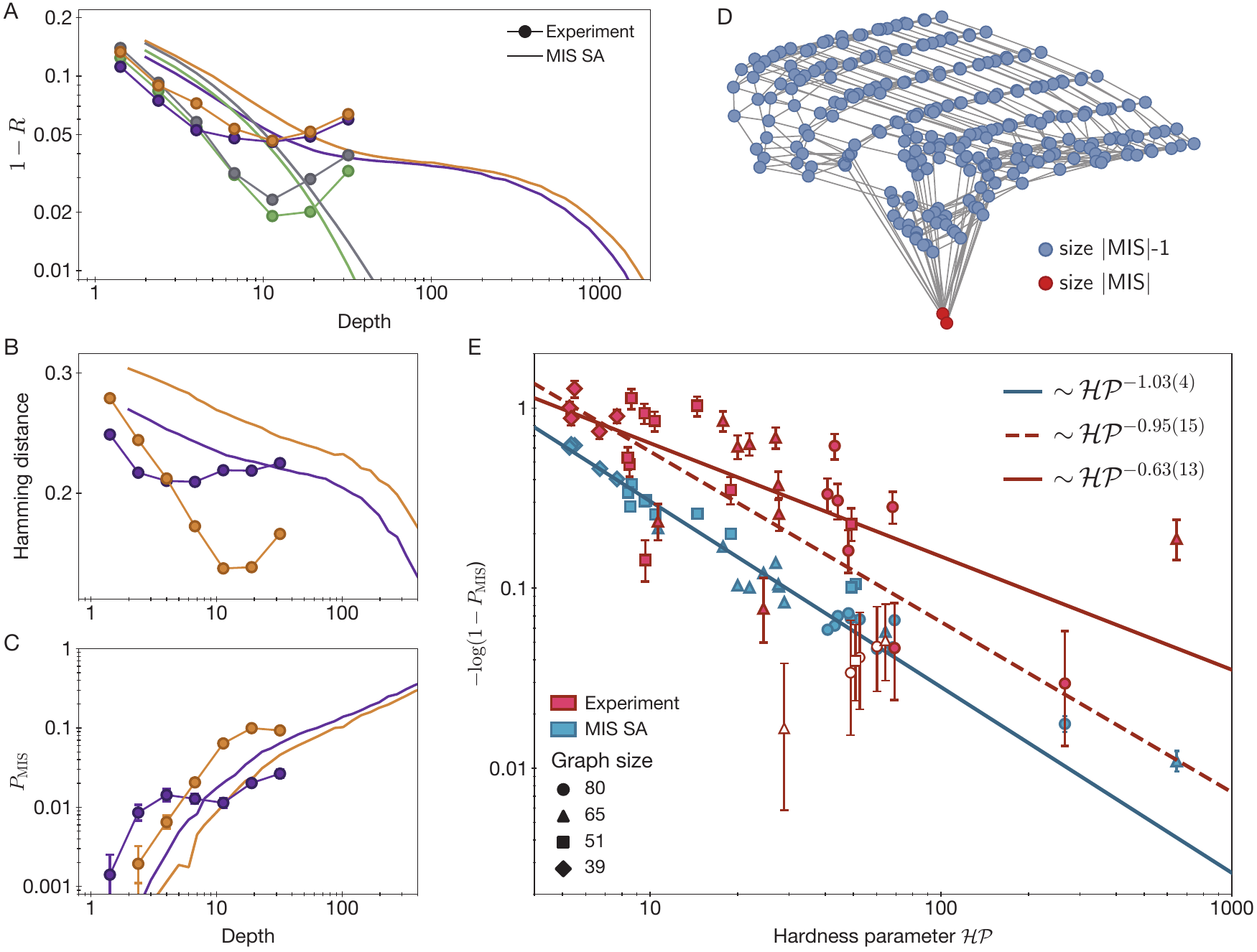}
	\centering
	\caption{\textbf{Benchmarking the quantum algorithm against classical simulated annealing.} 
		\textbf{A.} Performance of the quantum algorithm, and the optimized simulated annealing with the MIS Hamiltonian, shown as a function of depth ($\tilde{p}$ for quantum algorithm and $p_{\text{SA}}$ for simulated annealing) for four 80-vertex graphs. Green ($\mathcal{HP} = 1.8$, $\rho = 0.13$) and grey ($\mathcal{HP} = 2.1$, $\rho = 0.11$) graphs are easy for the quantum and classical algorithm; however, purple ($\mathcal{HP} = 69$,  $\rho  = 0.08$) and gold ($\mathcal{HP} = 68$, $\rho = 0.06$) are significantly harder and show a plateau at $R = (|\text{MIS}|-1)/|\text{MIS}|$, i.e., independent sets with one less vertex than the MIS.
		\textbf{B, C.} One of the hard graphs (gold) shows much better quantum scaling of average normalized Hamming distance to the closest MIS, and MIS probability ($\PMIS$) compared to the other graph (purple). In contrast, the performance of SA (lines) remains similar between the two graphs.
		\textbf{D.} Configuration graph of independent sets of size $|\text{MIS}|$ and $|\text{MIS}|-1$ for an example 39-vertex graph ($\mathcal{HP} = 5$), where the edges connect two configurations if they are separated by one or two steps of simulated annealing. At low temperatures, simulated annealing finds the MIS solutions by a random walk on this configuration graph. 
		\textbf{E.} $-\log(1-\PMIS)$ for instance-by-instance optimized quantum algorithm (crimson) and simulated annealing (teal) reached within a depth of $32$, for $36$ graphs selected from the top two percentile of hardness parameter $\mathcal{HP}$ for each size.
		Power-law fits to the SA (teal, $\sim \mathcal{HP}^{-1.03(4)}$) and the quantum data (dashed crimson  line, $\sim \mathcal{HP}^{-0.95(15)}$) are used to compare scaling performance with graph hardness $\mathcal{HP}$.
		If only graphs with minimum energy gaps large enough to be resolved in the duration of the quantum evolution are considered ($\delta_{\min} >1/T $, excluding hollow data points), the fit (solid crimson line) shows a superlinear  speedup $\sim \mathcal{HP}^{-0.63(13)}$ over optimized simulated annealing.
	}
	\label{fig_4}
\end{figure*}

We compare the quantum algorithm and SA on two metrics:  the approximation error $1-R$, and  the probability of sampling an exact solution $\PMIS$, which determines the inverse of time-to-solution. As shown in Figure~\ref{fig_4}A,  for relatively shallow depths and moderately hard graphs, optimized SA results in approximation errors similar to those observed on the quantum device. In particular, we find that the hardness in approximating the solution for short SA depths is also controlled by degeneracy density $\rho$ (Fig.~S18A,B). However, some graph instances appear to be considerably harder for SA compared to the quantum algorithm at higher depths (see e.g. gold and purple curves in Fig.~\ref{fig_4}A).

Detailed analysis of the SA dynamics for graphs with low degeneracy densities $\rho$ reveals that for some instances, the approximation ratio displays a plateau at $R = (|\text{MIS}|-1)/|\text{MIS}|$, corresponding to independent sets with one less vertex than the MIS (Fig.~\ref{fig_4}A, gold and purple solid lines). Graphs displaying this behaviour have a large number of local minima with independent set size $|\text{MIS}|-1$, in which SA can be trapped up to large depths. By analyzing the dynamics of SA at low temperatures as a random walk among $|\text{MIS}|-1$ and $|\text{MIS}|$ configurations (Fig.~\ref{fig_4}D), we show in \cite{SI} that the ability of SA to find a global optimum is limited by the ratio of the number of suboptimal independent sets of size $|\text{MIS}|-1$ to the number of ways to reach global minima, resulting in  a ``hardness parameter"  $\mathcal{HP} =  {\DMISminusone}/({|\text{MIS}| \DMIS})$ (Fig.~\ref{fig_4}E). This parameter determines the mixing time for the Markov chain describing the SA dynamics at low temperatures, and it appears to increase exponentially with the system size for the hardest graphs (Fig.~S11). This suggests that a large number of local minima causes SA to take an exponentially long time to find the exact MIS for the hardest cases as $N$ grows. \\

\begin{figure*}
	\centering
	\includegraphics[width=150mm]{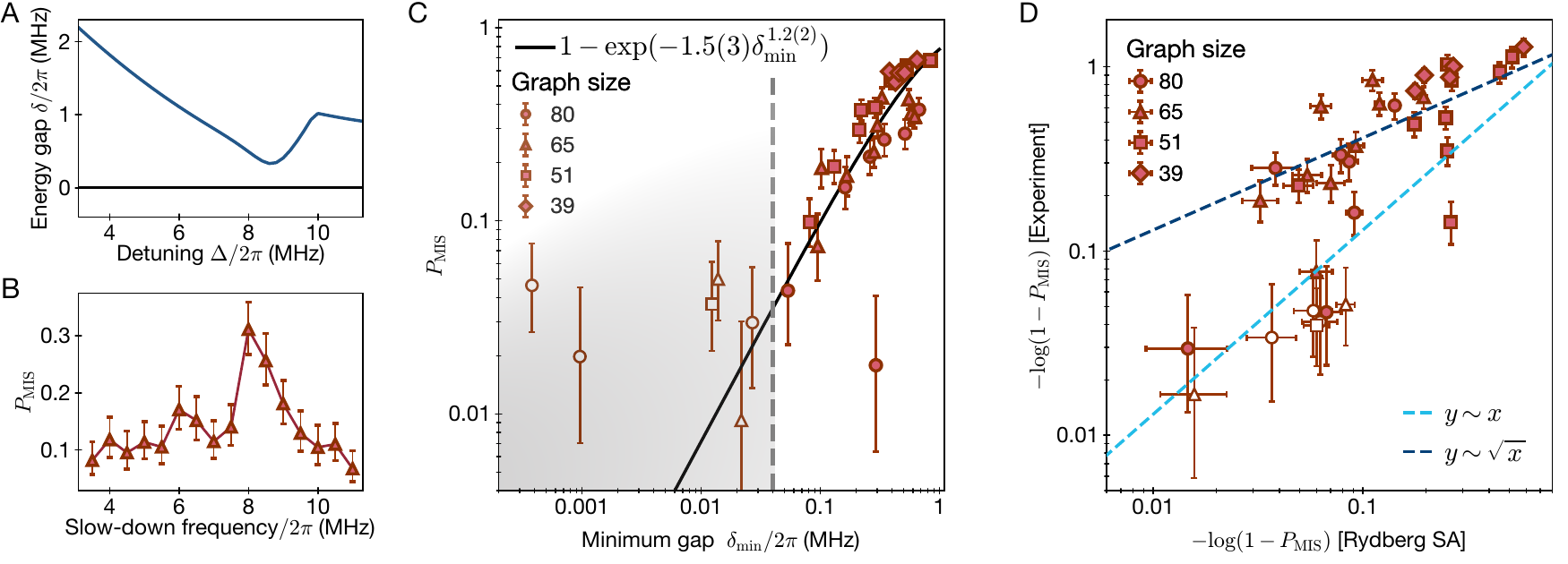}
	\caption{\textbf{Understanding hardness for the quantum algorithm.} \textbf{A.} Energy gap between the ground (black) and first-excited (blue) states, calculated using DMRG for a graph of $65$ atoms. \textbf{B.} To maximize $\PMIS$ for hard graphs, the frequency at which the detuning sweep is slowed down is varied  (see Fig.~S9). The largest $\PMIS$ corresponds to a slow-down frequency close to the location of the minimum gap. \textbf{C.} Measured $\PMIS$ for a fixed effective depth $\tilde{p} = 32$ as a function of the calculated minimum gap $\delta_{\min}$. For many instances the relation is well-described by the Landau-Zener prediction for quasi-adiabatic ground state preparation. The shaded region corresponds to when the gap is too small ($\delta_{\min} < 1/T$) to be properly resolved relative to the quantum evolution time, and points in this region are excluded from the fit both here and in the solid crimson line in Fig.~\ref{fig_4}E.
		\textbf{D.} Scaling of $-\log(1-\PMIS)$ observed in the experiment versus in simulated annealing under the classical Rydberg cost function, Eq.~(S14), for best $\PMIS$ reached within a depth of 32.
		These results are consistent with a nearly quadratic speedup for a subset of graphs where $\delta_{\min} > 1/T$.}
	\label{fig_5}
\end{figure*}

\noindent \textbf {Quantum speedup on the hardest graphs.}
We now turn to study the algorithms' ability to find exact solutions on the hardest graphs (with up to $N=80$), chosen from graphs in the top two percentile of the hardness parameter $\mathcal{HP}$ (Fig.~S11). We find that for some of these graphs (e.g. gold curves in Fig.~\ref{fig_4}A-C), the quantum algorithm quickly approaches the correct solutions, reducing the average Hamming distance (number of spin flips normalized by $N$) to the closest MIS and increasing $\PMIS$, while SA remains trapped in local minima at a large Hamming distance from any MIS. For other instances (e.g. purple curves in Fig.~\ref{fig_4}A-C) both the quantum algorithm and SA struggle to find the correct solution. Moreover, in contrast to our earlier observations  suggesting variational  parameter concentration for generic graphs, we find that for these hard instances, the quantum algorithm needs to be optimized for each graph individually by scanning the slow-down point of the detuning sweep $\Delta (t)$ to maximize $\PMIS$ (Fig.~\ref{fig_5}A,B, and S9 \cite{SI}).

Figure~\ref{fig_4}E shows the resulting highest $\PMIS$ reached within a depth of $32$ for each hard graph instance as a function of the classical hardness parameter $\mathcal{HP}$. For simulated annealing, we find the scaling $\PMIS = 1 - \exp(-C~\mathcal{HP}^{-1.03(4)})$, where $C$ is a positive fitted constant, which is in good agreement with theoretical expectations~\cite{SI}. While for many instances the quantum algorithm outperforms SA, there are significant instance-by-instance variations, and on average, we observe a similar scaling $\PMIS =  1 - \exp(-C~\mathcal{HP}^{-0.95(15)})$ (dashed red line). 

To understand these observations, we carried out detailed analyses of both classical and quantum algorithms' performance for hard graph instances. Specifically, in \cite{SI} we show that for a broad class of SA algorithms with both single-vertex and correlated updates, the scaling is at best $\PMIS = 1 -\exp(-C~\mathcal{HP}^{-1})$ (where $C$ generally could have polynomial dependence on the system size), indicating %that our version of SA scales optimally.
that the observed scaling of our version of SA is close to optimal. To gain insight into the origin of the quantum scaling, we numerically compute the minimum energy gap $\delta_{\text{min}}$ during the adiabatic evolution using density-matrix renormalization group (Fig.~\ref{fig_5}A, \cite{SI}). Figure~\ref{fig_5}C shows that the performance of the quantum algorithm is mostly well-described by quasi-adiabatic evolution with transition probability out of the ground state governed by the minimum energy gap, according to the Landau-Zener formula $\PMIS = 1-\exp{(-A \delta_{\min}^\eta)}$ for a constant $A$, and $\eta = 1.2(2)$\cite{landau_lifshitz}. This observation suggests that our quantum algorithm achieves near-maximum efficiency, consistent with the smallest possible value of $\eta = 1$ obtained for optimized adiabatic following \cite{Roland_optimized_Grover}. 

By focusing only on instances with large enough spectral gaps such that the evolution time $T$ obeys the ``speed limit'' determined by the uncertainty principle ($\delta_{\min}>1/T$) associated with Landau-Zener scaling \cite{landau_lifshitz}, we find an improved quantum algorithm scaling  $\PMIS = 1 - \exp(-C~\mathcal{HP}^{-0.63(13)})$ (Fig.~\ref{fig_4}E solid red line). Since $1/(-\log(1-\PMIS)) \approx 1/\PMIS$ is proportional to the runtime sufficient to find a solution by repeating the experiment, the smaller exponent observed in the scaling for quantum algorithm ($\sim \mathcal{HP}^{1.03(4)}$ for SA and $\sim \mathcal{HP}^{0.63(13)}$ for the quantum algorithm) suggests a superlinear (nearly quadratic) speedup in the runtime to find an MIS, 
%over the optimized SA 
for graphs where the deep-circuit-regime ($T > 1/\delta_{\min}$) is reached. We emphasize that achieving this speedup requires an effective depth large enough to probe the lowest-energy many-body states of the system; in contrast, no speedup is observed for graph instances where this depth condition is not fulfilled. \\

\noindent \textbf{Discussion and Outlook.}
Several mechanisms for quantum speedup in combinatorial optimization problems have been previously proposed. Grover-type algorithms are known to have a quadratic speedup in comparison to brute-force classical search over all possible solutions~\cite{grover_fast_1996, durr1999quantum}. 
%the latter is much less effective than more optimized algorithms such as SA. 
A quadratic quantum speedup has also been suggested for quantized SA based on discrete quantum walks ~\cite{sze04, SBBK2008QSA}. However, these methods utilize specifically constructed circuits, and are not directly applicable to the algorithms implemented here. In addition, the following mechanisms can contribute to the speedup observed in our system. 
The quantum algorithm's performance in the observed regime appears to be mostly governed by the minimum energy gap $\delta_{\min}$ (Fig.~\ref{fig_5}C). We show in~\cite{SI} that under certain conditions, one can achieve coherent quantum enhancement for minimum gap resulting in a quadratic speedup via $\delta_{\min} \sim \mathcal{HP}^{-1/2}$. In practice, however,  we find that the minimum energy gap does not always correlate with the classical hardness parameter $\mathcal{HP}$, as is evident in the spread of the quantum data in Fig.~\ref{fig_4}E (see also Fig.~S21). Some insights into these effects can be gained by a more direct comparison of the quantum algorithm with SA using the same cost function corresponding to the Rydberg Hamiltonian~\cite{SI} (Fig.~\ref{fig_5}D). While the observed power law scaling supports the possibility of a nearly quadratic speedup for instances in the deep circuit regime ($\delta_{\min}>1/T$), it is an open question if such a speedup can be extended, with a guarantee, on all instances. Finally, it is possible that $\delta_{\min}$ alone does not fully determine the quantum performance, as suggested by the data points that deviate from the Landau-Zener prediction in Fig.~\ref{fig_5}C, where enhancement through diabatic effects could be possible~\cite{crosson2014different, zhou_quantum_2020}.

While the scaling speedup observed here suggests a possibility of quantum advantage in runtime, to achieve practical runtime speedups over specialized state-of-the-art heuristic algorithms~(e.g. \cite{redumis2}), qubit coherence, system size, and the classical optimizer loop need to be improved. 
The useful depth accessible via quantum evolution is limited by Rydberg state lifetime and intermediate-state laser scattering, which can be suppressed by increasing the control laser intensity and intermediate-state detuning. Advanced error mitigation techniques such as STIRAP \cite{StirapReview} as well as error correction methods should also be explored to enable large-scale implementations.
%for further improvement. 
The classical optimization loop can be improved by speeding up the experimental cycle time, and by using more advanced classical optimizers. Larger atom arrays can be realized using improvements in vacuum-limited trap lifetimes and sorting fidelity.

%{\color{red} Demonstrating the potential of quantum systems for discovery of new algorithms, our results highlight a number of new scientific directions. }
Our results demonstrate the potential of quantum systems for the discovery of new algorithms and highlight a number of new scientific directions. 
%that can be explored. 
It would be interesting to investigate if instances with large Hamming distance between the local and global optima of independent set sizes $|\text{MIS}|-1$ and $|\text{MIS}|$ can be related to the overlap gap property of the solution space, which is associated with classical optimization hardness~\cite{Gamarnik_OGP}. In particular, our method can be applied to the optimization of   ``planted graphs,'' designed to maximize the Hamming distance between optimal and suboptimal solutions, which can provably limit the performance of local classical algorithms~\cite{GZ2019planted}. 
Our approach can also be extended to beyond unit disk graphs by using ancillary atoms, hyperfine qubit encoding, and a reconfigurable architecture based on coherent transport of entangled atoms \cite{bluvstein2021quantum}. Furthermore, local qubit addressing during the evolution can be used to both extend the range of optimization parameters and the types of optimization problems \cite{Lucas_Ising_Formulation}.  Further analysis could elucidate the origins of classical and quantum hardness, for example, by using  graph neural network approaches \cite{sohrabizadeh2021enabling}.   Finally, similar approaches can be used to explore realizations of other classes of quantum algorithm (see e.g.,~\cite{wild_quantum_2020}), enabling a broader range of potential applications.  \\

\textbf{Acknowledgments} We thank Ignacio Cirac, Jason Cong, Simon Evered, Marcin Kalinowski, Mao Lin, Tom Manovitz, Michael Murphy, Benjamin Schiffer, Juspreet Singh, Atefeh Sohrabizadeh, Jordi Tura, and Dominik Wild for illuminating discussions and feedback on the manuscript.

\textbf{Funding:} We acknowledge financial support from the DARPA ONISQ program (grant no.\ W911NF2010021), the Center for Ultracold Atoms, the National Science Foundation, the Vannevar Bush Faculty Fellowship, the U.S. Department of Energy (DE-SC0021013 and DOE Quantum Systems Accelerator Center (contract no.\ 7568717), 
%the Office of Naval Research, 
the Army Research Office MURI, QuEra Computing, and Amazon Web Services. M.C. acknowledges support from DOE CSG award fellowship (DE-SC0020347). H.L. acknowledges support from the National Defense Science and Engineering Graduate (NDSEG) fellowship. D.B. acknowledges support from the NSF Graduate Research Fellowship Program (grant DGE1745303) and the Fannie and John Hertz Foundation. G.S. and  X.G. acknowledges support from the Max Planck/Harvard Research Center for Quantum Optics fellowships. R.S. and S.S. were supported by the U.S. Department of Energy under grant DE-SC0019030. B.B. acknowledges support from  %DARPA grant W911NF2010021, 
a Simons investigator fellowship, NSF grants CCF 1565264 and DMS-2134157, and DOE grant DE-SC0022199. H.P. acknowledges support by the Army Research Office (grant no. W911NF-21-1-0367). The DMRG calculations in this paper were performed using the ITensor package \cite{itensor}, and both DMRG and simulated annealing were run on the FASRC Odyssey cluster supported by the FAS Division of Science Research Computing Group at Harvard University. 

%\textbf{Author contributions:} S.E., A.K., M.C., T.T.W., H.L., D.B., G.S., A.O., and J.-G. L. contributed to building the experimental setup, performing the measurements, and data analysis. M.C.,  J.-G. L., R. S., X.-Z. L., B. N., X. G.,  L. Z., S. C., H. P., and S.-T. W. contributed to theoretical analysis and interpretation.  B. B., E. F., S. S., and N. G. contributed to interpretation of the observations and benchmarking studies.  All authors discussed the results and contributed to the manuscript.       All work was supervised by   M.G., V.V., and M.D.L.

\textbf{Competing interests:} N.G., M.G., V.V., and M.D.L. are co-founders and shareholders of QuEra Computing. A.K. is a shareholder and an executive at QuEra Computing. A.O. and S.-T.W. are shareholders of QuEra Computing.

%\textbf{Data and materials availability:} All data needed to evaluate the conclusions in the paper are present in the paper and the supplementary materials. 

\newpage

\let\oldaddcontentsline\addcontentsline% Store \addcontentsline
\renewcommand{\addcontentsline}[3]{}% Make \addcontentsline a no-op
\bibliographystyle{Science}
\bibliography{References_MIS_main_and_SI.bib}
\let\addcontentsline\oldaddcontentsline% Restore \addcontentsline

%\bibliographystyle{Science}
%\bibliography{References_MIS_main_and_SI.bib}

\clearpage

%%%%%%%%%%%%%%%%%%%%%%%%%%%%%%%%%%%%%%%%%%%%%
%%%%%%%%%%%%FIGURES%%%%%%%%%%%%%%%%%%%%%%%%%
%%%%%%%%%%%%%%%%%%%%%%%%%%%%%%%%%%%%%%%%%%%%

\newpage

%\let\oldaddcontentsline\addcontentsline% Store \addcontentsline
%\renewcommand{\addcontentsline}[3]{}% Make \addcontentsline a no-op
%\bibliographystyle{Science}
%\bibliography{SLbib.bib}
%\let\addcontentsline\oldaddcontentsline% Restore \addcontentsline

\clearpage
\onecolumngrid
\begin{center}
	\textbf{\Large Supplementary Materials}
\end{center}
\normalsize

\setcounter{equation}{0}
\setcounter{figure}{0}
\setcounter{table}{0}
\makeatletter
\renewcommand{\theequation}{S\arabic{equation}}
\renewcommand{\thefigure}{S\arabic{figure}}
\setlength\tabcolsep{10pt}
\setcounter{secnumdepth}{2}
\renewcommand\thesection{\arabic{section}}

\newcommand\numberthis{\addtocounter{equation}{1}\tag{\theequation}}
\newcommand{\insertimage}[1]{\includegraphics[valign=c,width=0.04\columnwidth]{#1}}

\tableofcontents

\newpage

\section{NP-Completeness of Encoded Graphs \label{sec:graphs}}

Here we show that the (decision version) MIS problem is NP-complete on the ensemble of graphs encoded on our Rydberg programmable quantum simulator, consisting of vertices placed on a square lattice and with edges between nearest and next-nearest (diagonal) neighbours. The diagonal connections are crucial, since, otherwise, the MIS problem on the resulting bipartite graphs is known to be not NP-complete \cite{clark_unit_1990}.

To show that the MIS problem on the encoded graphs is NP-complete, we employ a variant of the argument used in Ref.~\cite{clark_unit_1990}. The idea is to reduce the MIS problem on planar graphs with maximum degree 3, which is proven to be NP-complete~\cite{garey_rectilinear_1977}, to the ensemble of graphs we are considering here. The reduction involves transforming a planar graph $G$ with maximum degree 3 to a graph $G'$ in our target ensemble in a way such that $G$ has an independent set of size $M\geq k$ if and only if $G'$ has an independent set of size $M'\geq k'$. This proves NP-completeness by establishing that MIS on our ensemble of graphs is as hard as MIS on planar graphs with maximum degree 3.

\begin{figure}
	\centering
	\includegraphics[width=\textwidth]{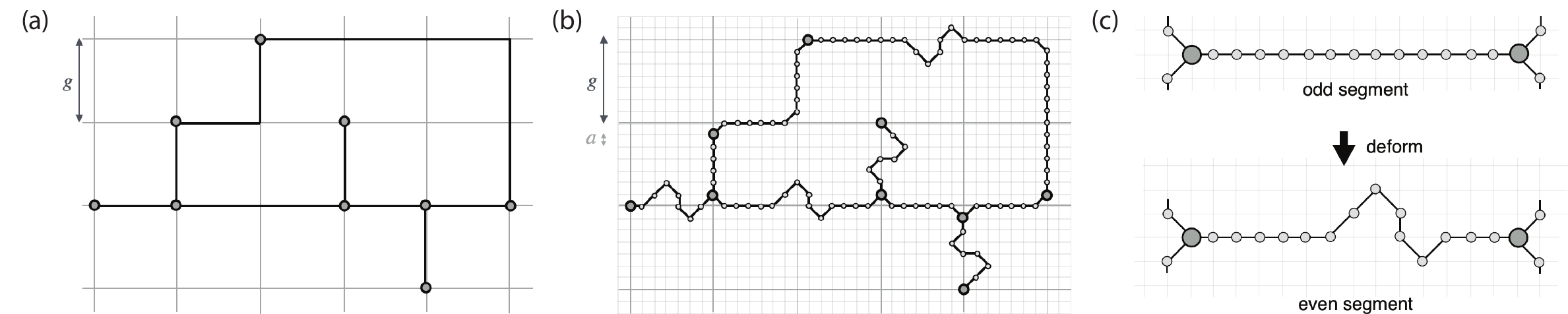}
	\caption{\textbf{Graph reduction procedure.} \textbf{A.} A planar graph $G$ with maximum degree 3 can be embedded on a square grid with lattice spacing $g$, and edges running along the grid lines that do not cross.  \textbf{B.} This graph $G$ can be reduced to a unit disk graph $G'$, with vertices placed on the grid of a square lattice (lattice constant $a$) and unit disk radius $\sqrt{2}a$. \textbf{C.} In order to ensure that each 1D chain connecting two vertices of $G$ contains an even number of vertices, one can use deformations such as those depicted to connect two vertices of $G$.}
	\label{fig:NPC}
\end{figure}

The details of the graph reduction argument are as follows: any planar graph $G=(V,E)$ with maximum degree 3 can be embedded on a square grid with spacing $g$ using $O(|V|^2)$ area, such that its vertices are located at integer coordinates on the grid and its edges are drawn as line segments on this square grid with no edge crossings~\cite{valiant_universality_1981}. We then replace each edge $\{u,v\}\in E$, by a path consisting of an even number of $2k_{u,v}$ ancillary vertices. To do this, we choose a finer grid, with length $a=g/12$. The $2k_{u,v}$ ancillary vertices replacing the edge $\{u,v\}$ are placed on lattice points of this finer grid and all vertices are connected by an edge if they are neighbors or next-nearest (diagonal) neighbors. More specifically, the ancillary vertices are placed in such a way that they form a one-dimensional chain between $u$ and $v$, where each ancillary vertex has exactly two neighbors. In addition, vertices of the graph $G'$ may be displaced by one lattice unit in order to preserve their degrees. This can always be achieved with the proper choice of $k_{u,v}$ (see Fig.~\ref{fig:NPC}). It is then straightforward to verify that $G$ has an independent set of size $M\geq k$ if and only if $G'$ has an independent set of size $M' \geq k' = k+\sum_{\{u,v\}\in E} k_{u,v}$. 

\section{Experimental Platform \label{Sec:expt}}

\subsection{Hardware-efficient encoding of the MIS problem}

Our experiments are performed on the 2D Rydberg programmable quantum simulator described previously in \cite{ebadi_quantum_2020}. Laser-cooled neutral $^{87}$Rb atoms are loaded into 2D arrays of optical tweezers with programmable, defect-free patterns. A two-photon excitation (420 and 1013~nm) couples the $5S_{1/2}$ electronic ground state of each atom $|0\rangle$ to a highly excited $70S_{1/2}$ Rydberg state $|1\rangle$ (lifetime $\tau_r=150$~$\mu$s) via the off-resonant $6P_{3/2}$ intermediate state (lifetime $\tau_e=0.11$~$\mu$s). 

The quantum evolution of our system is determined by the time variation of the global two-photon Rabi frequency (with time-varying phase) $\Omega(t)e^{i\phi(t)}$ and detuning $\Delta(t)$ of the atoms, along with fixed long-range interactions $V_{ij}$ between pairs of atoms in the Rydberg state $\ket{1}$. The laser excitation parameters $\Omega(t)$, $\phi(t)$, and $\Delta(t)$ are controlled with acousto-optical modulators driven by an arbitrary waveform generator that sets the amplitude, phase, and frequency of the 420~nm light. The corresponding values for the 1013~nm light are kept constant during the quantum evolution.

For each randomly generated graph instance, atoms are deterministically positioned by optical tweezers at target locations corresponding to the graph vertices. The Rydberg blockade mechanism permits only one Rydberg excitation within a blockade radius given by $R_b = (C_6/\hbar \Omega)^{1/6}$, where $C_6/h = 862,690$~MHz($\mu$m)$^6$ for the  $70S_{1/2}$ Rydberg state and the two-photon Rabi frequency $\Omega/ 2\pi =4.0$~MHz in the experiment. Choosing a lattice constant of $a = 4.5\,\mu$m for the underlying square lattice gives $R_b/a = 1.7$, resulting in Rydberg blockade extending to next-nearest (diagonal) neighbors and thus realizing the required connectivity of the target graphs. The relevant interaction energies between blockaded nearest and next-nearest neighbor atoms are $V_{\text{NN}}/ h=107$~MHz and $V_{\text{NNN}}/ h=13$~MHz, respectively. This strongly interacting quantum many-body system is used to encode unit-disk graphs corresponding to 80\% filling of a square lattice with next-nearest (diagonal) neighbor connectivity (Fig.~1). 

\subsection{Sources of decoherence}

The experimental parameters for two-photon laser excitation to the Rydberg state are summarized in Table \ref{table:rydberglasers}. Standard beams were used for graphs shown in Figs.~1--3 of the main text, while smaller beams with higher peak intensities were used for the smaller hard graphs in Figs.~4--5 in order to increase the decoherence timescale $T_e$ due to off-resonant intermediate state scattering while maintaining the same two-photon Rabi frequency $\Omega$.

Despite $T_e$ for the smaller beams being more than three times longer than for the larger beams, we did not see a significant increase in the performance of  the quantum algorithm when switching from larger to smaller beams. Other contributions to decoherence in the system include finite Rydberg lifetime (150~$\mu$s theoretically for 70S$_{1/2}$, 80~$\mu$s experimentally measured), finite atomic temperature (20 $\mu$K), and laser noise. The lower experimentally measured Rydberg lifetime may be due to Purcell enhancement of blackbody-induced decay  by the glass cell \cite{archimi2021measurements}.

\begin{table}
	\begin{center}
		\begin{tabular}{|c||c|c|} 
			\hline
			& \textbf{Standard beams} & \textbf{Smaller beams}\\
			\hline\hline
			$\Omega_{420} / 2\pi$ & 135~MHz & 305~MHz \\
			$\Omega_{1013}/ 2\pi$ & 60~MHz & 105~MHz \\
			$\Delta_e/ 2\pi$ & 1.0~GHz & 4.0~GHz \\
			\hline
			$\Omega/ 2\pi$ & 4.0~MHz & 4.0~MHz \\
			$T_e$ & $\sim 20~\mu$s & $\sim 70~\mu$s \\
			\hline
		\end{tabular}
	\end{center}
	\caption{\textbf{Laser parameters for Rydberg excitation.} Two-photon coupling with Rabi frequencies $\Omega_1$, $\Omega_2$ and intermediate state detuning $\Delta_e$ result in a two-photon Rabi frequency $\Omega$ and intermediate state scattering timescale $T_e$. Smaller beams were used for the smaller hard graphs to explore the effect of increasing the intermediate state scattering timescale. $T_e$ is estimated by assuming that scattering comes primarily from the 420~nm light acting on state $|0\rangle$.}
	\label{table:rydberglasers}
\end{table}

\subsection{Data post-processing \label{Sec:post}}

\begin{figure}[tb]
	\centering
	\includegraphics[width=0.9\linewidth]{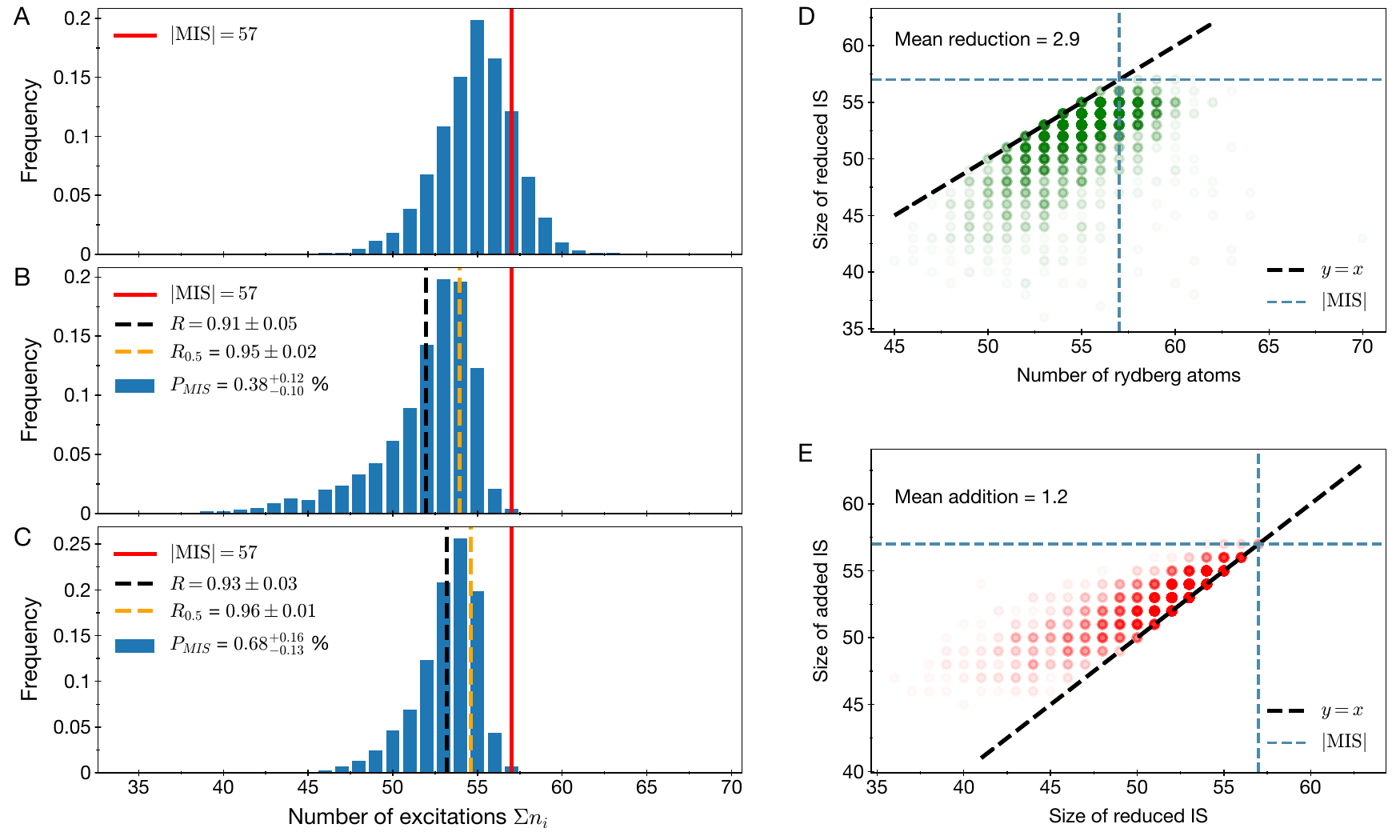}
	\caption{\textbf{Post-processing of experimental data.} \textbf{A.} Histogram of number of Rydberg excitations in readout  for a 179-atom graph (red line shows the MIS). \textbf{B.} Histogram after post-processing to remove Rydberg blockade violations. The approximation ratio is $R=\langle\sum n_i\rangle/|\textnormal{MIS}|$, and $R_{0.5}$ is the approximation ratio when averaging over the top half of the distribution to exclude long tails. $\PMIS$ is the MIS probability. \textbf{C.} Histogram after post-processing to locally add vertices to make the output independent sets maximal. \textbf{D.} Independent set size after vertex reduction vs.\ number of Rydberg excitations before post-processing. Horizontal distance of each point to the black dashed line indicates the magnitude of the reduction. Blue dashed lines indicate the exact MIS. \textbf{E.} Independent set size after vertex addition vs.\ before vertex addition, with the vertical distance of each point to the black dashed line indicating the magnitude of the addition. 
		%\lz{Should we call it Augmented IS instead of Added IS in Fig S1.E?}
	}
	\label{fig:postprocess}
\end{figure}

Single-site projective readout of the final many-body state of the system after quantum evolution is done using fluorescence imaging. Atoms in $|0\rangle$ are detected via fluorescence, while atoms in $|1\rangle$ are detected by the absence of fluorescence (and are hence not distinguishable from atom loss). Fig.~\ref{fig:postprocess}A shows an example histogram of the number of detected $|1\rangle$ atoms per image. A portion of the distribution lies above the MIS limit (maximum number of excitations allowed by the independent set (IS) constraint enforced by Rydberg blockade), which we attribute to a combination of detection errors, blockade violations due to finite interaction energy, quantum fluctuations, and potentially other mechanisms such as Rydberg antiblockade.\\

\noindent\textbf{Vertex reduction} We post-process all experimental data to remove Rydberg blockade violations and reduce the results to valid independent set (IS) solutions (see Sec.~\ref{Sec:sa} for post-processing on simulated annealing). The procedure starts by counting the number of blockade violations for each vertex. The vertex with the most number of violations is removed (flipped from $|1\rangle$ to $|0\rangle$), and ties are broken at random. This process of counting blockade violations and removing the worst vertex is repeated until no blockade violations remain. Fig.~\ref{fig:postprocess}B shows the resulting histogram of valid IS solutions, where the maximum number of possible excitations now corresponds to the MIS. In Fig.~\ref{fig:postprocess}D, we show a scatter plot of reduced IS size vs. the initial number of Rydberg excitations, demonstrating the magnitude of vertex reduction in post-processing.\\

\noindent\textbf{Vertex addition} Often, the resulting state after removing blockade violations is a not a maximal independent set, i.e., at least one vertex can still be added to the IS without violating Rydberg blockade. This can also be due to some combination of detection errors, nonadiabatic state preparation, and quantum fluctuations. In such cases, we can employ a constant overhead greedy algorithm to make the independent set maximal (until no more vertices can be added without violating blockade). An example of the results of such an algorithm is shown in Fig.~\ref{fig:postprocess}C, where the distribution has been shifted towards the MIS. The greedy algorithm for making a given IS maximal involves going through the graph vertex-by-vertex in a random order and flipping any vertex from $|0\rangle$ to $|1\rangle$ if it does not create blockade violations. This is done for ten random orderings in total, and at the end, the solution with the largest resulting independent set is used. Figure~\ref{fig:postprocess}E shows a scatter plot of the final IS size with vertex addition vs.\ the IS size before vertex addition, illustrating the magnitude of this post-processing. Apart from being a fixed-depth algorithm, vertex addition is also limited to local operations, which are insufficient to change the global graph ordering and cannot transform suboptimal solutions for the hardest graphs into an MIS. Consequently, in order to obtain good statistics on the scaling of the MIS probability $\PMIS$ in the presence of experimental imperfections, vertex addition was used for the data in Figs.~3--5. It was not used for the data from Fig.~2, in order for the classical optimizer to run exclusively on the performance of quantum many-body evolution, without any additional effects of classical post-processing. \\

\noindent\textbf{Perfect rearrangement} We can post-select on the perfect initialization of a given graph, the probability of which scales as $\sim 0.99^N$, where $N$ is the number of vertices in the graph. This was done for the smaller hard graphs (Fig.~4, 5) but not for the larger graphs (Fig.~2, 3), since the low post-selection probability for the latter would significantly increase the number of experimental repetitions required.
\\

\noindent\textbf{Limiting the number of vertex reductions}
As reflected in the histogram of the raw number of Rydberg excitations $\ket{1}$ without post-processing (Fig.~\ref{fig:postprocess}A), sometimes, a projective readout image contains a large number of blockade violations that must be removed. The reason for these large numbers of blockade violations is not yet clear, and can be potentially related to blackbody-induced Rydberg antiblockade observed in other experiments. In order to prevent the classical post-processing from solving too much of the MIS problem via vertex removal compared to the contribution from actual quantum evolution, we exclude results where the number of blockade violations exceeds 10\% of the graph size. As with post-selection on perfect rearrangement above, this post-selection on number of blockade violations was done for the smaller hard graphs (Fig.~4, 5) but not for the larger graphs (Fig.~2, 3).\\

\noindent\textbf{Figures of merit for MIS} After post-processing of experimental data, the results are analyzed using several possible figures of merit for the MIS problem. The first is the approximation ratio $R =\langle\sum{n_i}\rangle/|\textnormal{MIS}|$ (or the approximation error $1-R$),
defined as the mean IS size divided by the size of the MIS. For certain cases, where we are interested in the top 50\% of IS sizes, 
%(in closed loop optimization), 
the mean is taken over the top 50th percentile of IS sizes; this is denoted as $R_{0.5}$. Two other figures of merit used for the experiment are the probability of finding an MIS solution $\PMIS$, and the normalized Hamming distance HD (number of discrete spin flips divided by system size) from a given solution to the closest MIS solution.\\

\begin{figure}[tb]
	\centering
	\includegraphics[width=0.9\linewidth]{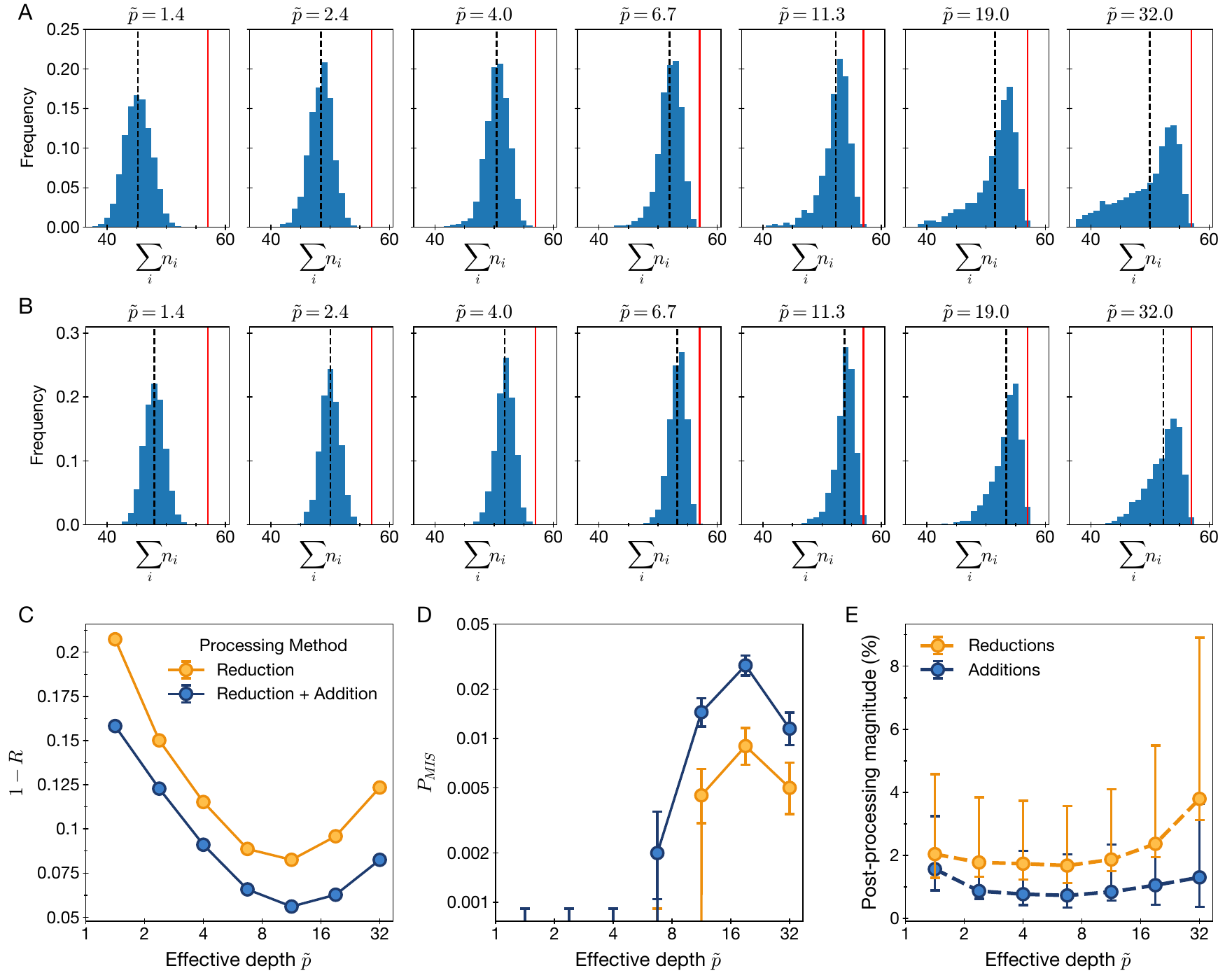}
	\caption{\textbf{Effect of vertex reduction and addition.} \textbf{A.} Histograms of outputs after vertex reduction for different durations $T$ of the quasi-adiabatic sweeps used in the experiment. \textbf{B.} Histograms of outputs after both reduction and addition for different sweep durations. \textbf{C.} Effect of post-processing on approximation error $1-R$, and on \textbf{D.} the MIS probability $\PMIS$ vs. sweep duration. \textbf{E.} Magnitude of post-processing for different sweep durations, expressed as a percentage of vertices on the graph. Data shown here is from the same 179 vertex graph as in the main text.}
	\label{fig:effectmagnitude}
\end{figure}

\noindent\textbf{Effect of post-processing} The effect of both vertex reduction and vertex addition in post-processing on experimental figures of merit is shown in Fig.~\ref{fig:effectmagnitude}. The effect of vertex addition on top of vertex reduction is a constant-factor improvement in both $1 - R$ (Fig.~\ref{fig:effectmagnitude}C) and $\PMIS$ (Fig.~\ref{fig:effectmagnitude}D), with little effect on the time-scaling. The fraction of the graph subject to this post-processing is small (Fig.~\ref{fig:effectmagnitude}E), and does not change significantly with pulse duration.

\begin{figure}[tb]
	\centering
	\includegraphics[width=0.9\linewidth]{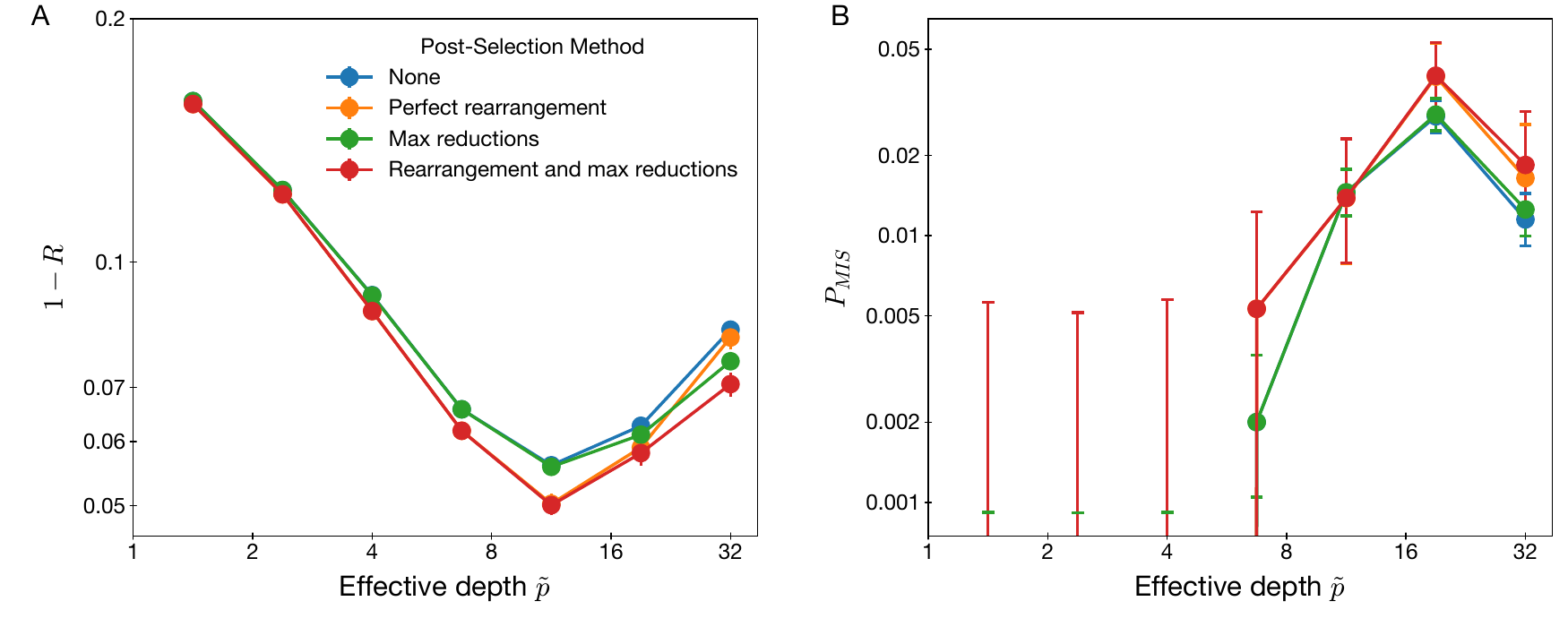}
	\caption{\textbf{Effect of data post-selection.} For both \textbf{A.} the approximation error $1-R$ and \textbf{B.} the MIS probability $\PMIS$, post-selection on perfect rearrangement gives a slight improvement (orange), while post-selection on a maximum number of vertex reductions gives negligible additional benefit (green, red). Data shown here is from the same 179 vertex graph as in the main text.}
	\label{fig:postselectdata}
\end{figure}

The effect of post-selecting on data with perfect rearrangement and on a maximum number of vertex reductions is shown in Fig.~\ref{fig:postselectdata}. For both $1-R$ and $\PMIS$, post-selection on a maximum number of vertex reductions has little effect at early times, while post-selection on perfect rearrangement gives a slight improvement.\\

\section{Closed Loop Quantum-Classical Optimization \label{Sec:optimizer}}

\subsection{Interface with experiment}

Closed-loop quantum-classical optimization is done by using a classical optimizer to find the best time-varying laser pulses characterized by $\Omega(t)$, $\phi(t)$, and $\Delta(t)$ that optimize the performance of the quantum machine. These laser pulses are parametrized by a small number variational parameters which implement the QAOA (Sec.~\ref{sec:qaoa}) or VQAA (Sec.~\ref{sec:qaa}) algorithms.  With a given set of variational parameters, the quantum machine is run with a target graph, and the final state is projectively read out. This is repeated (typically 50 times in our experiment) with the same variational parameters to gather sufficient statistics on the relevant figure of merit for the performance of the quantum platform. A classical optimizer takes the figure of merit output from the quantum machine and produces updated variational parameters to search for the parametrization that optimizes quantum performance. These parameters are then converted to values of $\Omega(t)$, $\phi(t)$, and $\Delta(t)$, and are fed back into the arbitrary waveform generator that controls the laser excitation for running the quantum machine to evaluate the figure of merit with the new control parameters.

\subsection{Classical optimizers}

In this work, we used the classical optimizers from the QuEra Stochastic Optimizers (QuESO) package provided by QuEra Computing for the close-loop optimization. QuESO includes a number of optimizer routines, including both gradient and non-gradient based algorithms. 

Since the total number of measurement shots one can take on a realistic time scale is limited, we balance the projection noise in estimating observables with the total number of optimization iterations within our allocated time budget. In the experiment, we tried non-gradient based algorithms such as the covariance matrix adaptation evolution strategy (CMA-ES) algorithm~\cite{hansen_cma_2006} and Nelder–Mead method~\cite{nelder_mead_1965} and gradient-based algorithms such as simultaneous perturbation stochastic approximation (SPSA)~\cite{spall_multivariate_1992}, Adam~\cite{kingma_adam_2017}, AdaGrad~\cite{duchi_adaptive_2011}, and AdaBound~\cite{luo_adaptive_2019}. Comparing different optimizers, we found that local gradient-based optimization algorithms perform better, due to the limited number of measurement and complex optimization landscape when the number of variational parameters becomes large (e.g. $\sim10$ variational parameters).

In our experiment, we find empirically that Adam and its variant AdaBound work the best on our device. Adam and AdaBound use momentum to accelerate the training, where the momentum in gradient-based optimization theory corresponds to accumulating gradient in previous steps with a proper damping factor. These optimizers are quite reliable even if the gradients are noisy, partly because the noise in different steps can compensate each other. AdaBound is an improvement on top of Adam that can prevent an extremely bad data point from ruining the training, so we eventually used AdaBound for the close-loop optimization data appearing in this work.

For the estimation of the gradients (in Adam, AdaGrad, and AdaBound), we implemented two methods. First, the finite-difference (FD) method, which measures the gradient with respect to each variational parameter by measuring two neighbouring points for each parameter, requiring $2p$ queries to the quantum machine. The second method is based on simultaneous perturbations (SP) \cite{spall_multivariate_1992}, which uses only two neighboring points in the p-dimensional space spanned by variational parameters to estimate the gradient at each iteration, independent of the number of variational parameters. The FD method is more accurate but requires many more measurements to estimate each gradient. We empirically find that the SP gradient estimator works better, given typical optimization runs of $\sim 100,000$ measurements ($15$ hours of continuous operation).

\begin{figure}[tb]
	\centering
	\includegraphics[width=0.9\linewidth]{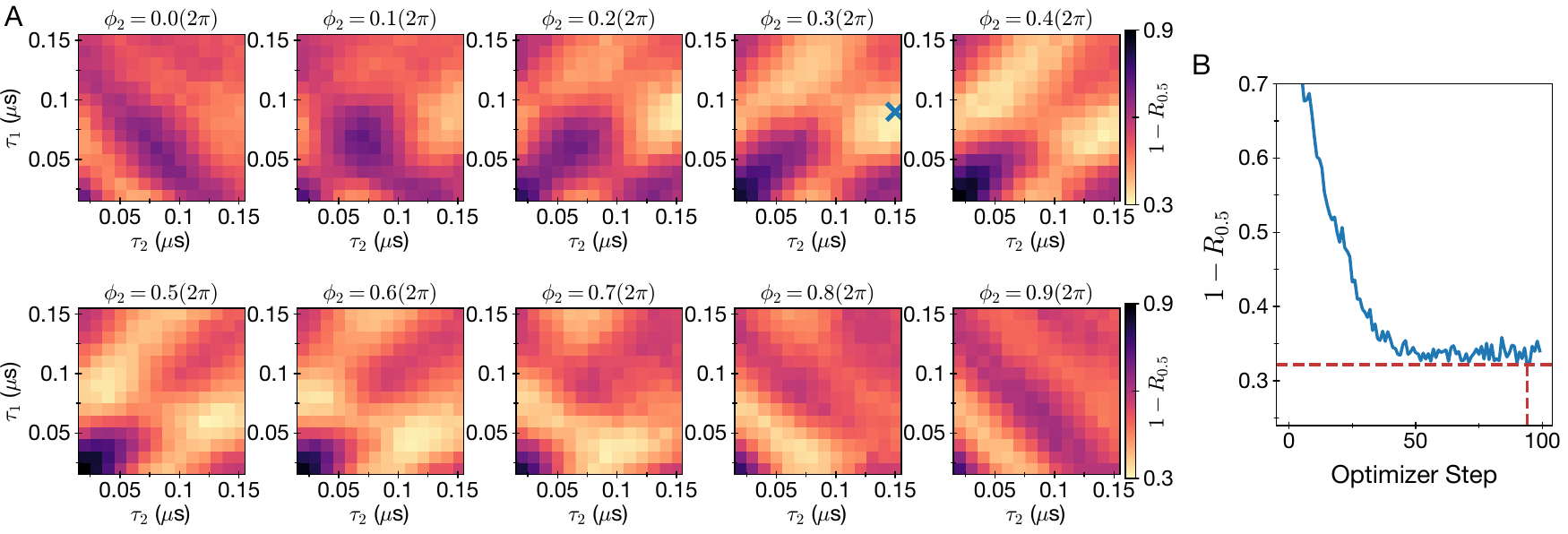}
	\caption{\textbf{QAOA depth $p=2$ direct search}. \textbf{A.} The phase of the first pulse was fixed at $\phi_1=0$, and the phase of the second pulse was varied in steps between $\phi_2=0$ and $\phi_2=0.9\times 2\pi$. At each value of $\phi_2$, 2D scans of pulse times $\tau_1$, $\tau_2$ were performed. The location of the global minimum in $1-R_{0.5}$ is indicated with the blue cross. \textbf{B.} The classical optimizer gives results comparable to the direct search, with fewer queries to the quantum machine.}
	\label{fig:qaoap2}
\end{figure}

\begin{figure}[tb]
	\centering
	\includegraphics[width=0.9\linewidth]{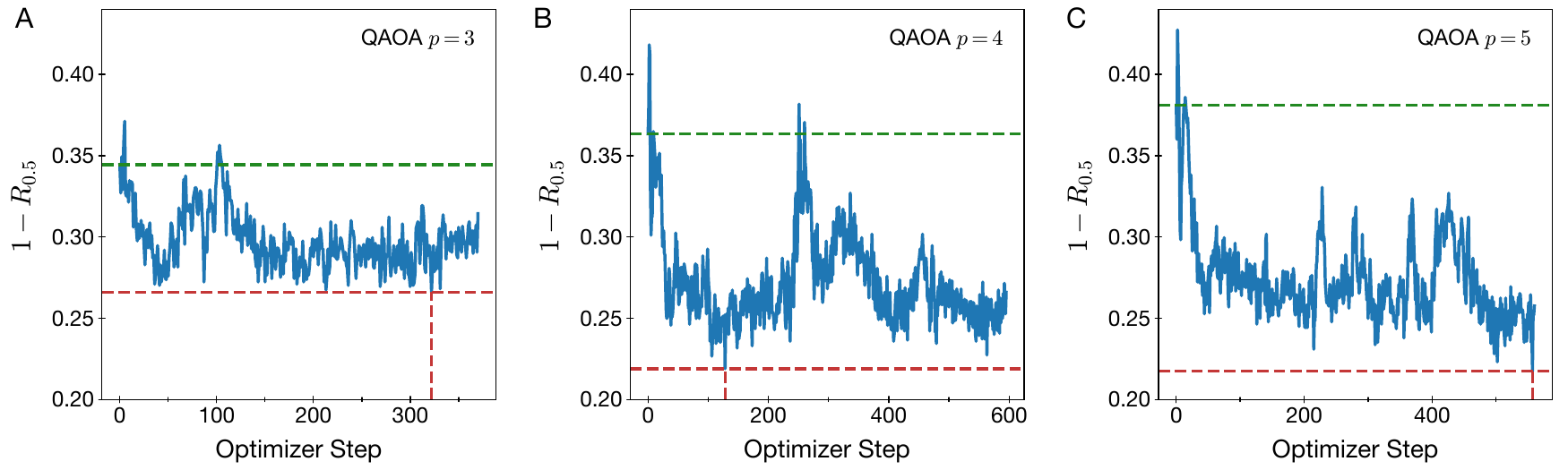}
	\caption{\textbf{Optimization of QAOA for depths $p=3, 4, 5$.} The variation in approximation error $1-R_{0.5}$ with each step of the classical optimzer is shown for depths \textbf{A.} $p=3$, \textbf{B.} $p=4$, and \textbf{C.} $p=5$. The green line indicates the figure-of-merit for the initial parameters, and the red lines indicate the found optimum at each depth.}
	\label{fig:qaoap345}
\end{figure}

\section{Quantum Approximate Optimization Algorithm (QAOA) \label{sec:qaoa}}

\subsection{Experimental parameterization}

The standard quantum approximate optimization algorithm (QAOA) is parametrized in terms of $p$ layers of time evolution under non-commuting Hamiltonians $H_B$ and $H_C$:
\begin{equation}
	| \psi \rangle = e^{-i H_B\beta_p}e^{-i H_C \gamma_p} ... e^{-iH_B \beta_2} e^{-i H_C\gamma_2}e^{-iH_B \beta_1} e^{-i H_C\gamma_1} | \psi_0 \rangle.
\end{equation}
Typically, a combinatorial optimization problem is encoded in the cost function Hamiltonian $H_C$. The system starts in $|\psi_0\rangle = |\psi_B\rangle$, an eigenstate of another non-commuting mixing Hamiltonian $H_B$, and the goal is to find good approximate solutions of the combinatorial optimization problem by minimizing the cost function $\langle \psi |H_C|\psi \rangle$ using the $2p$ variational parameters $(\gamma_1, \gamma_2, ..., \gamma_p)$ and $(\beta_1, \beta_2, ... , \beta_p)$.

In the case of MIS encoded on Rydberg atom arrays, our QAOA protocol is slightly different. The Hamiltonians $H_C$ and $H_B$ are defined as
\begin{equation}
	H_C = -\hbar \Delta \sum_{i} n_i\hspace{0.2in} H_B = \dfrac{\hbar}{2} \Omega \sum_{i}(\ket{0}_{i}\bra{1} + {\rm h.c.}) + \sum_{i<j}V_{ij}n_i n_j.
\end{equation}
The mixing term $H_B$ includes Rydberg interactions $V_{ij}$, which are always on during quantum evolution, meaning that in the ideal blockade approximation, $H_B$ only couples states in the independent set subspace. Instead of starting from an eigenstate of $H_B$, we start from the initial state $|\psi_0\rangle = | 00 ... 0 \rangle$, which is the ground state of the Hamiltonian $H_C$ for initial $\Delta < 0$.  In the independent set subspace, $H_C$ is the same as $H_{\text{cost}}$ from Eq.~1 in the main text, which for $\Delta > 0$ has a ground state corresponding to the MIS of the underlying unit disk graph. 

Evolution under $H_B$ is implemented as a variable duration laser drive with constant Rabi frequency $\Omega$ and detuning $\Delta = 0$, in the presence of Rydberg interactions $V_{ij}$. Evolution under $H_C$ constitutes a global Z rotation on excited atoms, and is implemented as a phase jump in the laser drive between resonant pulses. The variational parameters for the optimization are thus the evolution times $(\tau_1, ... \tau_p)$ under the mixing term $H_B$ %implementing $e^{-iH_B \tau_i}$, 
along with laser phases $(\phi_2, ... \phi_p)$ for each time step, which correspond to the global Z rotations $e^{-iH_C \gamma_i}$ (the initial phase $\phi_1$ is set to zero since $e^{-iH_C \gamma_1}$ has no effect).

\subsection{Variational optimization results}
The QAOA for depth $p=1$ is simply a single pulse of variable duration $\tau_1$, with $\phi_1=0$. Depth $p=2$ consists of three variational parameters $\tau_1$, $\tau_2$, and $\phi_2$, which were optimized by a brute-force direct search (Fig.~\ref{fig:qaoap2}A). The classical optimizer produces comparable results with fewer queries from the quantum machine (Fig.~\ref{fig:qaoap2}B). For QAOA depths $p=3,4$, and $5$, the classical optimizer was used to optimize the variational parameters. Figure~\ref{fig:qaoap345} shows the improvement in $1-R_{0.5}$ at each depth. The initial QAOA parameters at each depth $p$ consisted of the optimal parameters from depth $p-1$, plus a $p$-th pulse with a randomly chosen initial duration $\tau_p= 30-70$~ns, and the corresponding phase consisting of randmoly selecting either an increase or decrease to the previous phase with a randomly chosen magnitude. Pulse durations for each layer are constrained to $250$~ns, corresponding to an effective depth of $\tilde{p} = 2$.

\subsection{Performance limitations}
Our attempts to implement the QAOA resulted in a saturation of system performance beyond $p=4$ at a value that fell far below what was achieved with the piecewise linear, quasi-adiabatic parameterization of quantum evolution (see Section \ref{sec:qaa}).
Although, in principle, in the limit of infinite depth $p$, the QAOA is able to reproduce adiabatic evolution, there are various practical reasons 
%in practice 
that limit the performance of the QAOA in our experiments.
One reason is that at higher depth $p$, the number of parameters in the QAOA grows, and it becomes progressively more difficult to optimize them due to our limited experimental measurement budget for obtaining precise estimates of the objective function and particularly its gradient.
In contrast, it is easier to optimize the quasi-adiabatic algorithm at longer evolution times (effective depths $\tilde{p}$), where it can be described with fewer parameters compared to QAOA.

\begin{figure}[htb]
	\centering
	\includegraphics[height=2.4in]{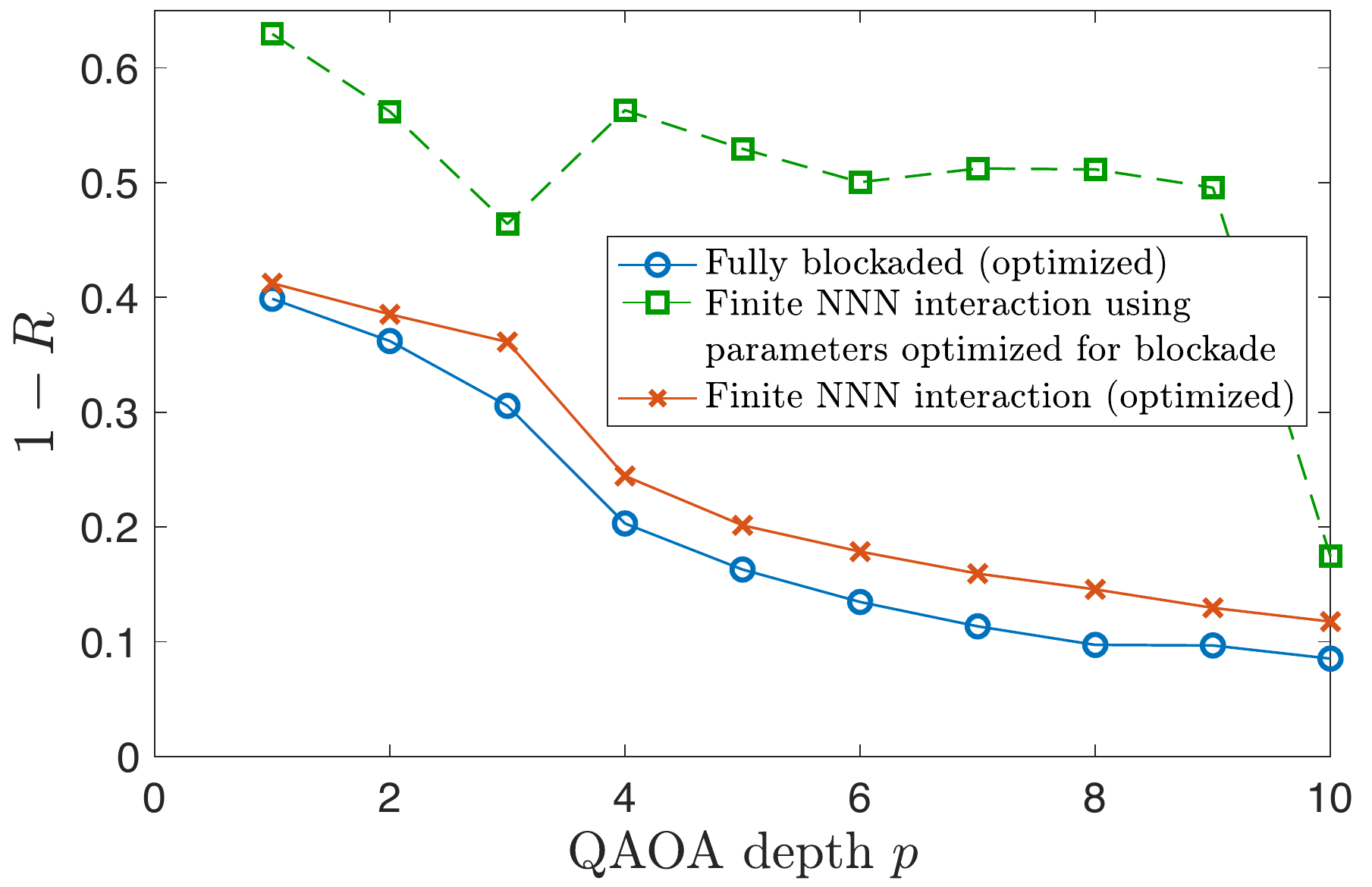}
	\caption{\textbf{Effects of imperfect blockade on QAOA.} The mean approximation error achieved by the QAOA at various depths $p$ on an example 24-vertex graph, in both the ideal (fully blockaded) case and the case allowing blockade violations with finite next-nearest-neighbor (NNN) interactions. For the finite NNN interaction case, we apply both the parameters optimized for the ideal fully blockaded interaction (green) as well as the parameters optimized for dynamics under finite NNN interactions (red). For both parameters, the finite NNN interaction performs worse than the ideal case (blue) due to leakage out of the independent set subspace.}
	\label{fig:QAOA-leak}
\end{figure}

Furthermore, the QAOA is implemented on our platform as a series of resonant laser pulses. Due to the modest next-nearest (diagonal) neighbor interaction $V_{\text{NNN}}/2\pi=13$~MHz relative to the two-photon Rabi frequency $\Omega/2\pi =4.0$~MHz, the resonant pulses introduce blockade violations and leakage out of the independent set subspace that cannot be compensated for in subsequent layers, thus limiting the overall performance of QAOA.
In Figure~\ref{fig:QAOA-leak}, we compare the performance of QAOA with and without blockade violations using numerical simulations on a small $N=24$ vertex graph. Here, the outputs with blockade violations are reduced to an independent set using vertex reduction post-processing. We see that the performance is worsened when including a finite interaction energy on the next-nearest neighbours, hence allowing blockade violations, as compared to the ideal case.

Finally, we note that pulse imperfections can arise since the laser pulses for QAOA are implemented using an acousto-optic modulator (AOM). Changing the phase in the drive tone of an AOM results in a discontinuity in the wavefront of the optical beam, hence reducing its intensity on the atoms for a brief period ($\sim10$~ns). This cross-talk between phase and intensity can lead to imperfect implementation of our pulses.

\section{Variational Quantum Adiabatic Algorithm (VQAA) \label{sec:qaa}}

\subsection{Experimental parametrization}
The variational quantum adiabatic algorithm (VQAA) aims to find the best quasi-adiabatic path that interpolates between an initial Hamiltonian with a trivial ground state and a final Hamiltonian whose ground state is the solution to the problem of interest. In our implementation, VQAA corresponds to optimizing a time-varying detuning profile $\Delta(t)$ from negative to positive values at a constant Rabi coupling $\Omega$ (Fig.~2b). The detuning profile $\Delta(t)$ is parametrized as a piecewise linear function, with $\Delta_0$ being the initial detuning and the full profile $\Delta(t)$ determined by the durations $(\tau_1, ... \tau_f)$ and end detunings $(\Delta_1, ... \Delta_f)$ of each of the $f$ linear segments. The coupling $\Omega$ is first linearly ramped on at constant $\Delta_0$ in time $\tau_\Omega$, and is also turned off at the end of the sweep in  time $\tau_\Omega$ while holding the detuning constant at $\Delta_f$. An additional global parameter low-pass filters $\Delta(t)$ with time constant $\tau_\Delta$, which along with $\tau_\Omega$ suppresses excitations from sharp changes of the Hamiltonian. The overall $2f+3$ variational parameters for the quantum adiabatic algorithm are thus $(\tau_1, ... \tau_p)$, $(\Delta_1, ... \Delta_f)$, and $(\Delta_0, \tau_\Delta, \tau_\Omega)$.

\subsection{Variational optimization results}
\begin{figure}[tb]
	\centering
	\includegraphics[width=0.7\linewidth]{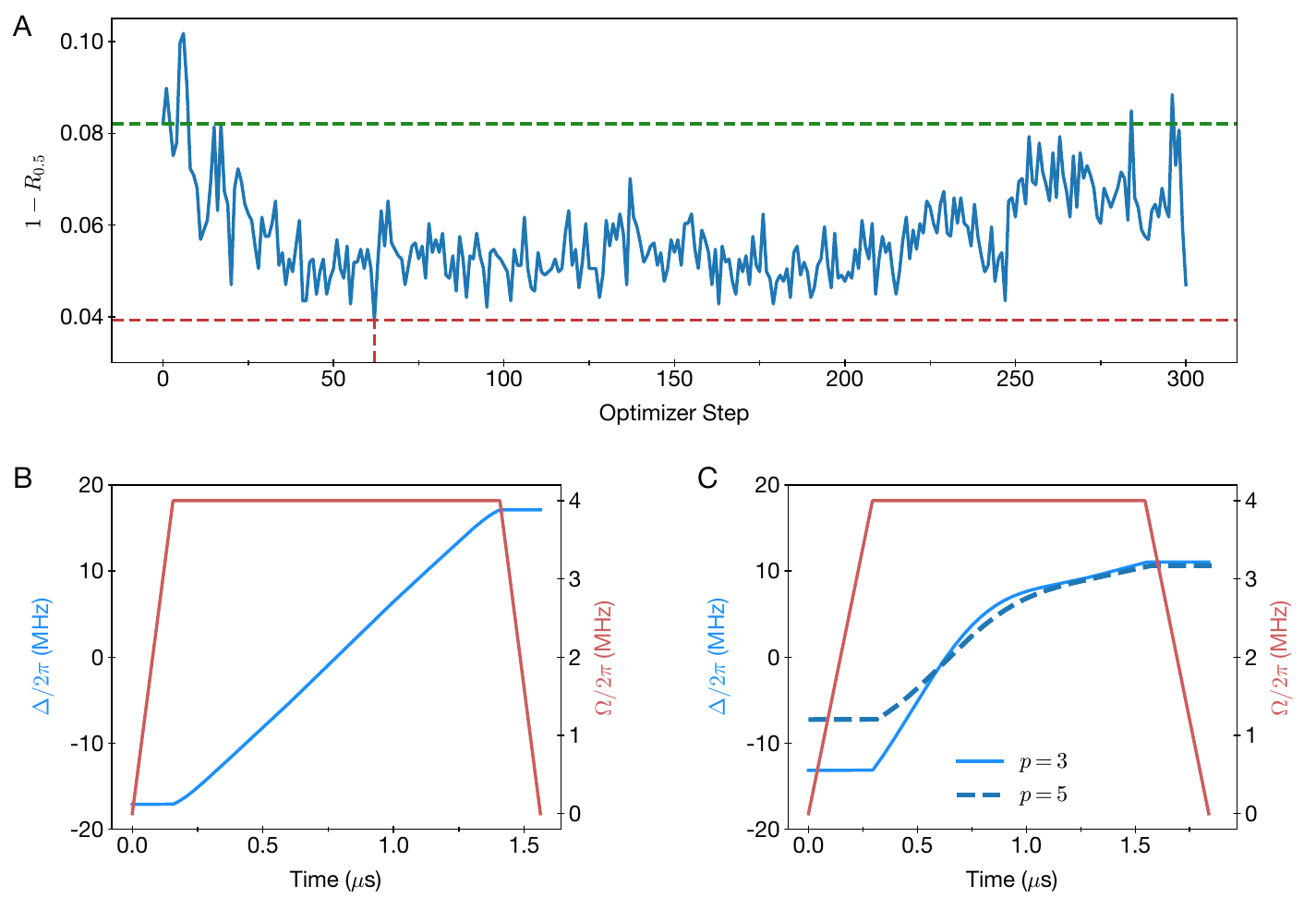}
	\caption{\textbf{Optimization of quasi-adiabatic detuning sweep.} \textbf{A.} Approximation error $1-R_{0.5}$ as a function of optimizer step for variational optimization of a three-segment piecewise linear detuning sweep. The green dashed line indicates the performance of the initial parameters, corresponding to a purely linear sweep. The red dashed line marks the optimum found by the optimizer. Total duration of sweep is fixed at $1.25~\mu$s ($\tilde{p} = 10$), and the turn on/off time $\tau_\Omega$ is limited to a maximum of $0.312~\mu$s.  \textbf{B.} Comparison of the initial simple linear (one-segment) detuning sweep and \textbf{C.} the optimized three-segment sweep (solid blue line). The dashed blue line shows an optimized five-segment sweep with comparable shape as well as performance.}
	\label{fig:p3pl}
\end{figure}

The results of variational optimization of piecewise linear quasi-adiabatic detuning sweeps are shown in Fig.~\ref{fig:p3pl}A for $f=3$ segments. In this optimization run, the optimizer starts from initial parameters corresponding to a purely linear sweep (Fig.~\ref{fig:p3pl}B), and finds an optimized sweep that is shown in Fig.~\ref{fig:p3pl}C. For time-scaling experiments, the pulse shape was optimized for a sweep duration $T = \tau_1 + \tau_2 + ... +\tau_f = 1.25~\mu$s ($\tilde{p}=10$), and subsequently rescaled for different sweep durations while keeping pulse turn-on/off time $\tau_\Omega$ constant. Increasing the number of segments for the piecewise linear detuning sweep (Fig.~\ref{fig:p3pl}C dashed line) resulted in similar shapes for the sweep as well as comparable performance to the three-segment sweep. 

%2021522-Exp-P3PL-1000-adabound-0002

\subsection{Additional manual optimization}

\begin{figure}[tb]
	\centering
	\includegraphics[width=0.5\linewidth]{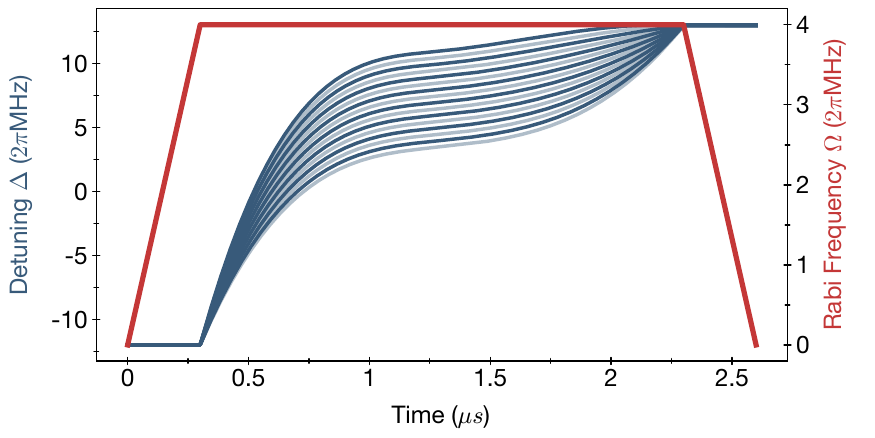}
	\caption{\textbf{Manual optimization of the quasi-adiabatic detuning sweep.} This family of cubic spline pulses ($\tilde{p} = 16$) are used for additional manual optimization. These pulses share an identical starting/ending detunings and minimum slope, but differ in the frequency where their inflection point occurs. The inflection point is manually scanned from $\Delta/2\pi = 3.5$~MHz to $11.0$~MHz in $0.5$~MHz steps, and the resulting pulses are shown.
	} 
	\label{fig:splinesweep}
\end{figure}

On most graphs, the same, three-segment piecewise linear detuning sweep yielded nearly-optimal experiment performance.
For most graphs, the three-segment piecewise linear detuning sweep (Fig.~\ref{fig:p3pl}C) yielded nearly optimal performance in reducing the approximation error $1-R$. However for the hardest graphs studied (Fig.~4 and 5), additional manual optimization resulted in a further increase in the MIS probability $\PMIS$. The optimization procedure consisted of parameterizing the detuning sweep as a cubic spline function that initially resembles the form of the classical optimizer output (Fig.~\ref{fig:splinesweep}). Subsequently, the detuning corresponding to the minimum slope is scanned to maximize $\PMIS$ (Fig.~5B). In the Landau-Zener picture of quantum many-body ground state preparation, the detuning that maximizes $\PMIS$ should correspond to the location of the minimum energy gap in the many-body spectrum (Fig.~5A, \cite{Roland_optimized_Grover}). Note that on several  graphs, the optimum detuning for the minimum slope was outside the range of the original detuning sweep from the classical optimizer output. Therefore, the final detuning of the sweep was extended (compared to the previously optimized pulse) to higher values to allow proper parametrization of the cubic spline interpolation.

\section{Characterizing Graphs using Tensor Network Algorithms \label{Sec:tensornetworks}}

In the main text, quantum and classical performance were analyzed in terms of the MIS degeneracy $D_{|\text{MIS}|}$ as well as the degeneracy $D_{|\text{MIS}-1|}$ of independent sets of size $|\text{MIS}|-1$. To obtain these properties for all the randomly generated graphs in this work, we use a generalized tensor network method.
%to compute the relevant properties. %up to a large size of $37 \times 37$. 
Here, we include a short description of the algorithms; more details can be found in Ref.~\cite{liu_tensor_2021}.

Tensor networks with real elements have been used in enumerating solutions of some combinatorial problems such as 3-coloring and satisfiability problems~\cite{Biamonte2017}. If the tensor elements are extended beyond just real and complex numbers, the same tensor network contraction algorithm can be adapted to find various properties of graphs, including the size of the MIS, the number of MIS solutions (MIS degeneracy), the total number of independent sets, and the independence polynomial. The algorithm can also be used to calculate the exact configurations of all MIS solutions as well as all independent sets of size $|\text{MIS}|-1$. We call these tensor networks with generic element types ``generic tensor networks.'' 

\begin{figure}
	\centerline{\includegraphics[]{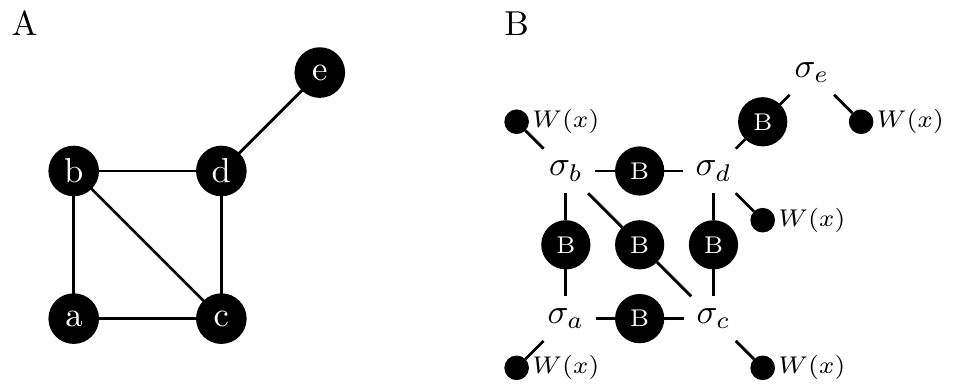}}
	\caption{\textbf{Tensor networks for characterizing graphs.} Mapping \textbf{A}. a graph  to \textbf{B} a tensor network. $\sigma_{a-e}$ denote tensor indices, while $W(x)$ and $B$ are tensors defined in Eq.~\eqref{eq:tensor}.}\label{fig:tensormap}
\end{figure}

Here, we briefly introduce how to use the method of generic tensor networks to find the number of independent sets (i.e.\ degeneracy) of a given size.
This problem is equivalent to finding the independence polynomial, which is defined, for a graph $G$, as 
\begin{equation}
	I(G, x) = \sum_{k=0}^{|\text{MIS}|} D_k x^k,
\end{equation}
where the polynomial coefficient $D_i$ denotes the degeneracy of independent sets of size $k$. Therefore, if we can compute the independence polynomial, the coefficients $\DMIS$ and $\DMISminusone$ tell us the degeneracies of the MIS and of the independent sets of size $|\text{MIS}|-1$, respectively. As shown in Fig.~\ref{fig:tensormap}, the independence polynomial of graph $G$ can be encoded in a tensor network, with a vertex tensor $W(x)$ placed on each vertex and an edge tensor $B$ placed on each edge%, defined respectively as 
\begin{equation}
	W(x) = \left(\begin{matrix}
		1 \\
		x
	\end{matrix}\right),
	\qquad \quad 
	B = \left(\begin{matrix}
		1  & 1\\
		1 & 0
	\end{matrix}\right). \label{eq:tensor}
\end{equation}
$W(x)$ is defined such that if a vertex belongs to a particular independent set, it contributes an $x$, and otherwise, it contributes a $1$. The edge tensor $B$ connects the vertices whenever the vertices are connected by an edge in the graph $G$ and the element $B_{11} = 0$ captures the independence set constraint. The contraction of the tensor network will produce the independence polynomial~\cite{liu_tensor_2021}:  
\begin{equation}
	I(G, x) = \sum\limits_{\sigma_1, \sigma_2, \ldots, \sigma_{|V|} = 0}^{1} \prod\limits_{i=1}^{|V|} W(x)_{\sigma_i} \prod\limits_{(i,j) \in E(G)} B_{\sigma_i \sigma_j}.
\end{equation}

Using recently developed
%Thanks to the recent development of 
contraction order optimization techniques
%in tensor-network based quantum circuit simulation tools
~\cite{Markov2008, pan_simulating_2021, kalachev_recursive_2021}, the contraction can be done efficiently on graphs with a small tree width. By labeling $x$ on different vertices, one can even enumerate all independent sets.
However, symbolic calculations are very slow. By changing the tensor element types in $W$ and $B$, one can calculate different properties of independent sets and significantly speed up the calculation for certain computations. 

To make the tensor network contraction results independent of the contraction order, we require the tensor elements to form a commutative semi-ring. For example, if we only need to calculate the MIS size, we can replace the tensor elements with the tropical algebra~\cite{liu_tropical_2021}: $x \oplus y = \max(x,y)$, $x \odot y = x + y$, $\mymathbb{0} = -\infty$, and $\mymathbb{1} = 0$, where the $0$ and $1$ elements in the tensors are replaced with the tropical $\mymathbb{0}$ and $\mymathbb{1}$, $x$ is replaced by $1$, and the $+$ and $\times$ operation in the tensor network contraction are replaced with the tropical algebra operations $\oplus$ and $\odot$. In addition, to compute the independence polynomial exactly, we use a polynomial fitting approach. To avoid the integer overflow problem for large graphs, we replace the element types with a finite field algebra and make use of the Chinese remainder theorem~\cite{liu_tensor_2021}. Lastly, by combining tensor elements with set operations, we also use the tensor network to enumerate all independent sets of different sizes, which we use to study the low-energy configurations and compute Hamming distances between the configurations.

\begin{figure}
	\centering
	\includegraphics[width=0.9\textwidth]{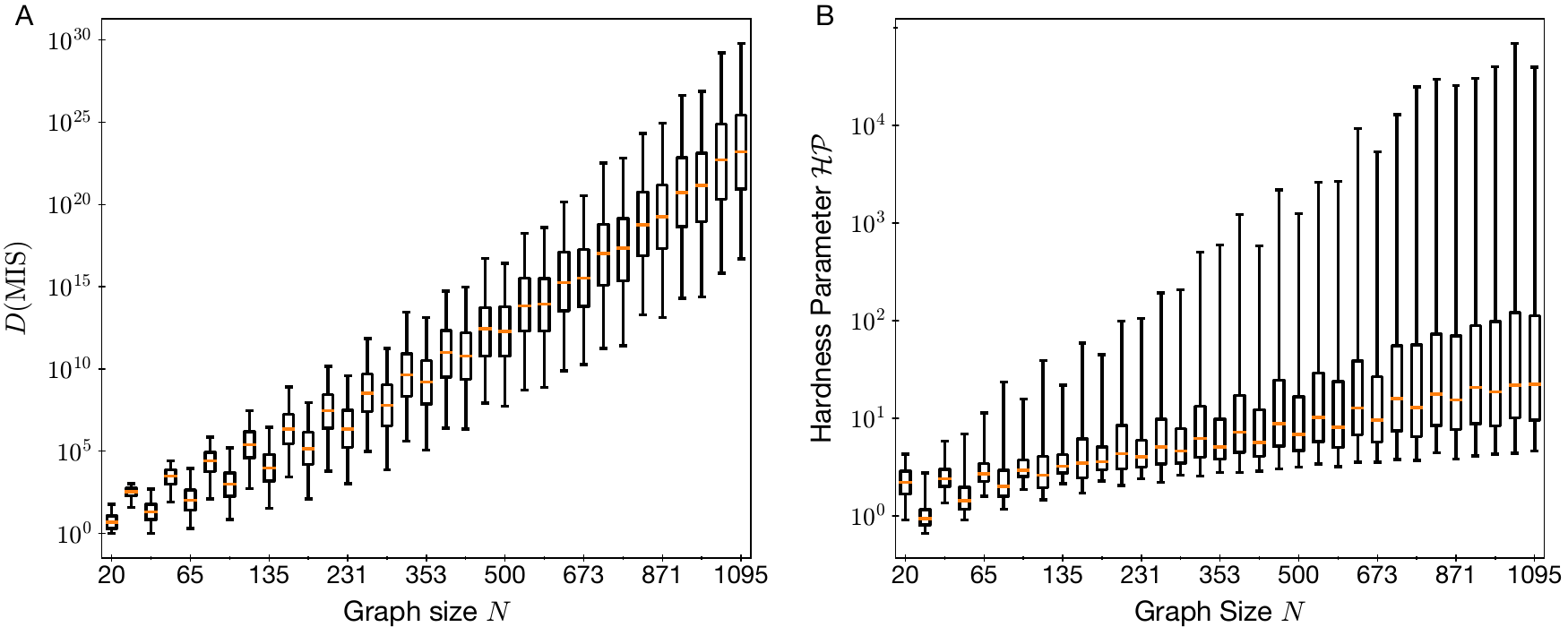}
	\caption{\textbf{Scaling of MIS degeneracy and hardness parameter with size.} For each lattice size, 1000 graphs with $80\%$ filling are generated randomly. Logarithmic box and whisker plot of \textbf{A}. $\DMIS$ and \textbf{B}.  $\mathcal{HP} = \DMISminusone/(|\text{MIS}|\DMIS)$   for graphs with up to $N = 1095$. Each box denotes the upper and lower quartiel, with the orange line showing the median. Whisker denote are the 2nd and the 98th percentile.}
	\label{fig:degeneracy_vs_L}
\end{figure}

With the generic tensor network algorithms, we computed $\DMIS$ and $\DMISminusone$ for graph sizes up to $N=1095$, corresponding to 80\% filling of a $37\times37$ square lattice. 
%The tensor network for maximal independent sets has a larger tree width than the one for independent sets; hence, we only compute maximal independent set properties for graphs on up to $16\times16$ lattices. 
We randomly generate $1000$ graphs with $80\%$ filling at each size $N$ and show the scaling of $\DMIS$ and $\mathcal{HP} = \DMISminusone/(|\text{MIS}|\DMIS)$ with increasing system sizes in Fig.~\ref{fig:degeneracy_vs_L}. One can see that for large $N$, in the worst case, both the MIS degeneracy and the hardness parameter $\mathcal{HP}$ seem to scale exponentially with the system size. The box plot also shows a wide range of MIS degeneracy and $\mathcal{HP}$ at each graph size $N$.

\section{Scaling of Quantum Approximation Error \label{Sec:coarsening}}

\subsection{Basic formalism}

In this section, we present a theory which describes the scaling behavior of the defect density in the Rydberg simulator's solution to the MIS problem and accounts for the main experimental observations. Based on generic ordering dynamics in (2+1)D, our starting point is the natural ansatz that after crossing the quantum critical point, the size of correlated regions grows with time as $\mc{R}(t)$\,$\sim$\,$t^{\,\mu}$. This dynamic growing correlation length $\mc{R}(t)$ will be a central player in our story. The exponent $\mu$ and the prefactor of the growth law are \textit{a priori} unknown and are governed by a combination of the quantum Kibble-Zurek mechanism \cite{zurek_dynamics_2005, PhysRevB.72.161201}, early-time coarsening \cite{PhysRevX.5.021015}, and ``standard'' late-time coarsening \cite{biroli2010kibble}.

The implications of this scaling hypothesis for the state(s) obtained as a solution to the MIS problem are straightforward. At an intermediate time $t$, the number of domains formed is given by $\mc{N}(t) \equiv A/(\pi \mc{R}^2(t))$, where $A$ is the geometric area of the graph. Then, the local deviations from the perfect solution (referred to as \textsl{defects} hereafter) arise from the boundaries where the different domains meet due to the possibly conflicting ordering between individual regions. Accordingly, the number of defects per domain scales as $\pi \mc{R}(t)$ because the error accumulates proportionately to the length of the domain wall. The total number of defects is therefore roughly $\mc{N} \mc{R}$\,$\sim$\,$1/\mc{R}(t)$. Taking $\mc{R}(t) \sim t^{\,\mu}$ as above, the approximation error $1 -R$ at the time of measurement, $T$, should go as \begin{equation}
	1-R \simeq \mc{N} (T)\, \mc{R}(T) \simeq \frac{1}{\mc{R}(T)} \sim T^{-\mu}.
\end{equation} 
Since $T$ is directly related to the evolution time, this simple calculation predicts a power-law decay of the error with the total sweep time, in agreement with the basic dependence seen experimentally. Physically, the healing process is driven by the interaction of the long domain walls with the bulk gapped quasiparticles about the ordered state within each domain \cite{chandran2013kibble}.

\subsection{Degeneracy-dependent corrections}

\begin{figure}[tb]
	\centering
	\includegraphics[width=0.65\linewidth]{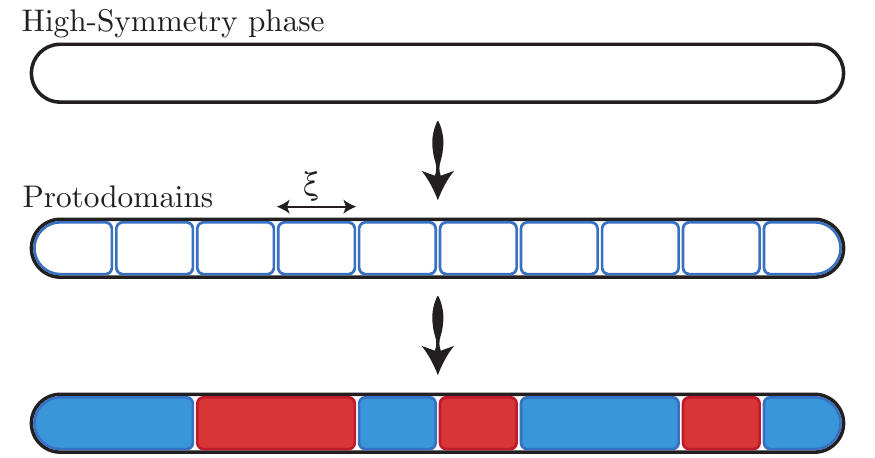}
	\caption{\textbf{Domain formation in 1D.} Schematic representation of domain formation in 1D. In the course of the ordering dynamics, the system is partitioned into protodomains of the size of the characteristic length scale. At the interface between adjacent protodomains, kinks are spontaneously formed with probability $p$, resulting in true domains. An analogous picture holds for 2D as well. Figure adapted from Ref.~\cite{mayo2021distribution}.}
	\label{fig:protodomains}
\end{figure}

We now turn to a more systematic calculation of the total number of defects $1-R$. To generalize the mechanism described above, we consider that the effect of the ordering dynamics is to initially partition a system of a given size into ``protodomains" (see Fig.~\ref{fig:protodomains}) of the same length scale over which the order parameter stabilizes  \cite{mayo2021distribution}. At the boundary between adjacent domains, kinks form with a given probability $p$. Conversely, with probability $(1$\,$-$\,$p)$, no kink is formed and the two adjacent protodomains coalesce to form a larger domain. Given $\mc{N}$ protodomains, the number of boundaries between them (which determines the number of stochastic events for kink formation) is $Z \mc{N}/2$, where $Z$ is the average coordination number of each protodomain. Note that here and henceforth, we have suppressed the explicit time dependence of $\mc{N}$ and related variables. Assuming that the success probability $p$ is the same at different locations of the graph, the probability distribution for the number of kinks, $k$, takes the binomial form
\begin{equation}
	\label{eq:dist}
	P(k) = \dbinom{Z \mc{N}/2}{k} \, p^k\, (1-p)^{Z \mc{N}/2-k}.
\end{equation}
The number of actual domains with $k$ kinks is $n$\,$=$\,$k/(Z/2)$. However, every time a kink fails to form, the average length of the domain walls decreases as
\begin{equation}
	\label{eq:r}
	r =  \frac{\pi \mc{N} \mc{R} - (\frac{Z \mc{N}}{2}-k)\, \mc{L}}{n},
\end{equation}
$\mc{L}$ being the length of the boundary between the two coalescing protodomains. In order to determine the number of errors, we have to calculate the average $\langle n r \rangle$ subject to the distribution \eqref{eq:dist}.

\subsection{Degeneracy of MIS states}

\begin{figure}[tb]
	\centering
	\includegraphics[width=0.9\linewidth]{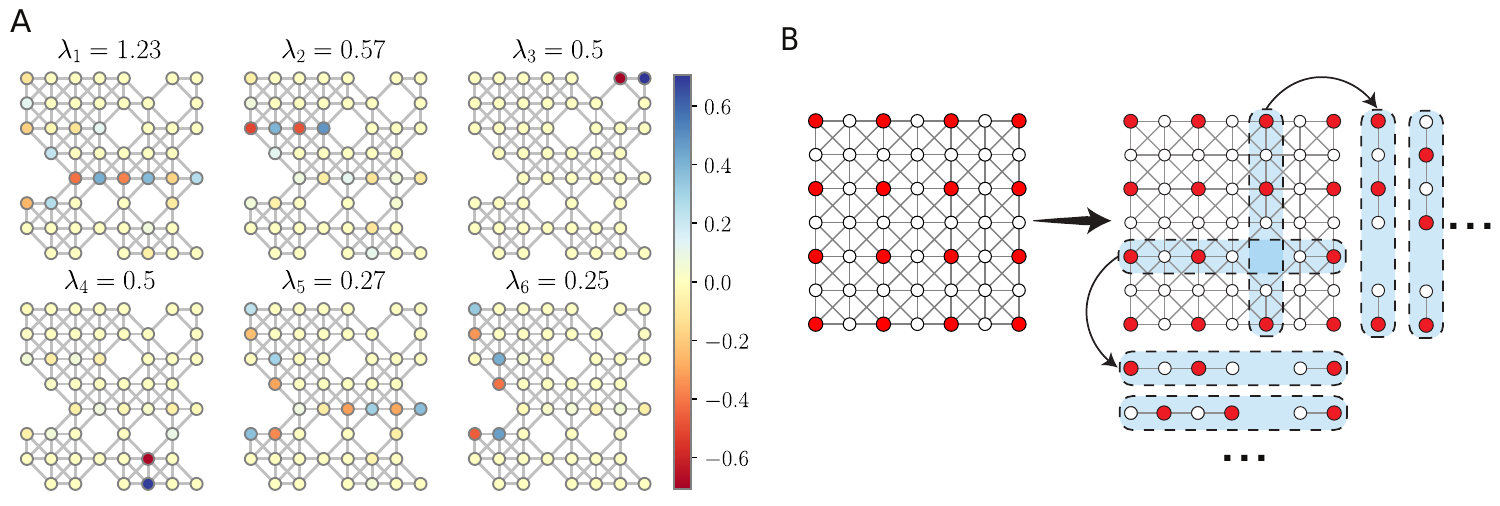}
	\caption{\textbf{Origin of degeneracy.} \textbf{A.} Principal component analysis (PCA) on experimental outcomes of a 51-atom graph shows the regions that change most between snapshots. Here, the top six principal components highlight distinct $1$D regions which contribute most to the degeneracy. \textbf{B.} Removing a vertex from a fully filled background lattice introduces degenerate linear regions emanating from the hole. This is due to the ability to freely slide excitations along columns and rows of the graphs in the vicinity of a hole.}
	\label{fig:hole}
\end{figure}

Before proceeding further, let us make a few observations about the degeneracy of the MIS solution space. From the principal component analysis shown in Fig.~\ref{fig:hole}A, we know that the degeneracy primarily originates from one-dimensional regions of the graph. One of the mechanisms by which such one-dimensional degeneracy can arise is via the presence of \textsl{holes}, defined as regions that have atoms missing compared to the perfect square lattice [see Fig.~\ref{fig:hole}B]. For instance, if a hole is located at a site $(x_1, y_1)$ in a $L$\,$\times$\,$L$ lattice, it contributes to a ``sliding degeneracy" \cite{fernandes2007monte} of approximately $[x ( L -x )$\,$+$\,$y (L - y)]/4$. A direct generalization of this argument shows that if there are two holes at positions $(x^{}_1, y^{}_1)$ and $(x^{}_1, y^{}_2)$, the degeneracy of the line segment \textit{between} them is $\sim \lvert y^{}_2$\,$-$\,$y^{}_1\rvert$. We emphasize though that the 1D ``strings'' of degeneracy need not always terminate in holes; in practice, their extent is also restricted by the interactions between multiple holes. However, the precise microscopic origin of these strings will not be important for our discussion. The key property of interest is that the existence of holes induces degeneracies along one-dimensional lines, and the degeneracy of each such segment is proportional to its linear length.

\begin{figure}[tb]
	\centering
	\includegraphics[width=0.65\linewidth]{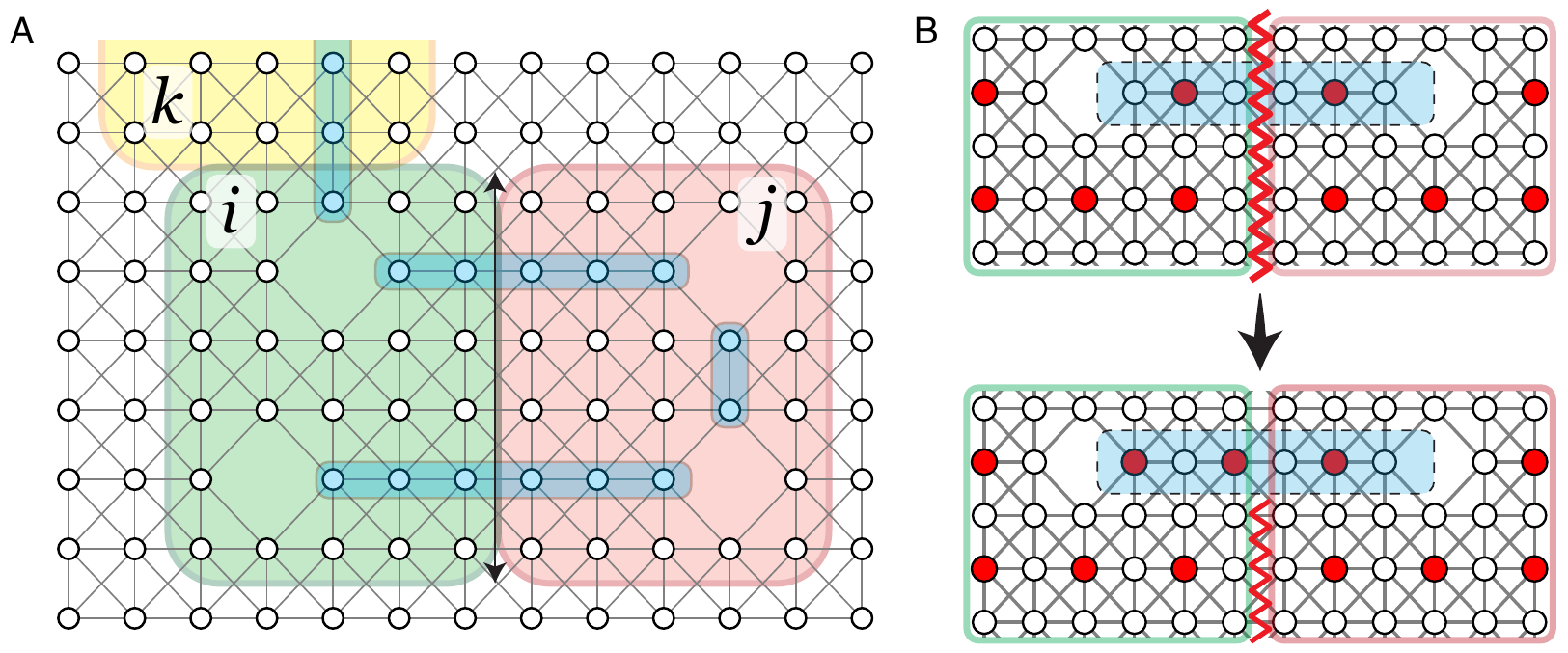}
	\caption{\textbf{A potential mechanism for healing domain walls.} \textbf{A}. Schematic illustration of one-dimensional strings which contribute to the degeneracy. These strings may either traverse the boundary between two (or more) protodomains (shaded in differing colors) or be confined to exclusively one protodomain. \textbf{B}. An example depicting how domain walls between independently seeded domains can be partially healed by sliding excitations along 1D lines. }
	\label{fig:count}
\end{figure}

While the exact distribution of holes (or larger vacancies) is a property of the individual graph, on average, the spacing between them is $2/\sqrt{\pi \rho} \equiv \zeta$ for a given density of holes $\rho$. For simplicity, we will also take $\zeta$ to be the characteristic length of the 1D strings due to reasons motivated above. Now, consider two protodomains, say, $i$ and $j$, as shown in Fig.~\ref{fig:count}: we will compute the degeneracy of the domain $i \cup j$ if the protodomains were to coalesce without kink formation. First, there will be a contribution from strings that lie entirely within each protodomain (such as the rightmost one in Fig.~\ref{fig:count}) given by $\delta_{i}\, \delta_{j}$, where $\delta_{\mu}$ denotes the intrinsic degeneracy in region $\mu$. Additionally, there is a second piece to the degeneracy stemming from the boundary between the two protodomains: this is determined by the product of the degeneracies of all the strings crossing the interface of length $\ell$ sites. Hence, the total degeneracy of the region $i \cup j$ is, to a good approximation, $d_{i,j} \simeq (\zeta)^{\gamma\, \ell} \delta_i \,\delta_j$, for some graph-dependent constant $\gamma$. Averaging over all such protodomains, we replace the $d_{i,j}$, $\delta_{i}$ by their averaged values and drop the associated site indices, leading to the useful estimate $\ell = [\log d - 2 \log \delta]/(\gamma \log \zeta)$.

\subsection{Scaling of the defect density}

Substituting this result in Eq.~\eqref{eq:r}, we find
\begin{alignat}{1}
	\nonumber\langle n r \rangle &= \sum_{k=0}^{Z \mc{N}/2 }{\dbinom{Z \mc{N}/2}{k}}\, p^k\, (1-p)^{Z \mc{N}/2-k} \left(\frac{k}{Z/2} \right)\frac{\pi \mc{N} \mc{R}-\Gamma  \mc{R}\, \left(\log d - 2 \log \delta\right)\, \left(\frac{Z \mc{N}}{2}-k\right)}{k/ (Z/2)}\\
	& = \frac{\mc{N} \mc{R}}{2} \left[2 \pi - (1-p)\,\Gamma\,Z\, \left(\log d - 2 \log \delta\right)\, \right],
\end{alignat}
where we have encapsulated all the nonuniversal graph-dependent properties in the coefficient $\Gamma$ and also reinstated a factor of $\mc{R}$ in the first line for dimensional consistency (since $\mc{L}$ grows with $\mc{R}$). Intuitively, the correction term means that when the degeneracy is higher, there are more ways for two protodomains, that may have been seeded independently, to merge smoothly without generating a domain wall---this is also one reason why graphs with large degeneracy are generically ``easier'' to solve. We now recognize that the total MIS degeneracy $\DMIS$ is 
\begin{equation}
	\DMIS \simeq \frac{\prod_{\langle i, j \rangle} d^{}_{i,j}}{\left(\prod_{i} \delta^{}_i \right)^{Z-1}} \quad \mbox{so, } \quad \log \DMIS \simeq \frac{Z \mc{N}}{2} \left(\log d - 2 \frac{Z-1}{Z} \log \delta \right) \simeq \frac{Z \mc{N}}{2} \left(\log d - 2 \log \delta \right),
\end{equation}
therefore
\begin{equation}
	\langle n r \rangle  = \mc{N} \mc{R} \left[\pi - (1-p)\, \Gamma \,\frac{\log \DMIS}{\mc{N}}\, \right].
\end{equation}
Noting that $\mc{N}$\,$\sim$\,$1/\mc{R}^2$, we can express the final result for the scaling of the net defect density as
\begin{equation}
	1-R \sim  T^{-\mu} \left[1 - (1-p)\, \Gamma \,\frac{\log \DMIS}{N}\,T^{\,2 \mu}\, \right],
	\label{eq:scaling}
\end{equation}
where $N$ is the total number of atoms and the (redefined) coefficient $\Gamma$ also absorbs the geometric factors relating $N$ and $A$.

\subsection{Comparison to the experiment}
\label{sec:exp}

\begin{figure}[htb]
	\centering
	\includegraphics[width=0.8\linewidth]{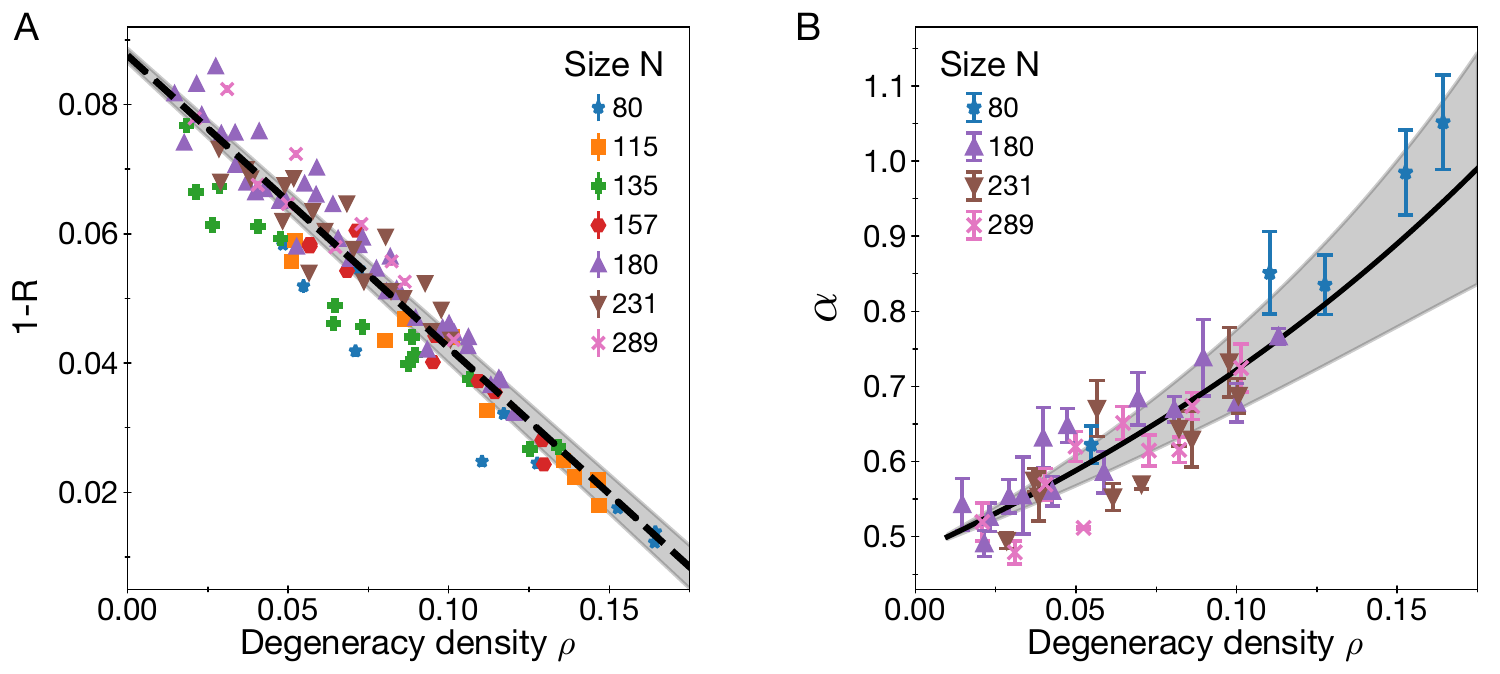}
	
	\caption{\textbf{Effect of degeneracy density on scaling dynamics.} \textbf{A.} Behavior of the approximation error $1-R$ as a function of degeneracy density $\rho \equiv (\log  \DMIS)/N$ for a fixed sweep duration, as predicted by the theoretical model of Eq.~\eqref{eq:scaling} with the parameters therein determined from a fit to experimental data. \textbf{B.} Effective power-law exponent $\alpha$ (corresponding to the effective time-scaling observed in Fig.~3~A of the main text) for graphs with different $\rho$. 
		This scaling behavior is captured by the theoretical model of Eq.~\eqref{eq:scaling} with a phenomenological value of $\mu$\,$=$\,$0.48(2)$ obtained from a fit as shown by the solid black line (the errors in $\mu$ are calculated through a bootstrap method).
		The grey shaded regions for both plots show the lower and upper bounds given the errors of the fit. 
	}
	\label{fig:data1}
\end{figure}

Assuming a phenomenological value of the exponent $\mu$, the scaling form \eqref{eq:scaling} potentially describes two key experimental observations:
\begin{itemize}
	\item For a fixed time, the error decreases linearly with degeneracy density $\rho \equiv (\log \DMIS)/N$ [Fig.~\ref{fig:data1}A]. The coefficient of this linear term (i.e., the slope obtained on plotting the approximation error $1-R$ as a function of $\rho$) becomes more negative with increasing $T$ for short depths.
	\item For varying sweep times, the correction from the term proportional to $\rho$ contributes an additional time dependence, so Eq.~\eqref{eq:scaling} does not describe a pure power law. However, one can still fit the data to a single \textit{effective} power law $T^{-\alpha}$; the exponent $\alpha$ thus obtained for different graphs increases with $\rho$ [Fig.~\ref{fig:data1}B].
\end{itemize}

A similar mechanism, albeit with different exponents, could potentially apply to the classical simulated annealing data in Fig.~\ref{fig:example_ratio_data} as well.

\section{Simulated Annealing \label{Sec:sa}}
We benchmark the experimental results against an optimized simulated annealing (SA) algorithm~\cite{Henderson2003, markov_chain_mixing}, which finds low-energy states of a cost Hamiltonian by imitating the cooling of a classical interacting spin system.  SA works by stochastically updating a spin configuration in  $\{\ket{0}, \ket{1}\}^N$ with probability based on a transition matrix $P$, which depends on a temperature $1/\beta$ and the system Hamiltonian. Here, $P_{s, s'}$ represents the probability of transitioning to $s'$ given that the current spin configuration is $s$. SA can be interpreted as a stochastic simulation of a  Markov chain on the space of all possible spin configurations. After many updates of the spin configuration, SA may converge to a stationary distribution $\pi\in\mathbb{R}^{2^N}$ satisfying $\pi = \pi P.$ The transition matrix can be designed to make SA converge to a desired stationary distribution by choosing the transition probabilities $P_{s, s'}$ to satisfy \textit{detailed balance} with respect to $\pi$, which means
\begin{align}
	% P_{s, s'}\pi_s = P_{s', s} \pi_s,
	\pi_s P_{s, s'} = \pi_{s'} P_{s', s},
	\label{eq:detailed_balance}
\end{align}
where $\pi_s$ is the population of the configuration $s$ in $\pi$. A Markov chain which satisfies detailed balance with respect to a $\pi$ is guaranteed to have $\pi$ as its unique stationary distribution, and will converge to $\pi$ at long times. Often, the Markov chain is structured so that $\pi$ is the Gibbs distribution of the cost Hamiltonian at some fixed temperature. In addition, our SA Markov chains will be designed to be \textit{ergodic}, which requires that it is possible to travel from any configuration $s$ to any other configuration $s'$ with a finite probability within a finite number of stochastic updates, and to be \textit{lazy}, which requires that the update rule leaves the configuration unchanged with probability at least $1/2$. Having SA algorithms with these properties will make it possible to prove lower bounds on the runtime of SA using results from the theory of Markov chains in Section~\ref{subsec:classical_scaling}.

\begin{figure}
	\centering
	\includegraphics[width=6.6in]{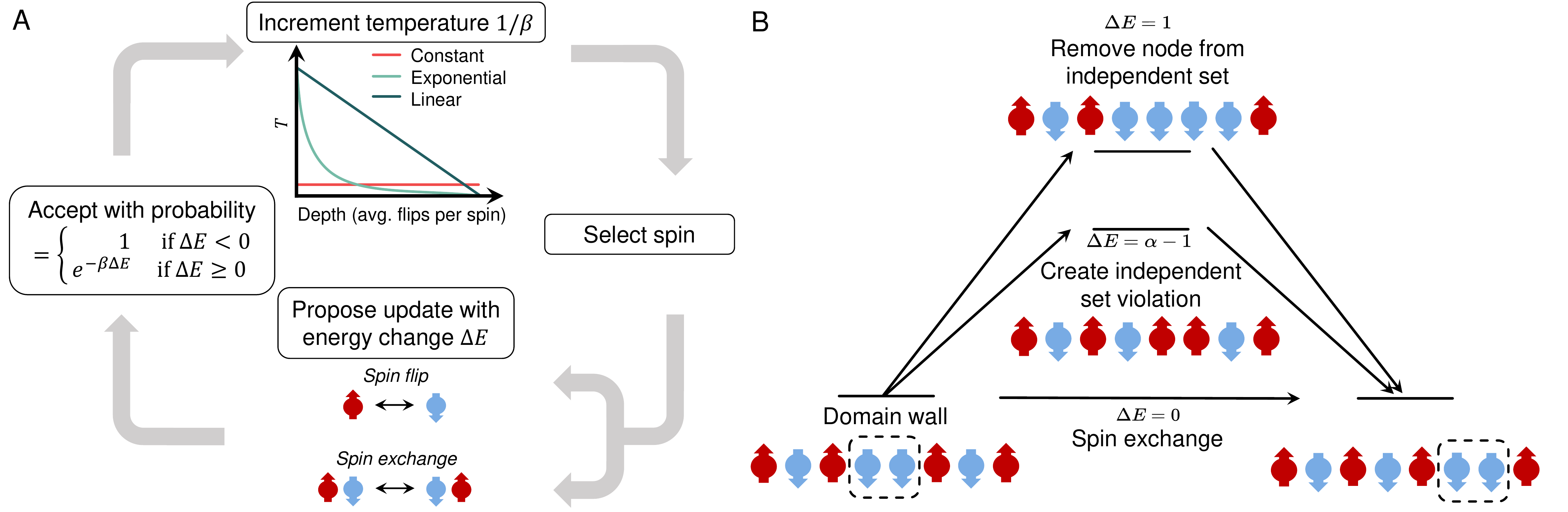}
	\caption{\textbf{Simulated annealing algorithm and dynamics. }\textbf{A.} Outline of the simulated annealing algorithm on a classical spin system. All spins are initialized in the state that corresponds to no vertices being in the independent set; then, a collective update is iteratively applied. \textbf{B.} In one dimension, local minima are caused by domain walls in the antiferromagnetic ordering. The local minima can be escaped by moving the domain wall to the edge of the system using one of two processes. The first process involves removing a vertex from the independent set and adding in a new vertex to the independent set. As it is energetically costly to remove a vertex ($\Delta E = 1$), this process occurs at a slow rate. As $\alpha-1\rightarrow 0$, it is more favorable to move the domain wall by adding a temporary blockade violation. Domain walls can be moved at zero energy cost via spin exchanges.}
	\label{fig:simulated_annealing_diagram}
\end{figure}

We implement two different classes of Hamiltonians encoding the MIS problem to compare with the experimental results. We say a vertex is in the set if it is in state $\ket{1}$ and out of the set if it is in state $\ket{0}$, such that each spin configuration in $\{\ket{0}, \ket{1}^N$ corresponds to a set of vertices.  The first Hamiltonian is  proportional to the Rydberg Hamiltonian with interaction energies $V_{ij}$ and an added constant detuning $\Delta$,
\begin{equation}
	H_{\text{cost}}^{\text{Ryd}} =-\sum_i n_i+ \sum_{i< j}\frac{V_{ij}}{\hbar\,\Delta}n_in_j\label{eq:HRyd}.
\end{equation}
We use an optimized value of $\Delta/2\pi = 11$\,MHz, which is similar to the final detuning used in the experimental pulse sequence.
For a graph $G=(V, E)$, the second Hamiltonian is the ``standard" MIS Hamiltonian
\begin{align}
	H_{\text{cost}}^{\text{MIS}} = -\sum_{i\in V} n_i+\sum_{(i, j)\in E}\alpha\, n_i n_j,\label{eq:HMIS}
\end{align}
where $\alpha$ is a uniform penalty on each edge. To guarantee that the ground state of Eq.~\eqref{eq:HMIS} corresponds to the MIS, we must have $\alpha > 1,$ so that it is strictly more energetically favorable to have at most one vertex per edge in state $\ket{1}$, as opposed to both vertices in state $\ket{1}$. Under these conditions, the ground state maximizes the number of spins in the corresponding set subject to the independent set constraint. \\ 

\subsection{Description of the SA algorithms\label{subsec:SA_description}}
For each of the two cost Hamiltonians, we use a specifically optimized variant of SA that follows the general outline of Figure~\ref{fig:simulated_annealing_diagram}A.
First, spins are initialized in a configuration corresponding to having no vertices in the independent set, and the temperature is set to $1/\beta_i$. Then, a vertex is selected from a fixed probability distribution that depends on the variant of SA (see below). An update of the local spin configuration around the chosen vertex is proposed and the change in energy $\Delta E$ under the cost Hamiltonian from the proposed update is computed. The update is accepted with probability
\begin{equation}
	\text{probability} = \begin{cases}
		e^{-\beta\Delta E}, & \Delta E \geq 0 \\
		1, & \Delta E < 0
	\end{cases}.\label{eq:acceptance_probability}
\end{equation}
Therefore, updates which lower the system's energy are accepted greedily whereas energetically unfavorable spin flips are accepted with a probability dependent on the energy increase and temperature. The temperature is incrementally lowered after each attempted update until the final temperature $1/\beta_f$ is reached. The average number of attempted updates per spin is called the \textit{depth} of the algorithm, $p_{\text{SA}}$.\\

\noindent\textbf{MIS SA algorithm.}
The variant of SA for the MIS Hamiltonian~\eqref{eq:HMIS} (MIS SA) uses an optimized Metropolis-Hastings~\cite{Hastings, Metropolis} update rule, which ensures that the algorithm converges to a unique stationary distribution under the detailed balance condition. The algorithm works by first selecting a vertex uniformly at random. If the vertex is in the set (corresponding to a spin state of $\ket{1}$), 
then the proposed updates are
\begin{itemize}
	\item Remove the vertex from the set with probability $\epsilon$;
	\item Spin exchange with a neighboring vertex,  with probability; $(1-\epsilon)/8$ for each neighbor
	\item No update with the remaining probability.
\end{itemize}
The proposed update is then accepted or rejected with probability given by Eq.~\eqref{eq:acceptance_probability}.
Note that due the geometric layout of the ensemble of unit disk graphs studied in this work, the maximum degree of any vertex is eight, so the probability of proposing no update is non-negative. If the vertex is not in the independent set, adding the vertex to the independent set is proposed with unit probability. 
We will optimize the MIS SA performance over $\epsilon$ in the following section.

The Markov chain associated with MIS SA is ergodic at finite $\epsilon$.
%We will show that
Furthermore, the Markov chain satisfies detailed balance, %condition,
which uniquely specifies its stationary distribution.
%of the Markov chain.
To see this,
consider two spin configurations $s, s'$ where $\Delta E = E_s-E_{s'}$ and $E_s$ is the energy of the spin configuration $s$. Based on the update rule described above and the detailed balance condition, the steady state populations are related by 
\begin{equation}
	\frac{\pi_s}{\pi_{s'}} = e^{-\beta \Delta E}\times \begin{cases}
		1 & s = \text{spin exchange}(s')\\
		\epsilon & s = \text{vertex removal}(s')\\
		\epsilon^{-1} & s = \text{vertex addition}(s').
	\end{cases}
\end{equation}
Putting everything together, the stationary distribution associated with the detailed balance, for a spin configuration $s$, is 
\begin{equation}
	\pi_s = \frac{1}{Z}\epsilon^{-|s|}e^{-\beta  E_s},\label{eq:sstate}
\end{equation}
where $|s|$ is the number of spins in $s$ in state $\ket{1}$, $Z$ is a normalization factor so that the sum of probabilities of the stationary distribution is one.
The stationary distribution is equal to the Gibbs distribution when $\epsilon=1$, and corresponds to a uniform mixture of MISs as $\epsilon \rightarrow 0$ or $1/\beta\rightarrow 0.$ In practice, the Markov chain is also lazy at low temperatures for large independent sets because more than half of the proposed spin exchange and vertex addition updates will not be accepted because the blockade penalty $\alpha$ makes most spin exchanges and node additions energetically unfavorable.  These features make MIS SA amenable to bounding the eigenvalue gap of its Markov chain in Section~\ref{subsec:classical_scaling}, which enables strict upper bounds on the performance of our implementation of MIS SA.\\

\noindent\textbf{Rydberg SA algorithm.} The update rule for SA with the Rydberg Hamiltonian (Rydberg SA) is further optimized at the expense of losing the detailed balance condition. We forego having detailed balance because we do not attempt to analytically bound the performance of Rydberg SA 
%in Section~\ref{Sec:scaling}
due to the more complicated energy spectrum of Eq.~\ref{eq:HRyd}.
%, so we do not need to know the details of its stationary distribution.
The update rule begins by selecting a spin uniformly from the set of free vertices (vertices in the set with no neighbors in the set) and vertices in the set. If a free vertex is selected, adding it to the set by changing its state to $\ket{1}$ is proposed. If a vertex in the set is selected, a spin exchange with each neighbor is proposed with probability $1/8$. With the remaining probability, a vertex removal by changing the spin state to $\ket{0}$ is proposed. By choosing to update only vertices in the independent set and free vertices, the dynamics of Rydberg SA can be accelerated because the remaining vertices can only be added to the independent set with large blockade interaction energy penalties, so these updates are almost always rejected in practice. \\

\noindent\textbf{Post-processing of simulated annealing data.} Once the chosen depth of SA is reached, we post-process the final spin configuration with a constant depth greedy algorithm, similar to the routine used to post-process the experimental outputs. During the greedy algorithm, independent set violations are greedily removed in descending order of the number of blockade violations per vertex. Then, vertices are greedily added back into the independent set in order of increasing degree.  Note that under Eq.~\eqref{eq:acceptance_probability}, independent set violations are removed and free vertices are added to the independent set greedily
during the course of the algorithm itself. Therefore, if the final temperature is very low compared to the cost of adding an independent set violation or removing a vertex from the independent set (local excitations), the probability of SA outputting anything other than a maximal independent set is highly suppressed, and the likelihood of post-processing a given vertex is extremely rare. This is the case for the low temperature limit of Rydberg SA and MIS SA, where in the following section we find that a large energy penalty on independent set violations is optimal ($\alpha \gg 1$ for MIS SA, $\Delta\lesssim V_{ij}$  for Rydberg SA, where $i, j$ are nearest or next-nearest neighbors). 

\subsection{Optimization of the SA algorithms \label{subsec:SA_optimization}}
In this section, we will optimize the performance of the SA algorithms over several parameters. For MIS SA, we will optimize over the independent set violation penalty $\alpha$ in Eq.~\eqref{eq:HMIS} as well as the probability of removing a vertex $\epsilon$ from the set in the update rule, described in the previous section. Finally, we will optimize how temperature is lowered with depth for both MIS SA and Rydberg SA.\\

\noindent \textbf{MIS SA optimization.}
We first optimize MIS SA over all $\alpha\geq 1$ and $\epsilon \in (0, 1].$ 
The effect of changing $\alpha$ can be understood intuitively in one dimension, where raising $\alpha$ introduces kinetic constraints that can make climbing out of local minima via single spin flips energetically unfavorable.
In a one-dimensional system, the global solution of the MIS problem corresponds to the antiferromagnetic arrangement of spin variables. The local optima correspond to having a few isolated domain wall configurations, i.e., two consecutive spins not in the independent set, as shown in the left part of Fig.~\ref{fig:simulated_annealing_diagram}B.
In order to obtain the exact global solution starting from one of the local optima, the SA process has to eliminate domain wall configurations by moving them and ultimately annihilating them by combining pairs of domain walls.
This process is limited by the effective speed of moving domain walls.

For MIS SA, a domain wall can be moved by three different  processes:
\begin{enumerate}[(1)]
	\item One of the Rydberg excitations is eliminated at the cost of unit energy $1$, and then a new excitation is created (Fig.~\ref{fig:simulated_annealing_diagram}B top path)
	\item A new excitation is created by violating the independent set  constraint at the cost of energy $\alpha-1$, and then the constraint is restored by removing a different excitation (Fig.~\ref{fig:simulated_annealing_diagram}B middle path)
	\item The spins are directly exchanged (Fig.~\ref{fig:simulated_annealing_diagram}B bottom path).
\end{enumerate}
The speed at which domain walls propagate  depends on  $\alpha$  and the probability of proposing a vertex removal $\epsilon$. The dynamics are slow when $\alpha$ and $\epsilon$ are both large, corresponding to the cases where independent set violations are energetically highly unfavorable and spin exchanges are unlikely to occur. Then, the dominant process is the process (1) which requires removing a vertex with zero blockade violations at energy penalty $\Delta E=1$. The dynamics are also slow when $\alpha$ is small and $\epsilon$ is small, because single blockade violations can be added but not removed, which makes subsequent spin exchanges with neighboring vertices more difficult because more vertices are in the set.  In contrast, when $\alpha$ is large and $\epsilon$ is small, the dynamics are fast and result from spin exchanges at no energy cost (3). When $\alpha\rightarrow 1$ and $\epsilon$ is small, spin exchanges (3) or spin flips (2)  enable fast dynamics, and the dominant process is decided by $\epsilon$.  

\begin{figure}
	\centering
	\includegraphics[width=4.5in]{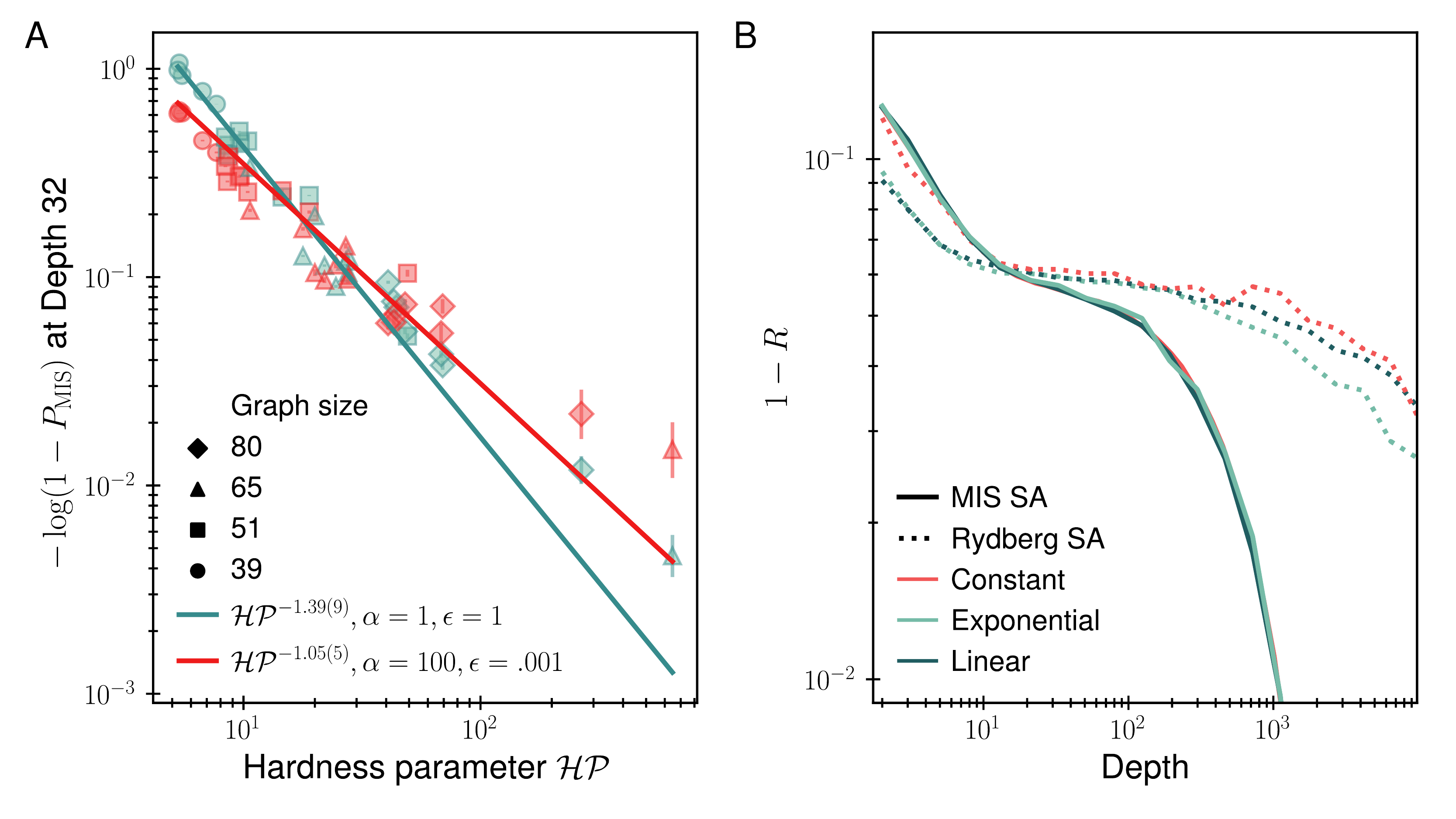}
	\caption{\textbf{Optimization over MIS SA parameters and temperature. }\textbf{A.} Using $\epsilon=1, \alpha=1$  yields worse scaling for $-\log(1-\PMIS)$ as a function of $\mathcal{HP}$ compared to $\epsilon=0.001, \alpha=100$. \textbf{B.} Approximation ratio for a single graph instance obtained with MIS SA for each different schedule (constant, exponential, and linear) at optimized initial and final temperatures as a function of depth. An exponential schedule is optimal for Rydberg SA whereas all three schedule types show similar performance for MIS SA. }
	\label{fig:temperature_optimization}
\end{figure}

Therefore, we optimize over two distinct regimes: $\alpha$ small and any $\epsilon$, and $\alpha$ large with $\epsilon$ small. We generate 80 instances on graph sizes between $N=39$--$80$ vertices in the top two percentiles  maximizing $\mathcal{HP}$ for each system size.
We see numerically that the performance of $\alpha = 100, \epsilon=0.001$ yields the best scaling of $\PMIS$ with $\mathcal{HP}$, so we use these parameters in the main text. We display example fits in Figure~\ref{fig:temperature_optimization}A: $\alpha =1, \epsilon=1$ and $\alpha = 100, \epsilon = 0.001.$ We attribute the difference in scaling to the fact that for $\alpha =1$ and intermediate or small $\epsilon$, there are many more accessible low energy states, corresponding to $|\MIS|$ and $|\MIS|-1$ configurations where each vertex has at most one blockade violation. Including these states likely introduces relatively more local minima than global minima, which would increase the difficulty of finding an $\MIS$. \\

\noindent \textbf{Temperature optimization. }
Next, we optimize the rate at which temperature is lowered as a function of depth. We benchmark three different temperature schedules as a function of depth: constant, exponentially lowered, and linearly lowered, as shown in  Fig.~\ref{fig:simulated_annealing_diagram}. For each schedule type, we optimize the initial and final temperatures using a grid search on four different graphs with between $N=51$--$180$ nodes. Two of the graphs were generated randomly and two of the graphs were chosen from the distribution of graphs maximizing $\mathcal{HP}$. We identify the optimal initial and final temperature for each temperature schedule via a grid search by averaging the approximation ratio over all graphs between depths $10^3$--$10^4$.

The performance of MIS SA is similar between different temperature schedules (Fig.~\ref{fig:temperature_optimization}B). This is because in practice, we set $\epsilon=0.001, \alpha=100$, so the dynamics are essentially restricted to spin configurations corresponding to valid independent sets. Because $\epsilon$ is so small, raising the energy via a node removal is highly unlikely, even at higher temperatures. Therefore, the dynamics for all three temperature schedules are comparable. In practice, we implement a constant, near-zero temperature schedule ($1/\beta=10^{-8}$).

For Rydberg SA, we find that the exponential schedule with initial and final temperatures $1/\beta_i = 0.32, 1/\beta_f = 0.03$ is optimal. To understand this, we summarize the dynamics of Rydberg SA.
The maximum blockade violation penalty for nearest neighbors is much larger than the initial temperature $1/\beta_i=0.32\ll V_{ij}/\Delta\sim 9.8$, so blockade violations with nearest neighbors are unlikely to occur even at early depths, and the dynamics are primarily driven by spin exchanges. The long Rydberg tails cause the energy landscape to be uneven among independent sets of the same size, so it is necessary to operate at higher temperatures during the algorithm to traverse between independent sets of the same size. There is a balance between annealing at low temperatures $1/\beta\rightarrow 0$, where the stationary distribution has large overlap with the MIS, and higher temperatures, where SA can easily climb out of local minima caused by the algebraically decaying long-range interactions via spin exchanges. This motivates why a temperature schedule operating at higher temperatures than MIS SA is optimal. 
We then use this optimized temperature schedule in the main text to benchmark against the quantum algorithm.

\begin{figure}
	\centering
	\includegraphics[width=6in]{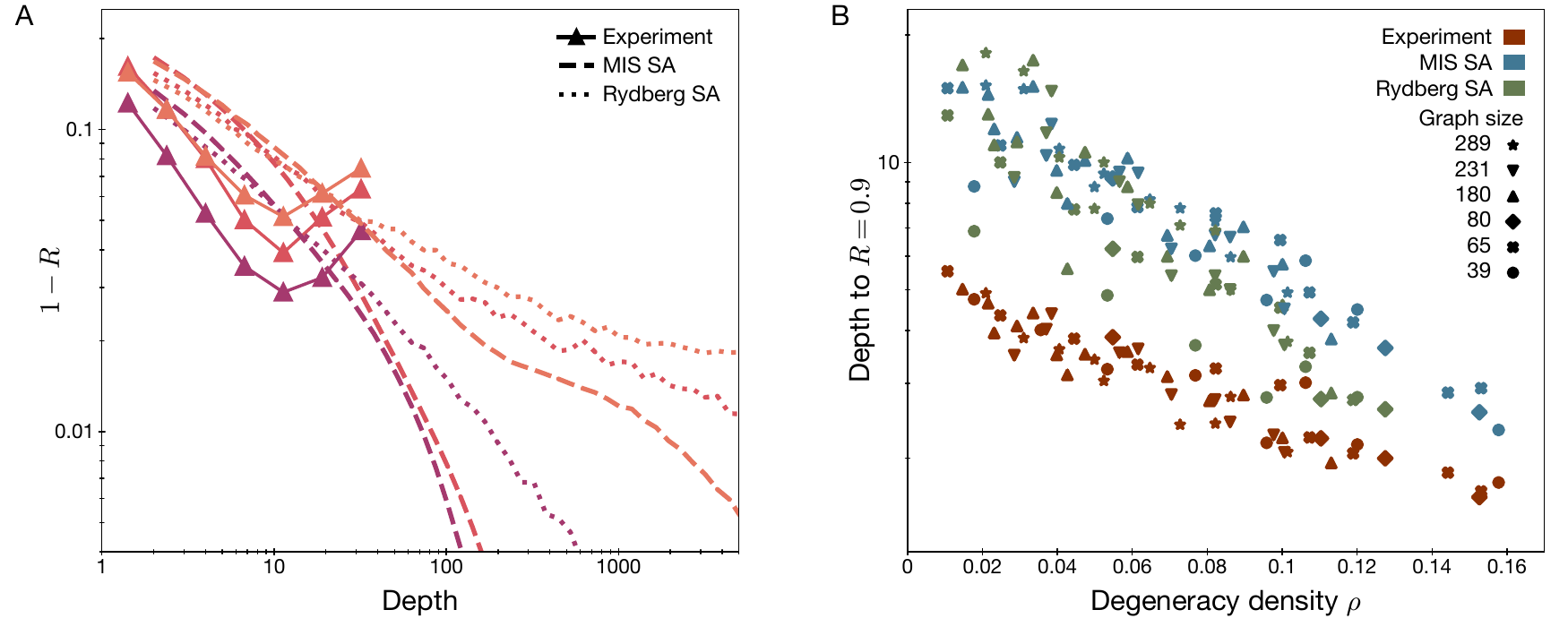}
	\caption{\textbf{Benchmarking approximation ratio. }\textbf{A.} Approximation ratio $1-R$ as a function of depth for the quantum algorithm (triangles), MIS SA (dashed lines), and Rydberg SA (dotted lines) on three 180-vertex instances (orange $\rho = 0.042$, red $\rho= 0.069$, purple $\rho= 0.113$) \textbf{B.} Depth to $R=0.90$ for instances on up to 231 vertices for the quantum algorithm (crimson), MIS SA (teal), and Rydberg SA (green) as a function of the degeneracy density $\rho$. We see a power law scaling in depth to $R=0.9$ as a function of degeneracy density, independent of system size, as motivated in Section~\ref{Sec:coarsening}. }
	\label{fig:example_ratio_data}
\end{figure}

Figure~\ref{fig:example_ratio_data}A shows example data for the quantum algorithm and both SA variants on three 180-vertex instances with different degeneracy densities $\rho$ (orange $\rho = 0.042$, red $\rho= 0.069$, purple $\rho= 0.113$). We find that at early depths, the experiment appears to have  slightly better power law scaling for $1-R$ versus depths over the SA algorithms. This can also be seen in Figure~\ref{fig:example_ratio_data}B, where we plot data for 115 instances between graph sizes of $N=39$--$231$, including both randomly selected instances and instances from the hardest 2\% of graphs maximizing $\mathcal{HP}$. As the degeneracy density decreases and the problem becomes more difficult, both SA variants take increasingly long to reach a fixed approximation ratio compared to the experiment.

\section{Scaling of Quantum and Classical MIS Probability \label{Sec:scaling}}

\subsection{Classical scaling\label{subsec:classical_scaling}}

In this section, we show how the performance of SA using the MIS Hamiltonian is related to $\mathcal{HP}$ at low temperatures. For the specific variant of MIS SA implemented numerically in this work with $\alpha\rightarrow\infty$, we show that the spectral gap of the Markov chain matrix is at most $2\mathcal{HP}^{-1}$. We then generalize this result to a larger class of SA algorithms which collectively update constant-sized clusters of spins using the MIS Hamiltonian, for all $\alpha \geq 2$ (a similar proof applies for $1<\alpha <2$, which we omit). In particular, we show that the spectral gap of a reversible, lazy, ergodic Markov chain matrix is $O(\text{poly}(N)\mathcal{HP}^{-1})$, where $\text{poly}(N)$ is some polynomial in $N$, provided that the stationary distribution is sufficiently ``close'' to the Gibbs distribution for the associated Hamiltonian, which includes not only Metropolis-Hastings update rules designed to perfectly sample from Gibbs distributions, but also the cases, for example, when the update rule only approximately satisfies detailed balance.
We then show how the Markov chain matrix spectral gap controls the  MIS probability as a function of depth for the implemented MIS SA algorithm, and argue that the hitting time for finding the MIS is at least $\Omega(\mathcal{HP})$ at zero temperature.
\\

\noindent\textbf{Upper bound on the spectral gap. }
\label{subsubsec:mixing_time}
We first upper bound the spectral gap $\delta_{\min}^{\text{SA}}$ of the Markov chain associated with our MIS SA algorithm when $\alpha\rightarrow\infty$. Because our MIS SA algorithm satisfies detailed balance, at high depths it is guaranteed to converge to a stationary distribution $\pi$ given by Eq.~\eqref{eq:sstate}.
We will then illustrate how the spectral gap controls the rate of convergence to the stationary distribution, which at $1/\beta\rightarrow 0$ is a uniform mixture of MISs. %We will also show how the spectral gap directly controls the MIS probability and analytically motivate the functional form fit in Figure~4D in the main text.
In the particular case of the implemented MIS SA algorithm, the bound we obtained is 
\begin{equation}
	\boxed{\delta_{\min}^{\text{SA}}\leq 2\mathcal{HP}^{-1}}.\label{eq:MIS_SA_bound}
\end{equation}
Our bounds on the spectral gap are all based on the Cheeger inequality \cite{cheeger_ref_diaconis}, which states
\begin{equation}
	\frac{\Phi^2}{2} \le \frac{\delta_{\min}^{\text{SA}}}{N}\leq 2\Phi,\label{eq:cheeger}
\end{equation}
where the Cheeger constant $\Phi$ is
\begin{equation}
	\Phi=\min_{S\subset\Omega,\pi_S<\frac{1}{2}}\frac{Q_{S,S^c}}{\pi_S},\quad Q_{S,S'}=\sum_{s\in S, s'\in S'}\pi_s P_{s, s'},
\end{equation}
where $P_{s, s'}$ is the probability of traveling from state $s$ to $s'$ under one iteration of a lazy, ergodic, and reversible Markov chain defined on the set of all possible spin configurations. Here, $S$ is a subset of all possible spin configurations $\Omega$, and $S^c$ is its complement. The factor of $\frac{1}{N}$ in Eq.~\eqref{eq:cheeger}, which does not appear in standard statements of the Cheeger bound, arises from the convention that SA depth is defined as the average number of spin flips per vertex. The total equilibrium population of set $S$ is 
\begin{align}
	\pi_S = \sum_{s\in S}\pi_s.
\end{align}
We can upper bound the Cheeger constant (and therefore the spectral gap) by identifying bottlenecks in the probability transfer between bipartitions of the state space. Here, these bottlenecks will occur between independent sets of size $\rm{|{MIS}|}-1$ and the MISs. \\

\begin{figure}
	\centering
	\includegraphics[width=6.5in]{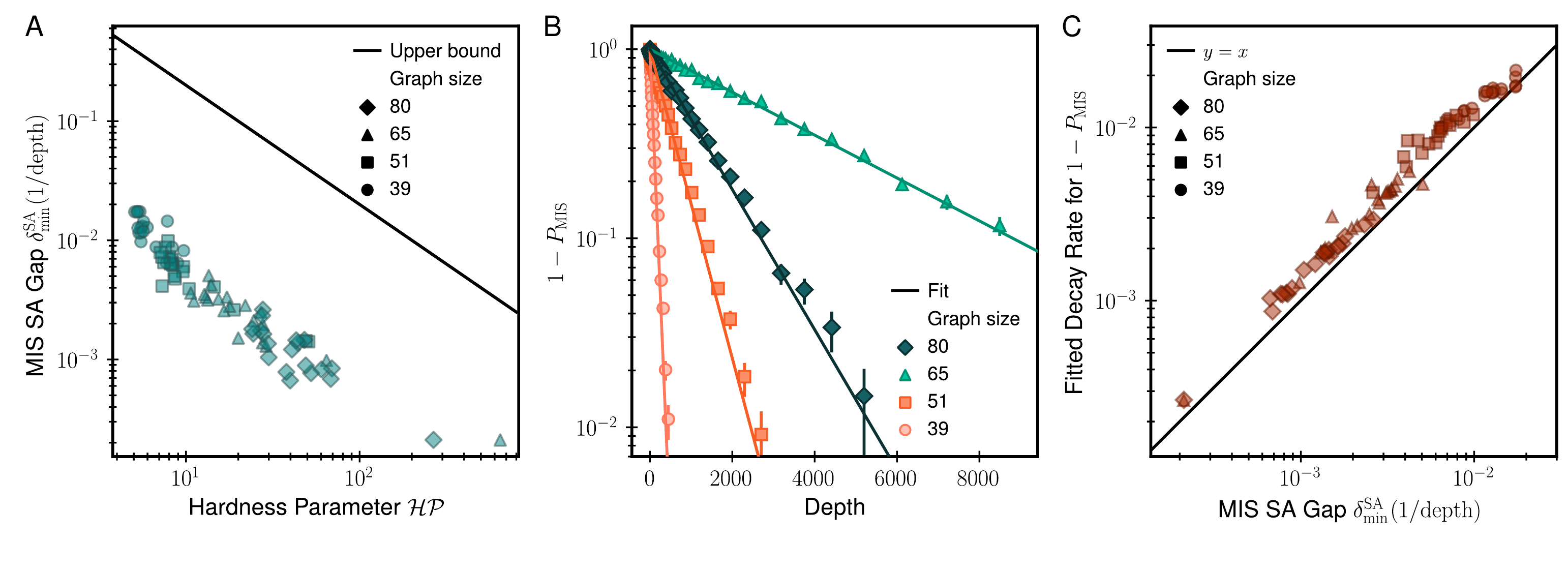}
	\caption{\textbf{Simulated annealing spectral gaps and MIS probability. }\textbf{A.} MIS SA Markov chain spectral gap for 80 instances drawn from the top $2\%$ of graphs maximizing $\mathcal{HP}$, as a function of hardness parameter $\mathcal{HP}$. The spectral gaps are proportional to $\mathcal{HP}^{-1}$, consistent with the analytic upper bound on the spectral gap (solid line) \textbf{B.} Data for four graphs fitting the functional form $1-\exp(\text{const.}\times p_{\text{SA}})$, where $p_{\text{SA}}$ is the depth for MIS SA. \textbf{C.} The fitted constant in the exponential decay is approximately equal to the spectral gap (solid line). Error bars on the fitted exponent are smaller than the marker size.}
	\label{fig:LZ_benchmarking}
\end{figure}

\noindent {\it Proof for MIS SA for $\alpha\rightarrow\infty$} ---
For MIS SA, if we set $S=S_{|\MIS|}^c$ to be the set of all states that are not MIS, then at sufficiently low temperatures, $\pi_S \leq 1/2.$ The equilibrium population of $S$ is
\begin{equation}
	\pi_S=1-\pi_{S_{|\MIS|}}>\pi_{S_{|\MIS|-1}}=\DMISminusone\frac{\epsilon^{-(|\MIS|-1)}e^{\beta(|\MIS |-1)}}{Z},\label{eq:MIS_population}
\end{equation}
where $Z$ is the partition function and $S_{|\MIS|}$ is the set of all MISs. We have substituted the derived stationary distribution for MIS SA (Eq.~\eqref{eq:sstate}). In the $\alpha \rightarrow \infty$ limit, the only states that connect to the MISs are nonmaximal independent sets of size $|\MIS|-1$, so we have
\begin{align}
	Q_{S,S^c}
	&\leq \DMIS|\MIS|\frac{\epsilon^{-(|\MIS|-1)}e^{\beta(|\MIS|-1)}}{Z}\frac{1}{N}.
\end{align}
The factor of $\DMIS|\MIS|$ is the number of possible transitions into MISs, and the factor of $\frac{1}{N}$ corresponds to the probability of selecting a specific vertex to update in the Markov chain. 
Putting everything together, we get
\begin{align}
	\frac{\delta_{\min}^{\text{SA}}}{N}\leq 2\Phi&\leq\frac{Q_{S,S^c}}{\pi_S} \leq \frac{2\mathcal{HP}^{-1}}{N},
\end{align}
recovering the desired bound. A similar calculation yields an identical bound at higher temperatures when $\pi_{S_{|\MIS|}}\leq 1/2$ by taking $S=S_{|\MIS|}.$
Figure~\ref{fig:LZ_benchmarking}A shows the numerically computed MIS SA spectral gaps at $1/\beta=0$, which go as $\mathcal{HP}^{-1}$, along with the Cheeger bound. The bound is somewhat loose, but it captures the expected scaling of the gap with $\mathcal{HP}$. \\

% Define Pi(MIS-1)
\noindent {\it Proof for general SA using the MIS Hamiltonian }-- 
We now show for any energy penalty $\alpha \geq 2$ and any SA algorithm which updates $O(1)$ spins at each step, the spectral gap of its Markov chain is bounded by
\begin{equation}
	\boxed{\delta_{\min}^{\text{SA}}\leq \text{poly}(N)\mathcal{HP}^{-1}}\label{eq:general_bound}
\end{equation}
at sufficiently low $1/\beta$, where $\text{poly}(N)$ is some polynomial in $N$. Because $\mathcal{HP}$ appears to grow at least superpolynomially in $N$ (Figure~\ref{fig:degeneracy_vs_L}B), the dominant contribution to the bound comes from the hardness parameter. Our proof relies on the assumption that the stationary distribution does not exponentially favor states related to the MIS by $O(1)$ bit flips over other states with the same energy. 

Despite the additional freedom afforded these algorithms by allowing more update rules and energy landscapes, we will show that the above bound still applies up to polynomial corrections in $N$.
We assume that the stationary distribution of our SA Markov chain is polynomially close to the Gibbs distribution,
\begin{align}
	\pi_s =\frac{1}{Z}\text{poly}(N) e^{- \beta E_s},
\end{align}
where as before $E_s$ is the energy of the configuration $s$ and $Z$ is the partition function. This assumption is reasonable since it ensures configurations with the same energy have comparable equilibrium probabilities, and the equilibrium population favors low energy states in the low temperature ($\beta\to \infty$) limit. Otherwise, one could in principle construct a SA algorithm which preferentially favors non-maximal states and avoids getting trapped in local minima bottlenecks. Most SA algorithms (e.g.\ the Metropolis-Hastings algorithm) can easily be designed to sample from the Gibbs distribution, so this proof includes a wide class of algorithms.

First, assume $\alpha\geq 2$, and let the set of states which can directly transition to MISs be denoted as $R.$ Again, we take $S=S_{|\MIS|}^c$ and focus on low temperatures where $\pi_S\leq 1/2.$ We have $\pi_S > \pi_{S_{|\MIS|-1}}$, and 
\begin{align}
	\frac{Q_{S, S^c}}{\pi_S} &= \frac{\sum_{s\in R, s'\in S_{|\MIS|}}\pi_s P_{s, s'}}{\pi_S}\nonumber\\
	&\leq \frac{\sum_{s\in R, s'\in S_{|\MIS|}}\pi_s}{\pi_{S_{|\MIS|-1}}}\label{eq:removep}\\
	&\leq \frac{O(\text{poly}(N))\pi_R}{\pi_{S_{|\MIS|-1}}}\label{eq:step1}
\end{align}% break it up into multiple lines!!
where $\pi_R$ is the total equilibrium population of all states in $R$, and in Eq.~\eqref{eq:removep} we have used that $P_{s, s'}\leq 1.$
In Eq.~\eqref{eq:step1}, we use the fact that each state in $R$ only can connect to a polynomially large number of MISs if it updates a constant number of spins. Simplifying, we have
\begin{align}
	&\leq \frac{O(\text{poly}(N))|R|e^{\beta(|\MIS|-1)}}{\pi_{S_{|\MIS|-1}}Z}\label{eq:step2}\\
	&\leq \frac{O(\text{poly}(N))|R|}{\DMISminusone}\\
	&\leq \frac{O(\text{poly}(N))\DMIS}{\DMISminusone}\label{eq:step3},
\end{align}
where $|R|$ denotes the size of the set $R$.
In Eq.~\eqref{eq:step2}, we use the fact that all states in $R$ have energy at least $-(|\MIS|-1)$. Note this relies on $\alpha\geq 2$; otherwise, we could have states violating the independent set condition in $R$ with energies between $-|\MIS|$ and $-(|\MIS|-1)$ (for example,  an MIS with an added vertex, which creates a single blockade violation with energy $\alpha-1$). In Eq.~\eqref{eq:step3}, we used the fact that $|R| = \text{poly}(N)\DMIS$. As a result, we have 
\begin{align}
	\delta_{\min}^{\text{SA}}\leq \text{poly}(N)\mathcal{HP}^{-1}.\label{eq:general_bound_repeat}
\end{align}
As before, similar arguments work at high temperatures by taking $S=S_{|\MIS|}.$ The key point here is at most a polynomial number of states can lead into each MIS, and the equilibrium population of these states is polynomially related to that of maximal independent sets of size $|\MIS| -1$.

While these proofs apply to $\alpha \geq 2$, some subtleties arise as described above when considering $1<\alpha < 2$ (where the ground state is still guaranteed to encode the MISs). The proof is very similar, but we omit the full detail here. Essentially, one obtains the correct bound by partitioning the states corresponding to MIS with added vertices that incur at most one blockade violation per addition into $S^c$ along with the MISs. Using this approach, one can show that the bound in Eq.~\eqref{eq:general_bound_repeat} can be recovered at low temperatures where $\beta \gg\frac{1}{\alpha-1}\ln\text{poly}(N)$.   \\

\noindent\textbf{Hitting time lower bound.}
Our results on the spectral gap of the MIS SA algorithm  translate directly into lower bounds of $\Omega(\mathcal{HP})$ on the expected hitting time, the average time for SA to first find an MIS.  At $1/\beta=0$, the Markov chain is absorbing, so once the algorithm reaches an MIS it can no longer escape. In the $|\MIS|$ and $|\MIS|-1$ subspaces,
%the SA Markov chain is absorbing into the subspace of MIS states, and
the corresponding transition matrix $M$ is given by
\begin{equation}
	M = \begin{bmatrix}
		M_{|\MIS| -1} & M_0 \label{eq:markov_chain_matrix}\\
		0 & M_{|\MIS| }
	\end{bmatrix},
\end{equation} 
where $M_0$ encodes the transition probabilities from the $|\MIS|-1$ subspace to the $|\MIS|$ subspace. At zero temperature, there is zero probability of exciting from an MIS, so the lower left quadrant of $M$ is zero. 
Markov chains of this exact form are considered in~\cite{sze04}, which relates the Markov chain eigenvalues to the hitting time of states in $M_{|\MIS|}$. It follows from Lemma 5~\cite{sze04} that if $M_{|\MIS|-1}$ has an eigenvector (with eigenvalue $1-\delta_{u}^{\text{SA}}$) which has a uniform component whose length is bounded below by a constant, then the hitting time is $\Omega(N^{-1}(\delta_{u}^{\text{SA}})^{-1})$, where the factor of system size comes from our definition of depth. Here, the uniform component is the overlap of the eigenvector with the uniform vector $u = \frac{1}{\DMISminusone}[1, \dots, 1]$.

We can show that $M_{|\MIS| -1}$ has an eigenvector with high overlap with the uniform distribution over the space of independent sets of size $|\MIS|-1$ for our implemented MIS SA algorithm. Using the MIS SA update rule described in Section~\ref{subsec:SA_description}, we can compute the action of $M_{|\MIS| -1}$ on $u$. Note that $M_{|\MIS| -1}$ is symmetric, because $P_{s, s'}\neq 0$ between two $|\MIS| -1$ configurations if and only if they are linked by spin exchanges, in which case $P_{s, s'}=P_{s', s}=\frac{1-\epsilon}{8N}$. Therefore, for maximal $|\MIS| -1$ configurations, the corresponding columns and rows in $M_{|\MIS| -1}$ sum to one by conservation of probability. For non-maximal $|\MIS| -1$ configurations, the rows (and therefore columns) sum to $1-O(1/N)$, because probability can leak to MISs via $M_0.$  Putting everything together, the $(uM_{|\MIS| -1})_s$ component of the resulting vector, corresponding to spin configuration $s$, is given by
\begin{align}
	(uM_{|\MIS| -1})_s &= \frac{1}{{{\DMISminusone}}}\begin{cases} 1 & s \text{ maximal}\\
		1-O(\frac{1}{N}) & s \text{ non-maximal}
	\end{cases}\\
	&= u_s + O(N^{-1}\mathcal{HP}^{-1}),
\end{align}
where we have used the fact that the correction is given by the of non-maximal states to $\DMISminusone$ is $O(\mathcal{HP}^{-1})$. Therefore, $u$ is an eigenvector of $M_{|\MIS| -1}$ up to $O(N^{-1}\mathcal{HP}^{-1})$ corrections. 
The corresponding eigenvalue is therefore $1-O(N^{-1}\mathcal{HP}^{-1})$, so $\delta_{u}^{\text{SA}} = O(N^{-1}\mathcal{HP}^{-1})$. Therefore, we conclude that the hitting time is $\Omega(\mathcal{HP})$, where the factor of system size is absorbed into the definition of depth, consistent with the Main Text.  \\ 

\noindent\textbf{Functional form for MIS probability. }
Numerically, we find that the equation $\PMIS=1-e^{-\text{const.}\times p_{\text{SA}}}$, where $p_{\text{SA}}$ is the depth of SA, is a very good fit to the MIS SA data. Example fits to the data can be seen in Figure~\ref{fig:LZ_benchmarking}B. We find that the constant in this expression is, to good approximation, the numerically computed minimum energy gap of the MIS SA Markov chain in Figure~\ref{fig:LZ_benchmarking}C.

This functional form can be motivated at zero temperature, assuming SA enters the $|\MIS|-1$ subspace at a random independent set of size $|\MIS|-1$. Then, the initial distribution has $1-O(\mathcal{HP}^{-1})$ overlap with the eigenvector of $M_{|\MIS|-1}$ discussed in the previous section and $O(\mathcal{HP}^{-1})$ overlap with other eigenvectors.
The general form for $P_\MIS$ at depth $p_{\text{SA}}$ is given by 
\begin{equation}
	\PMIS = 1-vM^{p_{\text{SA}}}\mathbf{1}_{|\MIS|-1},
\end{equation}
where $v$ is a vector representing the initial configuration and $\mathbf{1}_{|\MIS|-1}$ is a vector of all ones in the $|\MIS|-1$ subspace and all zeros in the $|\MIS|$ subspace. 
This equation comes from taking the initial state $v$, evolving for $p_{\text{SA}}$ steps under the Markov Chain, then taking the resulting overlap with the $|\MIS|-1$ subspace to get $1-\PMIS.$
In the specific case where $v$ is uniform in the $|\MIS|-1$ subspace and has high overlap with the eigenvector from the previous section, this reduces to
$\PMIS = 1-e^{-\delta_{u}^{\text{SA}}p_{\text{SA}}}$
up to small $O(\mathcal{HP}^{-1})$ corrections, motivating the functional form fit in Figure~\ref{fig:LZ_benchmarking}B. All dynamics will therefore be exponential relaxation with a single timescale given by $\delta_{u}^{\text{SA}}$.

Numerical evidence confirms that the principal eigenvector of $M_{|\MIS|-1}$ has high overlap with the uniform distribution, so $\delta_{u}^{\text{SA}}=\delta_{\min}^{\text{SA}}$. This is consistent with the following analytic argument that the uniform distribution $u$ should be close to the principal eigenvector of $M_{|\MIS|-1}$. Namely, $M_{|\MIS|-1}$ is a $O(\mathcal{HP}^{-1})$ perturbation of the stochastic matrix $M_{|\MIS|-1}'$, which comes from taking  $M_{|\MIS|-1}$ and adding probability to the diagonal entries of non-maximal $|\MIS|-1$ states such that all rows sum to one. Because $M_{|\MIS|-1}'$ is stochastic and symmetric, its principal eigenvector is uniform with eigenvalue one. Therefore, the principal eigenvector of $M_{|\MIS|-1}$ is uniform up to $O(\mathcal{HP}^{-1})$ corrections unless there are crossings in the eigenvalues as the perturbation is added.
This suggests that the MIS probability is given by \begin{equation}
	\PMIS = 1-e^{-\delta_{\min}^{\text{SA}}p_{\text{SA}}}\label{eq:pmis_vs_gap}
\end{equation}
up to $O(\mathcal{HP}^{-1})$ corrections, assuming that MIS SA enters the $|\MIS|-1$ subspace uniformly at random. This functional form is well-supported numerically in Figure~\ref{fig:LZ_benchmarking}C.

\begin{figure}
	\centering
	\includegraphics[width=5.8in]{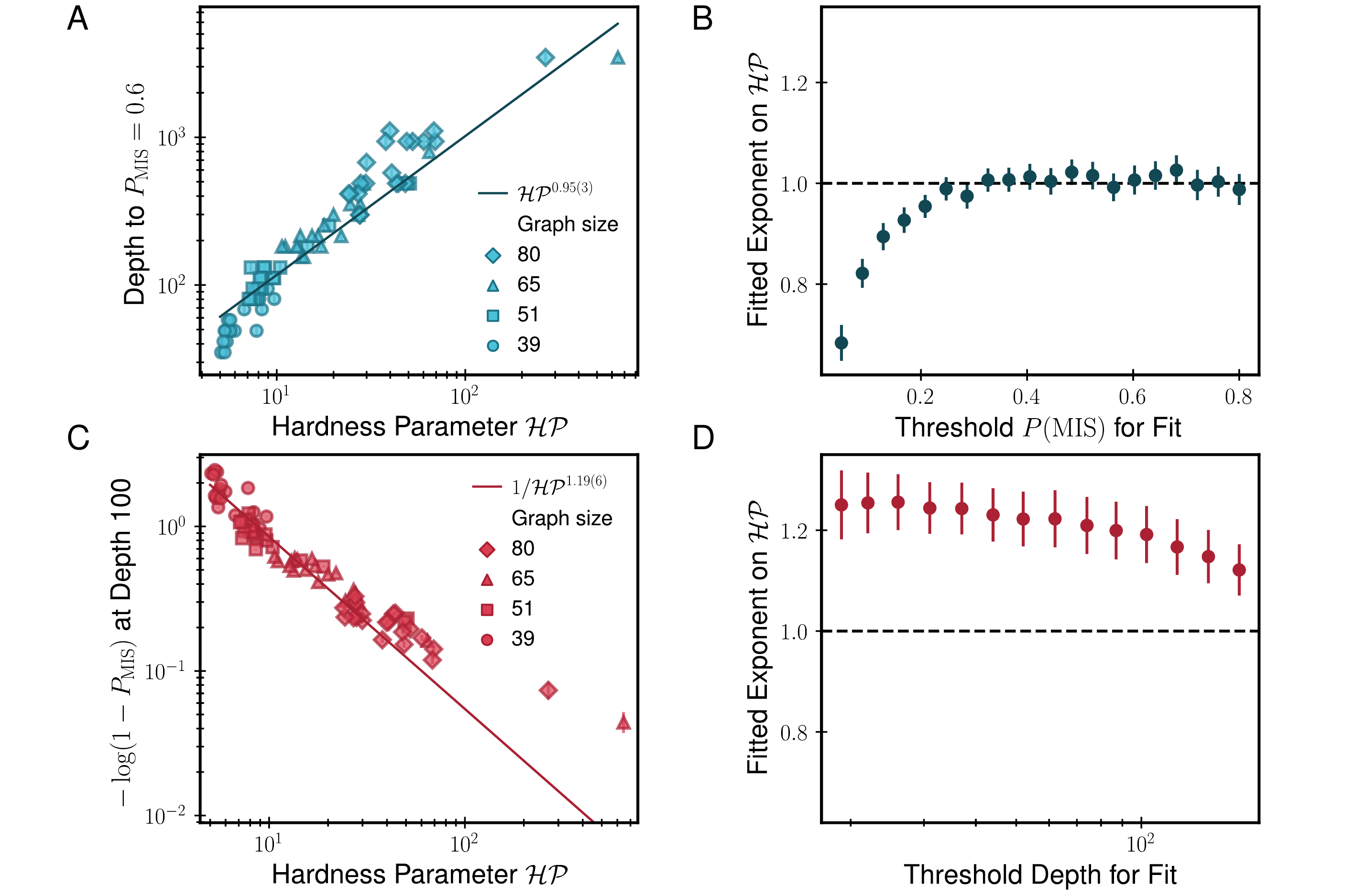}
	\caption{\textbf{Fit stability for MIS probability versus hardness parameter. }\textbf{A.} Depth $p_{\text{SA}}$ required to reach the threshold $\PMIS=0.6$ for MIS SA versus the hardness parameter $\mathcal{HP}$. The data is fitted to an exponential functional form where the spectral gap is assumed to be a power law in $\mathcal{HP}$,  $\PMIS=1-e^{-a\,\mathcal{HP}^{-b}\,p_{\text{SA}}}$. \textbf{B.} The exponent $b$ is then plotted for a range of threshold $\PMIS$s. \textbf{C.} A different way to fit the same functional form, now fitting $-\log(1-\PMIS)$ at a fixed threshold depth ($p_{\text{SA}}=100$) versus $\mathcal{HP}$. \textbf{D.} The fitted exponent $b$ versus the threshold depth used for the fit.}
	\label{fig:SA_fit_benchmarking}
\end{figure}

Motivated by the upper bound of $\mathcal{HP}^{-1}$ on the spectral gap, we phenomenologically assume that the spectral gap is a power law in $\mathcal{HP}$, $a \mathcal{HP}^b$.  Figure~\ref{fig:LZ_benchmarking}A shows that this is a reasonable assumption, albeit with some scatter in the trend. In the main text, we determine the scaling of MIS SA performance with $\mathcal{HP}$ by fitting the parameters $a, b$, and comparing $b$ for the quantum and classical algorithm. We investigate the quality of these fits in
Figure \ref{fig:SA_fit_benchmarking}. Figure \ref{fig:SA_fit_benchmarking}A, C show two different ways to fit the  functional form
\begin{equation}
	\PMIS=1-e^{-a\,\mathcal{HP}^{-b}\,p_{\text{SA}}}.\label{eq:full_functional_LZ}
\end{equation}
In~\ref{fig:SA_fit_benchmarking}A, we  fix $\PMIS=0.6$ and fit $p_{\text{SA}}$ versus $\mathcal{HP}$. The errorbars on the data points are the difference in sampled time points between which the threshold $\PMIS$ is reached. We then fit the data for a range of threshold $\PMIS$ in ~\ref{fig:SA_fit_benchmarking}B to test the robustness of the fit to different threshold depths. We see that at sufficiently large threshold $\PMIS$, the fit approaches a stable linear dependence on $\mathcal{HP}$, with $b=1.$ %The deviation from this trend at smaller $\PMIS$ could be due to the fact that corrections to Eq.~\eqref{eq:pmis_vs_gap} 
In~\ref{fig:SA_fit_benchmarking}C, we  fix depth and fit $\PMIS$ versus $\mathcal{HP}$. In ~\ref{fig:SA_fit_benchmarking}D, we find that the fitted values for $b$ are larger than one, but become closer to one at larger threshold depths. We attribute this discrepancy in the fits to the fact that we do not account for errors due to model uncertainties in Eq.~\eqref{eq:full_functional_LZ}, which primarily stem from the scatter in $\mathcal{HP}$ in MIS SA performance and spectral gap.

\subsection{Quantum scaling}

\noindent \textbf{Density-matrix renormalization group.}
In order to find the ground states of a quantum Hamiltonian $H$, we employ the density-matrix renormalization group (DMRG) algorithm \cite{white_density_1992, white_density-matrix_1993}, which we implement using the ITensor package \cite{itensor}. The desired wavefunction can be represented as a matrix product state (MPS) \cite{mcculloch2007density, verstraete2008matrix} of the form
\begin{equation}
	\vert \Psi \rangle = \sum_{\sigma_1\ldots\sigma_n} \sum_{l_1\ldots l_{n-1}} \mc{T}_{l_1}^{\sigma_1}\,\mc{T}_{l_1l_2}^{\sigma_2}\,\mc{T}_{l_2l_3}^{\sigma_3}\cdots \mc{T}_{l_{n-1}}^{\sigma_n}\, \vert \sigma_1,\ldots,\sigma_n \rangle,
\end{equation}
where $\mc{T}$ denote tensors with physical indices $\sigma$ and link indices $l$. DMRG then provides an efficient method to find the optimal MPS representation of the many-body state \cite{schollwock_density-matrix_2011-1}.

In this work, we obtain the low-lying eigenstates for a variety of graphs with system sizes ranging from $N=39$ to $80$ atoms using MPSs of bond dimensions $d = 200$--$1600$, with $d$ progressively increased as necessary till convergence. The system is regarded to have converged to its true ground state once the truncation error falls below a threshold value of $10^{-7}$, and in practice, this criterion was usually found to be satisfied after $\sim \mathcal{O}(10^2)$ sweeps. For further details of our sweeping procedure and DMRG parameters, we direct the reader to Ref.~\cite{samajdar_complex_2020}. Once a ground state $\rvert \psi_0 \rangle$ is obtained in this manner, we can also target the first-excited state by repeating this procedure but with the Hamiltonian $H' = H + w P_0$, where $P_0 = \rvert \psi_0 \rangle \langle \psi_0 \lvert$ is an operator that projects onto the ground state and $w$ is an energy penalty. The gap at any point in the $(\Delta/\Omega, R_b/a)$-parameter space is obtained from the difference in the energies of the first-excited and ground state; scanning all possible values of $\Delta/\Omega$ for a fixed $R_b/a=1.73$, we record the minimum gap thus obtained as the adiabatic gap relevant for the Landau-Zener transition.\\

\noindent\textbf{Effect of finite blockade and long-ranged interactions.}
Employing the above-mentioned procedure, we now calculate the minimum energy gaps for several for three different Hamiltonians. 

\begin{figure}[tb]
	\centering
	\includegraphics[width=1.\linewidth]{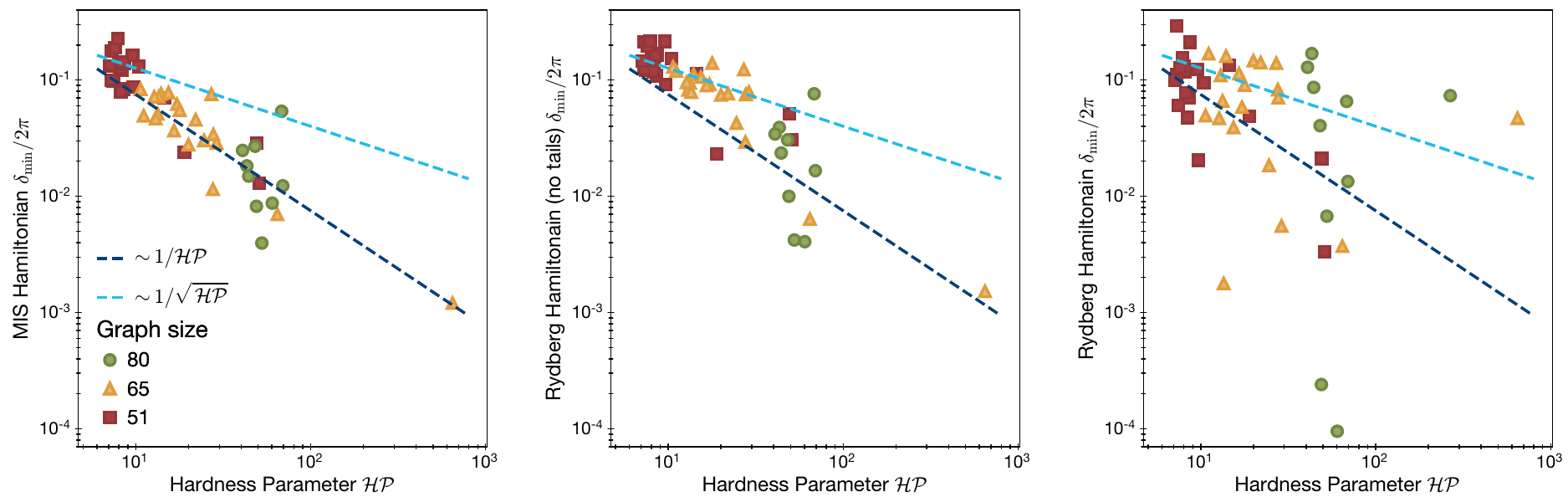}
	\caption{\textbf{Minimum energy gap versus hardness parameter.}
		The minimum quantum gap calculated using DMRG for
		\textbf{A.} The ``hard blockade" model of the  Hamiltonian in Eq.~\eqref{eq:H_approx} with first- and second-nearest neighbors strongly blockaded and no tails beyond them.
		\textbf{B.} The realistic Rydberg Hamiltonian, but with the interaction truncated to only first- and second-nearest neighbors and no tails beyond them. 
		\textbf{C.} The realistic Rydberg Hamiltonian with long-ranged tails described in Eq.~(1) of the main text.  }
	\label{fig:GapVSHardness}
\end{figure}

First, in Fig.~\ref{fig:GapVSHardness}A, we present the minimum quantum gap calculated for a ``hard blockade" Hamiltonian \textit{without} long-ranged tails, in which both the first- and second-nearest neighbors are strongly blockaded:
\begin{equation}
	\label{eq:H_approx}
	\tilde{H} = \sum_{i}\left(\frac{\Omega}{2} \ket{0}^{}_{i}\bra{1}^{}_{i} + {\rm h.c.}\right) - \Delta \sum_{i} n^{}_i
	+ \frac{1}{2}\sum_{\langle i,j\rangle}V^{}_{0}n^{}_i n^{}_j
	+ \frac{1}{2}\sum_{\langle\langle i,j\rangle\rangle}V^{}_{0}n^{}_i n^{}_j,
\end{equation}
where $\langle \cdots \rangle$ and $\langle\langle \cdots \rangle\rangle$ represent nearest and next-nearest neighbors, respectively, and we have set $\hbar=1$.
For this Hamiltonian, which may be regarded as an approximation to the hard-core, infinite Rydberg blockade for $V_0=V_1 \approx 27,\, \Omega = 1$, we find that %much of the scatter in Fig.~\ref{fig:GapVSHardness}B,C is reduced and that 
the quantum hardness (the inverse gap) correlates well with the classical hardness parameter $\mathcal{HP}$, consistent with a scaling of gap $\sim 1/\mathcal{HP}$. 

Next, we show the minimum energy gap for a more realistic soft blockade Hamiltonian without long-ranged tails in Figure~\ref{fig:GapVSHardness}B:
\begin{equation}
	\label{eq:H_approx}
	\tilde{H} = \sum_{i}\left(\frac{\Omega}{2} \ket{0}^{}_{i}\bra{1}^{}_{i} + {\rm h.c.}\right) - \Delta \sum_{i} n^{}_i
	+ \frac{1}{2}\sum_{\langle i,j\rangle}V^{}_{0}n^{}_i n^{}_j
	+ \frac{1}{2}\sum_{\langle\langle i,j\rangle\rangle}V^{}_{1}n^{}_i n^{}_j,
\end{equation}
where now $V_0$ and $V_1$ are the interaction energies between nearest and next-nearest neighbors in the Rydberg Hamiltonian, but all longer range interactions are removed. This is an ``intermediate'' model between the hard blockade model described above and the full Rydberg Hamiltonian. Notably, for many instances the adiabatic gap becomes larger in this model compared to the hard blockade mode,  with most instances falling in between  $1/\mathcal{HP}$ and $1/\sqrt{\mathcal{HP}}$ scaling.

Lastly, we consider the full Rydberg Hamiltonian introduced in Eq.~(1) of the main text; for this case, we retain the long-ranged tails of the $1/r^6$ van der Waals interaction up to a distance of 4$a$, which was shown to be sufficient for convergence of phase boundaries on the square lattice \cite{kalinowski2021bulk}. Figure~\ref{fig:GapVSHardness}C shows the minimum quantum gap plotted as a function of the classical hardness parameter $\mathcal{HP}$.
The significant scatter precludes the observation of any meaningful trend based on this data alone. However, the comparison also shows that the a combination of soft blockade and long-ranged tails help facilitate a better performance for many instances.\\

\noindent\textbf{Sufficient conditions for quadratic speedup. }
Although the numerics discussed above do not provide a definitive conclusion for the scaling of the gap  with hardness parameter, 
%$\sim\mathcal{HP}^{-1}$ on the graphs we study, 
it is interesting to consider the conditions under which they  scale as $O(\text{poly}(1/N)\mathcal{HP}^{-1/2})$, realizing a Grover-like speedup over SA, up to polynomial factors in the system size.
To this end, we denote the instantaneous eigenstates of the system as $\ket{1},\dots, \ket{2^N}$ ordered by eigenenergies ($E_1\leq E_2\leq \dots \leq E_{2^N}$), so that 
the minimum adiabatic gap is
\begin{equation}
	\delta_{\min}=\min_{t\in[0,T]}\left(E_2(t)-E_1(t)\right).
\end{equation}
The adiabaticity criterion is then that the total evolution time $T$ satisfies $T\gg 1/\delta_{\min}$.

For some hard combinatorial optimization problems,
in the limit of large system sizes, this minimum gap is expected to coincide with a first-order phase transition, where the ground state suddenly changes character across the transition point~\cite{young_2010}.
We will
assume that this is the case
and parametrize the adiabatic ramp by the drive-to-detuning ratio $\lambda=\frac{\Omega}{2\delta}$, denoting its value at this phase transition as $\lambda_{\rm{crit}}$.

\begin{figure}
	\centering
	\includegraphics[width=2in]{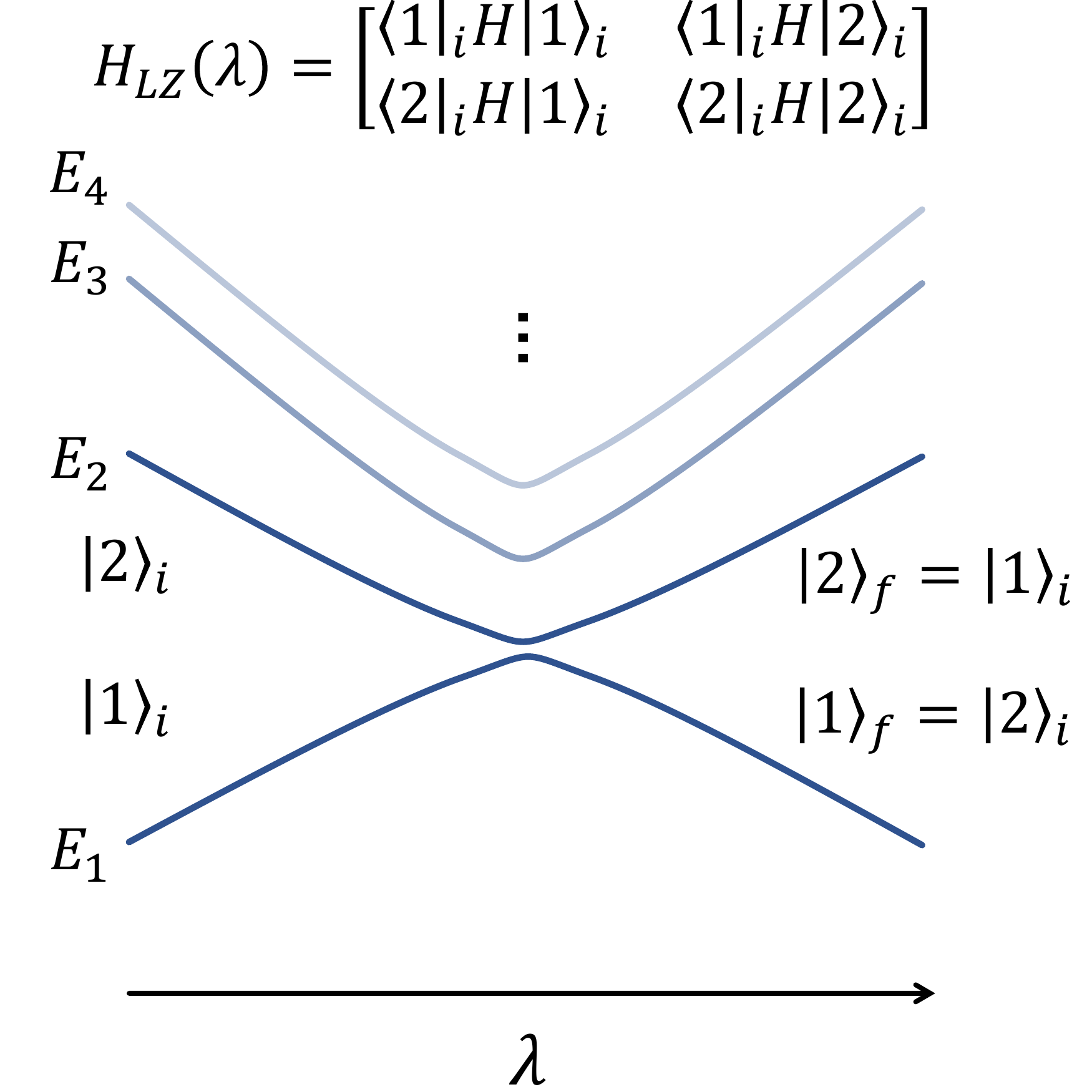}
	\caption{\textbf{Landau-Zener physics in many-body systems.} A level anti-crossing in a many-body system, where the instantaneous eigenstates swap at the gap closing point.  }
	\label{fig:LZ_crossing}
\end{figure}

We will also assume that the two lowest eigenstates are energetically well-isolated from higher excited states ($E_2-E_1\ll E_3-E_2$ near the gap closing point). In this case, the system's dynamics near the gap closing are well-described (up to corrections of order $(E_2-E_1)/(E_3-E_2)$) by a process with
Landau-Zener physics between the lowest two eigenstates, $\ket{1}$ and $\ket{2}$.
Figure~\ref{fig:LZ_crossing} shows such a scenario, where the lowest two eigenstates $\ket{1}_i, \ket{2}_i$ before the level crossing swap places at  $\lambda_{\rm{crit}}$, such that the states after the crossing are $\ket{1}_f = \ket{2}_i$ and $\ket{2}_f = \ket{1}_i$. For $\lambda=\lambda_{\rm{crit}}$, the eigenstates are the hybridized states $\ket{1}=\frac{1}{\sqrt{2}}(\ket{1}_i+\ket{2}_i), \ket{2}=\frac{1}{\sqrt{2}}(\ket{1}_i-\ket{2}_i)$. The system's dynamics in the $\ket{1}_i,\ket{2}_i$ subspace are governed by an effective Landau-Zener Hamiltonian
\begin{equation}
	H_{\rm{LZ}}=\delta_{\rm{eff}}\left(\lambda\right)\sigma_z^{\rm{eff}}+\Omega_{\rm{eff}}\left(\lambda\right)\sigma_x^{\rm{eff}},
\end{equation}
where $\sigma_x^{\rm{eff}}=\ket{1}_i\bra{2}_i+\mathrm{h.c.}$,  $\sigma_z^{\rm{eff}}=\ket{1}_i\bra{1}_i-\ket{2}_i\bra{2}_i$, and $\delta_{\rm{eff}}(\lambda),\Omega_{\rm{eff}}(\lambda)$ are unknown functions, such that $\delta_{\rm{eff}}(\lambda_{\text{crit}})=0$.

Under these assumptions, we find the adiabatic gap as 
\begin{equation}
	\delta_{\rm{\min}}=\Omega_{\rm{eff}}\left(\lambda_{\rm{crit}}\right)\left[1+O\left(\frac{E_2-E_1}{E_3-E_2}\right)\right]
\end{equation} 
where
\begin{equation}
	\Omega_{\rm{eff}}\left(\lambda_{\rm{crit}}\right)=\bra{1}_i H_{\rm{LZ}}\left(\lambda_{\rm{crit}}\right)\ket{2}_i=\bra{1}_i H\left(\lambda_{\rm{crit}}\right)\ket{2}_i\left[1+O\left(\frac{E_2-E_1}{E_3-E_2}\right)\right],
\end{equation}
where $H(\lambda)$ is the system Hamiltonian, and the $(E_2-E_1)/(E_3-E_2)$ corrections in the rightmost expression appear because $H_{\rm{LZ}}$ only approximately represents $H$ in the $\ket{1}_i,\ket{2}_i$ subspace.
Ignoring these corrections, 
we see that the size of the gap is determined by the overlap of the two asymptotic Landau-Zener states of the bare system Hamiltonian $H$ at the critical point.
This Hamiltonian is just a sum of local one- and two-body terms, while the asymptotic Landau-Zener states $\ket{1}_i,\ket{2}_i$ can be highly entangled superpositions of many independent sets. The size of the gap is thus controlled mainly by how much population in $\ket{1}_i$ is close in Hamming distance to population in $\ket{2}_i$, rather than by the operators appearing inside the matrix element.

We can now envision a concrete situation where the adiabatic algorithm exhibits a quadratic speedup over SA. For hard instances, the minimum gap typically occurs close to the end of the adiabatic evolution. The ground state after the critical point ($\ket{1}_f$) is composed of mostly a superposition of the MISs and the first excited state ($\ket{2}_f$) is a superposition of independent sets of size $|\MIS|-1$. Suppose these asymptotic Landau-Zener states form equal superpositions of the respective independent set states:
\begin{align}
	&\ket{1}_i=\ket{2}_f=\frac{1}{\sqrt{\DMISminusone}}\sum_{i\,\in\,\{\text{IS of size } |\MIS|-1\}}\ket{i}, \\
	&\ket{2}_i=\ket{1}_f=\frac{1}{\sqrt{\DMIS}}\sum_{i\,\in\,\{\rm{MIS}\}}\ket{i}.
\end{align}
%With this assumption, at the gap closing, the ground state transitions from a superposition of all MIS-1s to a superposition of all MISs.
In this situation, the gap that results from the above formula is
\begin{equation}
	\delta_{\min}=\Omega\left(\lambda_{\text{crit}}\right)|\text{MIS}|\sqrt{\frac{\DMIS}{\DMISminusone}}.
\end{equation}
Here, only the driver terms in $H$ contribute to the matrix element, since $\ket{1}_{i}$ and $\ket{2}_{i}$ have different Hamming weights. %The factor of $\frac{\Omega}{2\delta}_{\rm{crit}}$ is due to the prefactor of the driver. 
The factor of $|\rm{MIS}|$ appears because the driver connects every MIS in $\ket{2}_i$ to $|\rm{MIS}|$ independent sets of size $|\MIS|-1$ in $\ket{1}_i$. The factor under the square root comes partly from the normalization of the two wavefunctions, which contributes $1/\sqrt{\DMIS \DMISminusone}$, and also from the fact that there are $\DMIS$ nonzero, equal magnitude terms that add constructively. This last effect is a coherent enhancement of $\delta_{\min}$ which stems from the coherence of the superpositions $\ket{1}_i,\ket{2}_i$ that we have assumed. If we assume further that $\lambda_{\rm{crit}}$ scales polynomially or slower with system size, then the gap would go as $\text{poly}(1/N)\mathcal{HP}^{-1/2}$.
%Because of this enhancement, 
In this hypothetical case, the adiabatic algorithm would exhibit a quadratic speedup in $\mathcal{HP}$ over SA, which has a spectral gap of $\text{poly}(N)\mathcal{HP}^{-1}$.

For any particular graph instance, the two lowest eigenstates at the gap closing point are unlikely to exactly equal the fully symmetric superpositions assumed above because the Hamiltonian is not symmetric among the MISs or the independent sets of size $|\MIS|-1$.
%At this point, however, it is evident
It is evident, however,
that the gap
%scaling will be similar if
is mainly determined by the extent to which 
$\ket{2}_i$ is ``delocalized'' across the space of MISs, and by how much overlap the state $\ket{1}_i=\ket{2}_f$ has with the states immediately accessible from $\ket{2}_i$ via the driver. Therefore, a speedup over SA is possible if the eigenstates involved in the gap closing of the adiabatic algorithm are sufficiently delocalized across the solution space, or if $\ket{1}_i$ is localized near nonmaximal independent sets of size $|\MIS|-1$.
Assessing whether this type of speedup can be obtained requires further theoretical analysis of the low-energy states of this Hamiltonian.

%\newpage
%\setcounter{page}{13}
%\textbf{\LARGE References}
%\doublespacing
%\bibliographystyle{Science}
%\bibliography{References_MIS_main_and_SI.bib}
%\pagenumbering{gobble}

\end{document}